\documentclass[10pt,twocolumn]{article}%
\usepackage[utf8]{inputenc}
\usepackage[T1]{fontenc}
\usepackage[english]{babel}
\usepackage{amsmath,amssymb,amsfonts,amsthm}
\usepackage{graphicx}
\usepackage{fancyhdr}
\usepackage{hyperref}
\usepackage{longtable}
\usepackage{listings}
\usepackage{color}
\usepackage{multirow}
\usepackage{wrapfig}
\usepackage{enumitem}
\usepackage{footnote}
\usepackage{tikz,times}
\usepackage{pdflscape}
\usepackage{bbm}
\usepackage{multirow}
\usepackage{comment}
\usepackage{hhline}
\usepackage{graphicx}
\usepackage{caption,subcaption}
\usepackage{cleveref}
\Crefname{equation}{Eq.}{Eqs.}
\Crefname{figure}{Fig.}{Figs.}
\Crefname{tabular}{Tab.}{Tabs.}
\Crefname{table}{Tab.}{Tabs.}

\usepackage{colortbl}
\usepackage{siunitx}
\usepackage{booktabs}
\usepackage{csvsimple}
\usepackage{datatool}
\usepackage{array}
\usepackage{multibib}
\newcites{SM}{SM References}
\newcolumntype{L}[1]{>{\raggedright\let\newline\\\arraybackslash\hspace{0pt}}m{#1}}
\newcolumntype{C}[1]{>{\centering\let\newline\\\arraybackslash\hspace{0pt}}m{#1}}
\newcolumntype{R}[1]{>{\raggedleft\let\newline\\\arraybackslash\hspace{0pt}}m{#1}}

\usetikzlibrary{mindmap,backgrounds}

\makesavenoteenv{tabular}
\makesavenoteenv{table}

\newcommand{\lenet}{\texttt{LeNet}}
\newcommand{\xresnet}{\texttt{XResNet}}

\definecolor{1color}{HTML}{1F77B4}
\definecolor{2color}{HTML}{FF7F0E}
\definecolor{3color}{HTML}{2CA02C}
\definecolor{4color}{HTML}{D62728}
\definecolor{5color}{HTML}{9467BD}

\definecolor{codegreen}{HTML}{2CA02C}
\definecolor{codegray}{rgb}{0.01,0.01,0.01}
\definecolor{codepurple}{HTML}{1F77B4}
\definecolor{backcolour}{rgb}{0.95,0.95,0.95}

\newcommand\heading[1]{{\textbf{#1}}}

\lstdefinestyle{mystyle}{
	language=Python,
    backgroundcolor=\color{backcolour},   
    commentstyle=\bfseries\color{codegreen},
    keywordstyle=\bfseries\color{codepurple},
    numberstyle=\tiny\color{codegray},
    stringstyle=\bfseries\color{codegreen},
    basicstyle=\footnotesize\ttfamily,
    breakatwhitespace=false,         
    breaklines=true,                 
    captionpos=b,                    
    keepspaces=true,                 
    numbers=none,                    
    numbersep=5pt,                  
    showspaces=false,                
    showstringspaces=false,
    showtabs=false,                  
    tabsize=2,
    frame=tb
}
 
\usepackage{array}

\usepackage{lineno}
\usepackage{tikz}
\usetikzlibrary{shapes, arrows.meta, positioning, decorations.pathreplacing, arrows}
\usepackage[margin=0.75in]{geometry}
\usepackage[affil-it]{authblk} 
\usepackage{wrapfig}
\usepackage{soul}
\usepackage{xcolor}
\usepackage[normalem]{ulem}

\newif\ifdraft
\draftfalse 
\ifdraft
    \newcommand\COMMENTEDIT[2]{{{{#2}}}}
    \newcommand\REVIEWERTWO[2]{{#2}}
    \newcommand\REVIEWERFOUR[2]{{\color{orange!80!black}\sout{#1}} {\color{orange!80!black}{{#2}}}}
\else
    \newcommand\COMMENTEDIT[2]{{{{#2}}}}
    \newcommand\REVIEWERTWO[2]{{#2}}
    \newcommand\REVIEWERFOUR[2]{{#2}}
\fi

\author[1]{Patrick Wagner}
\author[1, 2]{Temesgen Mehari}
\author[3]{Wilhelm Haverkamp}
\author[4, *]{Nils Strodthoff}
\affil[1]{Fraunhofer Heinrich Hertz Institute, Berlin, Germany\\
    \texttt{patrick.wagner@hhi.fraunhofer.de}}
\affil[2]{Physikalisch-Technische Bundesanstalt, Berlin, Germany\\
    \texttt{temesgen.mehari@ptb.de}}
\affil[3]{Charit\'e Universit\"atsmedizin Berlin, Berlin, Germany\\
      \texttt{wilhelm.haverkamp@dhzc-charite.de}}
\affil[4]{Carl von Ossietzky Universit\"at Oldenburg, Oldenburg, Germany\\
     \texttt{nils.strodthoff@uol.de}}
\affil[*]{Corresponding author}

\begin{document}

\title{Explaining Deep Learning for ECG Analysis: Building Blocks for Auditing and Knowledge Discovery}

\twocolumn[
  \begin{@twocolumnfalse}
    \maketitle
    \begin{abstract}

\textbf{Deep neural networks have become increasingly popular for analyzing ECG data because of their ability to accurately identify cardiac conditions and hidden clinical factors. However, the lack of transparency due to the black box nature of these models is a common concern. To address this issue, explainable AI (XAI) methods can be employed. In this study, we present a comprehensive analysis of post-hoc XAI methods, investigating the \REVIEWERTWO{local (attributions per sample)}{glocal (aggregated local attributions over multiple samples)} and global \REVIEWERTWO{(based on domain expert concepts)}{ (concept based XAI)} perspectives. We have established a set of sanity checks to identify saliency as the most sensible attribution method. We provide a dataset-wide analysis across entire patient subgroups, which goes beyond anecdotal evidence, to establish the first quantitative evidence for the alignment of model behavior with cardiologists' decision rules. Furthermore, we demonstrate how these XAI techniques can be utilized for knowledge discovery, such as identifying subtypes of myocardial infarction. We believe that these proposed methods can serve as building blocks for a complementary assessment of the internal validity during a certification process, as well as for knowledge discovery in the field of ECG analysis.}
\end{abstract}
  \end{@twocolumnfalse}
]

\section{Introduction}
\label{sec:introduction}

\heading{AI-enhanced ECG} The electrocardiogram (ECG) is one of the most frequently performed diagnostic procedures \cite{NACMS2016} and plays a unique role in the first-in-line assessment of a patient's cardiac state. Until this day, most ECGs are assessed manually with only limited support through rule-based algorithms implemented in ECG devices subject to well-known limitations \cite{schlapfer2017computer}. In the last few years, this analysis paradigm has started to change drastically as a consequence of evidence from an increasing number of studies, which demonstrate the enormous diagnostic potential of the ECG in combination with modern deep learning methods, see \cite{TOPOL2021785} for a recent perspective. Most remarkably, this does not only apply to the prediction of diagnostic statements as routinely assessed also by cardiologists, from myocardial infarctions \cite{strodthoff2018detecting} over comprehensive prediction of ECG statements \cite{Strodthoff:2020Deep,Kashou2020}, and rhythm abnormalities \cite{Hannun2019} to challenging conditions such as hypertrophic cardiomyopathy \cite{Tison2019}, but includes conditions that are hard or even impossible to infer from an ECG for human experts. For example, a number of recent works demonstrated the ability to infer age and sex \cite{Attia2019}, ejection fraction \cite{Attia2019b}, atrial fibrillation during sinus rhythm \cite{attia2019artificial}, anemia \cite{Kwon2020} or even non-cardiac conditions such as diabetes \cite{Kulkarnibmjinnov-2021-000759} or cirrhosis \cite{Ahn2021}. These very promising findings will eventually transform the role the ECG plays in routine diagnostics with enormous potential for cost savings and improved patient journeys through more precise decisions for follow-up diagnostics or treatments.

\begin{figure}[t!]
    \centering
\begin{tikzpicture}[node distance=0.2cm]

    \newcommand\MINSIZE{.5cm}
    \newcommand\BOXSIZE{2.25cm}
    \newcommand\SEP{2mm}

    \newcommand\RED{red!20}
    \newcommand\GREEN{green!20}
    \newcommand\BLUE{blue!20}
    \newcommand\YELLOW{yellow!20}

    \tikzset{
        styleA/.style={rectangle, draw=black, fill=black!10, thick, text width=\BOXSIZE, rounded corners=3, inner sep=2mm, font=\small},
        styleB/.style={rectangle, draw=black, fill=black!20, thick, minimum size=\MINSIZE, minimum width=\BOXSIZE-0.5cm, rounded corners=3, align=center, font=\footnotesize}
    }

    \newcommand\HEIGHTXONE{5cm}
    \node[styleA, minimum height = \HEIGHTXONE] (X1) at (0,0) {};
    \node[below right, font=\bfseries, text width=\BOXSIZE] (x1) at (X1.north west) {\footnotesize{\Cref{sec:sanity}\&\ref{sec:glocal_xai}: Glocal XAI \\~}};
    \node[styleB, anchor=west] (X2) at ([shift={(\SEP,-.5cm)}]x1.south west) {Attributions}; 
    \node[styleB] (X3) [below=of X2] {Aggregations}; 
    \node[styleB, fill=orange!20, text width=\BOXSIZE-1cm] (X4) [below=of X3] {Verification \& Sanity Check}; 

    \draw [-{Latex[length=1.5mm]}, font=\footnotesize] (X2.east) to [bend left=30] node [above, sloped] (TextNodeX1) {} (X3.east);
    \node [font=\footnotesize] (bla) [right of=TextNodeX1] {\rotatebox{-90}{Segment}};
    \draw [-{Latex[length=1.5mm]}] (X3.south) to  node [above, sloped] (TextNodeX2) {} (X4.north);

    \node[styleA, minimum height = \HEIGHTXONE, anchor=north west] (Y1) at ([shift={(\SEP,0cm)}]X1.north east) {};
    \node[below right, font=\bfseries, text width=\BOXSIZE] (y1) at (Y1.north west) {\footnotesize{\Cref{sec:discovery_methods}: Knowledge Discovery}};
    \node[styleB, anchor=west] (Y2) at ([shift={(\SEP,-.5cm)}]y1.south west) {Attributions}; 
    \node[styleB] (Y3) [below=of Y2] {Beats}; 
    \node[styleB] (Y4) [below=of Y3] {Clustering}; 
    \node[styleB, fill=orange!20] (Y5) [below=of Y4] {Discovery}; 
    
    \draw [-{Latex[length=1.5mm]}, font=\footnotesize] (Y2.east) to [bend left=30] node [above, sloped] (TextNodeY1) {} (Y3.east);
    \node [font=\footnotesize] (bla) [right of=TextNodeY1] {\rotatebox{-90}{Aggreagte~~~~~~~}};
    \draw [-{Latex[length=1.5mm]}, font=\footnotesize] (Y3.east) to [bend left=30] node [above, sloped] (TextNodeY2) {} (Y4.east);
    \node [font=\footnotesize] (bla) [right of=TextNodeY2] {\rotatebox{-90}{~~~~~~~Explore}};

    \draw [-{Latex[length=1.5mm]}] (Y4.south) to  node [above, sloped] (TextNodeY2) {} (Y5.north);

    \node[styleA, minimum height = \HEIGHTXONE, anchor=north west] (Z1) at ([shift={(\SEP,0cm)}]Y1.north east) {};
    \node[below right, font=\bfseries, text width=\BOXSIZE] (z1) at (Z1.north west) {\footnotesize{\Cref{sec:tcav}: Global XAI with Concepts}};
    \node[styleB, anchor=west] (Z2) at ([shift={(\SEP,-.5cm)}]z1.south west) {ECG Features}; 
    \node[styleB] (Z3) [below=of Z2] {Concepts}; 
    \node[styleB] (Z4) [below=of Z3] {CAV\&TCAV}; 
    \node[styleB, fill=orange!20, text width=\BOXSIZE-.5cm] (Z5) [below=of Z4] {Concept Verification}; 
    \draw [-{Latex[length=1.5mm]}, font=\footnotesize] (Z2.east) to [bend left=30] node [above, sloped] (TextNodeZ1) {} (Z3.east);
    \node [font=\footnotesize, xshift=0.25cm] (bla1) [right of=TextNodeZ1] {\rotatebox{-90}{Define~~~}};
    \draw [-{Latex[length=1.5mm]}, font=\footnotesize] (Z3.east) to [bend left=30] node [above, sloped] (TextNodeZ2) {} (Z4.east);
    \node [font=\footnotesize] (bla) [right of=TextNodeZ2] 
    {\rotatebox{-90}{~~~Apply}};
    \draw [-{Latex[length=1.5mm]}] (Z4.south) to  node [above, sloped] (TextNodeZ3) {} (Z5.north);

    \newcommand\HEIGHTXTWO{2.6cm}
    \node[styleA, minimum height = \HEIGHTXTWO, anchor=south east, text width=\BOXSIZE-2*\SEP] (A1) at ([shift={(0,\SEP)}]X1.north east) {};
    \node[styleA, minimum height = \HEIGHTXTWO, anchor=north east, text width=\BOXSIZE-2*\SEP] (AA1) at ([shift={(-\SEP,\SEP)}]A1.north east) {};
    \node[styleA, minimum height = \HEIGHTXTWO, anchor=north east, text width=\BOXSIZE-2*\SEP] (AAA1) at ([shift={(-\SEP,\SEP)}]AA1.north east) {};
    \node[below right, font=\bfseries, text width=\BOXSIZE] (a1) at (AAA1.north west) {\footnotesize{\Cref{sec:local_xai}: Local XAI}};
    \node[styleB, anchor=west, minimum width=\BOXSIZE-0.5cm-2*\SEP] (A2) at ([shift={(\SEP,-.5cm)}]a1.south west) {Sample}; 
    \node[styleB, minimum width=\BOXSIZE-0.5cm-2*\SEP] (A3) [below=of A2] {Attribution}; 
    \draw [-{Latex[length=1.5mm]}, font=\footnotesize] (A2.east) to [bend left=30] node [above, sloped] (TextNodeA1) {} (A3.east);
    \node [font=\footnotesize] (bla) [right of=TextNodeA1] {\rotatebox{-90}{Explain}};
    
    \node[styleA, minimum height = \HEIGHTXTWO-.6cm, anchor=south west, text width=2*\BOXSIZE+3*\SEP] (B1) at ([shift={(0,3*\SEP)}]Y1.north west) {};
    \node[below right, font=\bfseries] (b1) at (B1.north west) {\footnotesize{\Cref{sec:data_models}: AI (Deep Learning)}};
    \node[styleB, anchor=west, minimum width=\BOXSIZE-1cm] (B2) at ([shift={(\SEP,-.5cm)}]b1.south west) {Data}; 
    \node[styleB, minimum width=\BOXSIZE-1cm] (B3) [right=of B2] {Model}; 
    \node[styleB, minimum width=\BOXSIZE-.5cm] (B4) [right=of B3] {Performance}; 
    \draw [-{Latex[length=1.5mm]}, font=\footnotesize] (B2.south) to [bend right=15] node [below, sloped] (TextNodeA1) {Train} (B3.south);
    \draw [-{Latex[length=1.5mm]}, font=\footnotesize] (B3.south) to [bend right=15] node [below, sloped] (TextNodeA1) {Evaluate} (B4.south);

    \draw [-{Latex[length=1.5mm]}] ([shift={(1.45cm,0)}]B1.south) to node [above, sloped] (bla) {} (Z1.north);

    \draw [-{Latex[length=1.5mm]}] (B1.west) to node [above, sloped] (bla) {} (A1.east);
    \draw [-{Latex[length=1.5mm]}] (B1.west) to node [above, sloped] (bla) {} (AA1.east);
    \draw [-{Latex[length=1.5mm]}] (B1.west) to node [above, sloped] (bla) {} (AAA1.east);

    \draw [-{Latex[length=1.5mm]}] (A1.south) to node [above, sloped] (bla) {} (X1.north);
    \draw [-{Latex[length=1.5mm]}] (AA1.south) to node [above, sloped] (bla) {} (X1.north);
    \draw [-{Latex[length=1.5mm]}] (AAA1.south) to node [above, sloped] (bla) {} (X1.north);
    
    \draw [-{Latex[length=1.5mm]}] (A1.south east) to node [above, sloped] (bla) {} (Y1.north);
    \draw [-{Latex[length=1.5mm]}] (AA1.south east) to node [above, sloped] (bla) {} (Y1.north);
    \draw [-{Latex[length=1.5mm]}] (AAA1.south east) to node [above, sloped] (bla) {} (Y1.north);
    
\end{tikzpicture}
\caption{Conceptual summary of the XAI study for ECG: We discuss two different ways of investigating consistent model behavior (1) through aggregation of local attribution maps across entire patient groups in the form of so-called glocal attribution maps, which can also be effectively used for knowledge discovery and (2) by using the global XAI method to verify if cardiologists' expert concepts are consistently exploited.}
    \label{fig:method}
\end{figure}


\heading{Need for XAI} Modern deep learning models owe their superior performance to their large number of millions to billions of parameters, which allow to capture \COMMENTEDIT{complicated combinations}{complex non-linear dependencies} of input features. However, this flexibility comes at the price of the impossibility to interpret such models on a parameter basis, which is why these are often perceived as black boxes. This led to the emergence of explainable AI (XAI) as a subfield of machine learning, which tries to shed light on the decision process implemented by neural networks, see \cite{covert2021explaining,lundberg2017unified,Samek2021, saporta2022benchmarking} for general reviews. Here it is important to keep in mind that different use cases pose different requirements for the combined ML+XAI system (see \Cref{fig:method}\COMMENTEDIT{}{ and \Cref{fig:test}} for an overview). These range from (1) Providing side-information to medical experts (2) Auditing of the ML system before deployment, e.g., to ensure that the models avoid excessive exploitation of spurious correlations or rely on undesired features/principles, see \COMMENTEDIT{\cite{DeGrave2021}}{\cite{DeGrave2021, lapuschkin2019unmasking}} for a recent application (3) Scientific discovery through using the ML model as a proxy for the relations between input and output in the data. While the first use-case is very important, it can eventually only be assessed within user studies \REVIEWERTWO{}{with cardiologists}. \REVIEWERTWO{In particular, we do not attempt to assess whether explanations are insightful for human experts and/or support real world decision making.}{In particular, we do not attempt to assess whether single local explanations are insightful as side-information for human experts as part of a clinical decision support system.} Instead, we focus primarily on the last two aspects, building on attributions as measures for feature relevance, and provide building blocks that can be employed in these contexts. Both topics relate to the frequent question of the alignment of neural network decisions and cardiologists' decision rules\REVIEWERTWO{, which is rarely assessed systematically beyond single hand-picked examples}{}. At this point, we present two methods that provide quantitative evidence of whether a neural network systematically exploits specific decision rules and which segments of the signal are most relevant for the classification decision across all samples with a particular pathology across an entire dataset. In both cases, we find good agreement across a diverse set of four pathologies.

\heading{XAI for ECG} Various XAI methods have been applied to deep learning models trained on ECG data, see \cite{Ayano2022} for a recent review, mostly in the form of ad-hoc adaptations of existing methods \COMMENTEDIT{mostly from}{applied in} computer vision. \COMMENTEDIT{With very exceptions, such as \cite{YAO2020174} or \cite{Elul2021}, who relied on model-inherent attention weights, most approaches involved so-called post-hoc XAI methods applied to trained models}{Most methods utilized post-hoc XAI techniques on trained models, with the exception of \cite{YAO2020174} or \cite{Elul2021}, which relied on the model's inherent attention weights.} \COMMENTEDIT{The clearly most popular approach considered in the ECG literature is GradCAM , used for example by .}{Gradient-weighted Class Activation Mapping (GradCAM)}\cite{Selvaraju2019}\COMMENTEDIT{}{ is considered the most popular approach }\cite{Raghunath2020,Kwon2020,van2021discovering,goodfellow2018towards,hicks2021explaining,lu2024decoding}\COMMENTEDIT{. GradCAM is}{,} followed in popularity by saliency maps \cite{simonyan2013deep}, see \cite{kwon2020deep,Jones9288132,lima2021deep,cho2020artificial}. Other choices include \COMMENTEDIT{LIME}{\textit{Local Interpretable Model-Agnostic Explanations} (LIME)} \cite{Hughes2021}. These techniques provide visually appealing attribution maps that are in most cases used to argue whether the decision criteria of deep neural networks align with human expert knowledge. 

\heading{Main contributions} The main contributions of our work align with four subtopics, which we describe in detail in the following paragraphs:

\begin{enumerate}
\item \heading{Sanity checks}
Prior approaches rarely provided arguments why a particular XAI method was chosen or if it was appropriate for this purpose, which is, however, a crucial prerequisite for the application of XAI methods in the first place. We argue that the ECG is particularly suited for a structured evaluation due to its periodic structure and ECG features as well-defined signal features. This makes it possible to set up precise sanity checks for XAI methods, allowing to assess whether they attribute consistently, both in a temporal as well as spatial manner. As one exemplary finding, we demonstrate that most existing approaches including GradCAM with the exception of saliency maps, fail to attribute in a temporally focused manner, see \Cref{fig:sanitycheck}, which clearly puts into question existing approaches in the field.

\begin{figure*}[!t]
    \centering
    \includegraphics[width=\textwidth]{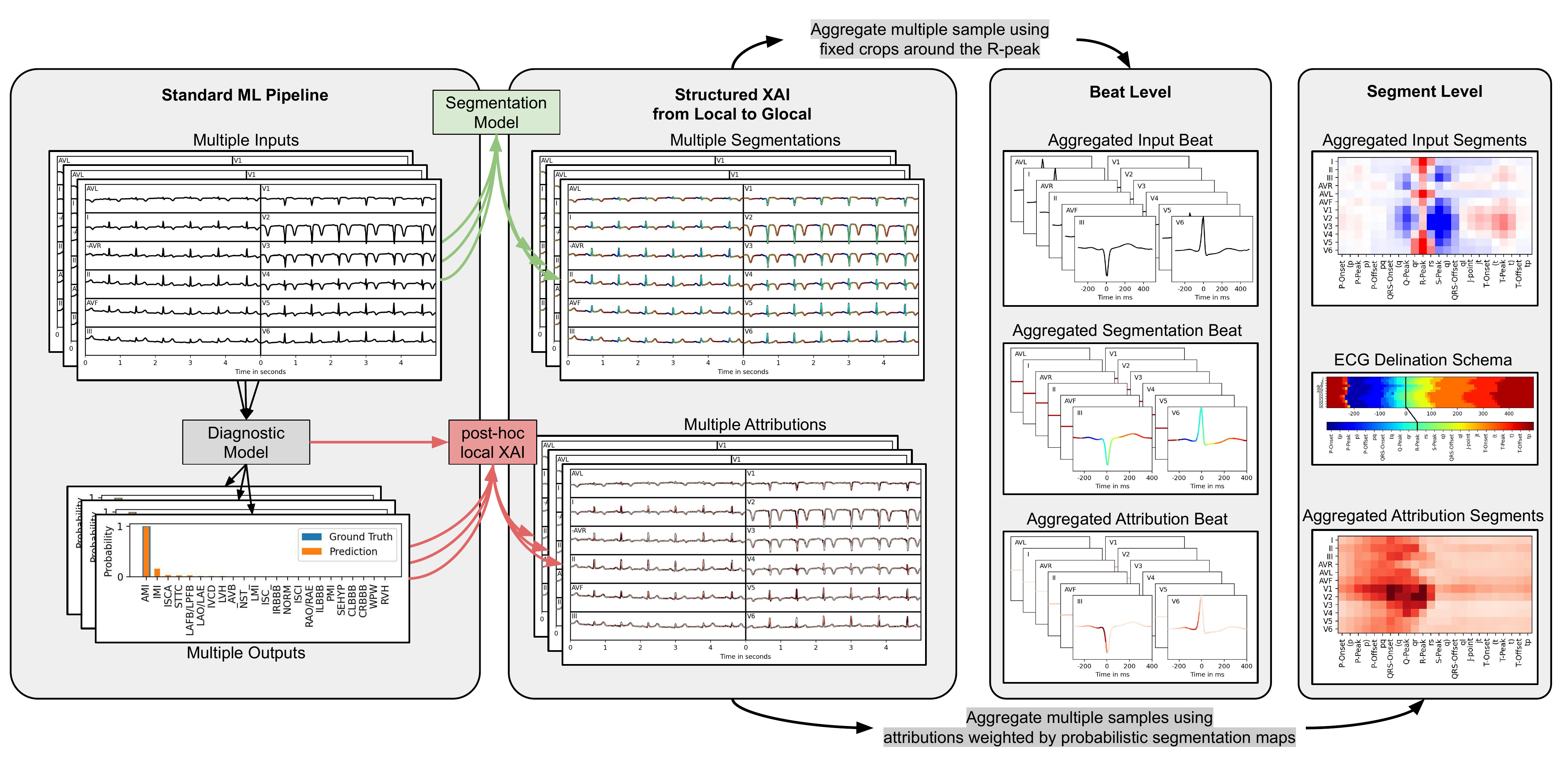}
    \caption{Graphical overview of the proposed aggregation process of local into glocal attributions. First column: Starting point is a deep learning model trained to predict diagnostic classes from raw ECG time series as input. Second column: Local attribution methods provide lead- and timestep-specific attributions for every input sample (bottom). A segmentation model provides probabilistic segmentation maps of the time series into well-established ECG segments. Third column: First possibility to define aligned attributions across several beats and patients is to align signal and corresponding attributions based on fixed crops around the R-peak. Fourth column: Second possibility is to use probabilistic segmentation maps as weighting factors for attribution maps. Performing a dot product between attribution and probabilistic segmentation map along the temporal dimension yields (after appropriate normalization) an aggregated attribution map (bottom) that resolves relevance according to leads and EGC segments. This allows for unprecedented qualitative and quantitative insights into consistent model behavior after aggregation across entire patient subgroups.
    }
    \label{fig:test}
\end{figure*}

\item \heading{``Glocal'' XAI} As second main contribution, we strongly argue for the automated analysis of attributions in a beat-aggregated form, in line with earlier proposals in the literature \cite{Jones9288132,van2021discovering,bender2022analysis}. \REVIEWERTWO{}{In an ECG, a "beat" refers to one complete cardiac cycle, including the P-wave, QRS complex, and T-wave, representing the heart's electrical and mechanical activity.} We exploit the periodic structure of the signal to provide attribution maps on the level of median beats or segments, see \Cref{fig:test} for a graphical overview. These can be compared in a meaningful way across samples and patients. In this way, attribution maps can be aggregated across patients (with similar pathologies) to derive dataset-wide patterns from sample-specific attribution maps. We showcase this approach for three major diagnostic classes and provide the first quantitative evidence for alignment of model behavior with cardiolologists' decision rules on the level of ECG segments and leads, see \Cref{fig:glocal_analysis_xresnet}. 

\item \heading{Global XAI} As third contribution, we aim to stress opportunities for XAI methods beyond attribution maps. To this end, we apply \COMMENTEDIT{TCAV}{\textit{Testing Concept Activation Vectors} (TCAV)} \cite{kim2018interpretability}, which allows to assess if a model exploits a given concept, defined via examples. Again, we argue that the ECG domain represents an ideal application domain for such techniques due to the availability of high-quality ECG features and concepts as defined via cardiologists' decision rules. This represents the first comprehensive assessment of diagnostic concepts in the context of global XAI for ECG. Indeed, for all considered pathologies, we find that expert rules are consistently exploited by the ML models, see \Cref{fig:tcav}.

\item \heading{Knowledge discovery} Finally, through unsupervised learning, we demonstrate that attributions are more informative for subgroup discovery than the respective input signals or high-level model features, see \Cref{fig:discovery}, which highlights a greater level of exploitative information of attribution maps. As an exploratory study, we use this insight to identify clinical meaningful subgroups of anteroseptal myocardial infarctions, see \Cref{fig:discovery_asmi}, which, ultimately, relates back to the diagnostic criteria used to annotate the dataset.
\end{enumerate}
\section{Materials and Methods}

\subsection{Data \& Models}
\label{sec:data_models}
\heading{Data} We base our experiments on the PTB-XL data set \cite{Wagner:2020PTBXL, Wagner2020:ptbxlphysionet,Goldberger2020:physionet}, which covers 21799 ECGs from 18869 patients annotated with ECG statements from a broad set of 71 statements covering diagnostic statements and form- and rhythm-related statements. We focus on the 23 statements at the sub-diagnostic level to allow a sufficiently finegrained analysis without adding the complexity of the full set of 44 diagnostic labels, some of which are only sparsely populated. We follow the benchmarking protocol for PTB-XL proposed in \cite{Strodthoff:2020Deep}, which uses eight of the ten stratified folds for training and the remaining two for validation (model selection) and testing. The ECG features that are used for the concept-based explainability methods (described in \cref{sec:tcav} and shown in \cref{sec:results_tcav}) are extracted by the University of Glasgow ECG analysis program \cite{Glasgow}. These features were made available as part of the PTB-XL+ dataset \cite{Strodthoff:2023PTBXLplus}.

\heading{Models}
In terms of model architectures, we work with convolutional neural networks as they are to date the predominantly used model architecture in the field, see e.g.,\cite{Hannun2019,Ribeiro2020,Attia2019b}, even though recent works suggest that recurrent architectures \cite{Mehari:2021Self} or structured state space models \cite{mehari2022advancing} might further improve over the convolutional state-of-the-art\REVIEWERTWO{.  We}{, most likely due to inductive biases of these models that are better suited to capture the sequential nature of ECG data. In this work,} we consider both a shallow model inspired by the \lenet{} architecture \cite{lecun1989backpropagation} and a ResNet-based \cite{he2016deep} architecture, which was shown to lead to competitive results \cite{Strodthoff:2020Deep} for ECG classification tasks on PTB-XL. The two model architectures are described in more detail below:

\lenet{}: shallow model inspired by the \lenet{} architecture \cite{lecun1989backpropagation}, which is composed of 3 one-dimensional Convolutional Layers (with kernel size 5, stride 2 and output channels 32, 64 and 128, respectively), interleaved with BatchNorm, ReLU and pooling layers (the first two pooling layers are MaxPool and the last one AvgPool). This is followed by two fully-connected layers, again interleaved with ReLU as activation.

\xresnet{}: ResNet-based \cite{he2016deep} architecture, which were shown to lead to competitive results \cite{Strodthoff:2020Deep} for ECG classification tasks on PTB-XL. More specifically, we use a one-dimensional adaptation of the XResNet architecture\cite{he2019bag}, which is described in detail in \cite{Strodthoff:2020Deep}. In our experiments, we use a xresnet1d50 architecture.

\heading{Prediction task} The task is framed as a multi-label classification task and consequently, \REVIEWERTWO{}{we use  class-wise sigmoid activations and} binary cross-entropy \REVIEWERTWO{is used}{} as optimization objective. We work at a sampling frequency of 100~Hz and use randomly cropped input sizes of \REVIEWERFOUR{250}{$T=250$} tokens, corresponding to 2.5 seconds, which were shown to lead to the best results for the given task \cite{mehari2022advancing}. 

\subsection{Post-hoc local XAI with attribution maps}
\label{sec:local_xai} 
\heading{Local XAI} The XAI community has put forward a range of post-hoc interpretability methods, which can be used to assess the attribution  of input features. These methods commonly provide an attribution (map), which shares the shape of the input and indicates the relevance of particular parts of the input sequence for a classification decision at hand. This allows us to temporally resolve the most relevant parts of the sequence but also to identify the most discriminative leads for a particular condition \REVIEWERFOUR{}{ \cite{Kwon2020,goodfellow2018towards,hicks2021explaining,lu2024decoding,kwon2020deep,lima2021deep,cho2020artificial,Hughes2021}}.
We consider and compare four different XAI methods that are either predominantly used in the ECG community or are popular choices in the XAI community. More specifically, we consider GradCAM \cite{Selvaraju2019}, saliency maps \cite{simonyan2013deep}, integrated gradients (IG) \cite{sundararajan2017axiomatic} and layer-wise relevance  propagation (LRP) \cite{Bach2015}. \textit{Saliency} maps explain model predictions by using the norm of their respective input gradients. \textit{GradCAM} attribution maps consist of a (feature depended) weighting of the activation gradient of the respective model prediction. \textit{Integrated Gradients (IG)} creates attribution maps by integrating the input gradient (of a chosen output neuron) from a predefined baseline input to the sample under consideration. \textit{Layer-wise Relevance Propagation (LRP)} propagates the model prediction from the output neuron back to the input. In every layer, each neuron is assigned an attribution value using the LRP rules, adhering to the conservation rule that ensures that the sum of attributions is (approximately) maintained across layers.

\subsection{Sanity checks for attribution methods}
\label{sec:sanity}

\REVIEWERTWO{}{In the field of Explainable Artificial Intelligence (XAI), the incorporation of sanity checks is paramount to maintaining the trustworthiness and authenticity of explanations of AI models. These sanity checks can also be described as XAI metrics \cite{schwalbe2023comprehensive} designed to ensure the quality of explanations. In \cite{doumard2023quantitative}, e.g., the authors propose six metrics to evaluate popular local additive explanation methods. Such assessments play a central role in validating that the explanations generated by XAI accurately represent the model's decision-making process. This validation is critical because inaccuracies in the explanations could potentially mislead users and lead to misunderstandings about the operating principles of the model. In addition, sanity checks are essential to identify any biases within an explanation model, particularly those that may favor certain attributes or characteristics.}

\heading{Proposed experiment} The sanity check \REVIEWERTWO{}{in this study} relies on ECG parameter regression. The intuition behind this experiment is the expectation that a model that was trained to regress a particular amplitude in a particular lead from the raw signal should focus spatially on this particular lead and temporally on the corresponding segment. To implement this experiment, we fit a regression model \REVIEWERTWO{}{$R:\mathbb{R}^{T \times 12}\rightarrow \mathbb{R}^{12}$} to a feature present in each lead (P-wave, R-wave and T-wave amplitude), i.e. a model with 12 output neurons, one for each lead feature. For this, we used the same model architectures introduced above \REVIEWERTWO{}{but with linear activation} and use the mean squared error as optimization objective. As regression targets we use ECG parameters extracted by the University of Glasgow  ECG analysis program \cite{Glasgow}, a commercial ECG analysis software, as available from PTB-XL+\cite{Strodthoff:2023PTBXLplus}. 

\REVIEWERTWO{For a given sample and lead $L$ which we use for parameter regression, we compute an attribution map $(\mathbf{a}^L)_{tl}$ (where $t=1\ldots T$ and $l=1\ldots 12$). In order to compare across different methods, the attribution maps are analyzed both spatially as well as temporally.}{
In order to compare across different attribution methods, we compute attribution maps for all samples and all output leads and analyze them both spatially as well as temporally. That is, for a given sample $\mathbf{x}\in\mathbbm{R}^{T \times 12}$ and a given output lead $l\in\{1\ldots 12\}$ we compute the respective attribution map $\mathbf{a^l}\in\mathbbm{R}^{T \times 12}$. Based on this, we consider two complementary aspects noted as spatial and temporal specificity.}

\REVIEWERTWO{\heading{Spatial specificity}
We define a measure $(s^L)_l$ for the \textit{spatial specificity} as follows:
First, we compute temporally aggregated attributions $(\mathbf{\tilde a}^L)_{l}$ by computing euclidean norm of the attribution map $(\mathbf{a}^L)_{tl}$ along the temporal direction ($t$).
We define the spatial specificity $s^L_l$ as the ratio between temporally aggregated attribution in the target lead, $(\mathbf{\tilde a}^L)_{L}$ and the euclidean norm of the temporally aggregated attributions $(\mathbf{\tilde a}^L)_{l}$ along the remaining channel dimension ($l$). We visualize the median and quantiles over all samples and obtain in this way a spatial specificity score for each lead $L$, which can be visualized in terms of boxplots.       This ratio is close to one if the attribution is specific to the lead in question. We expect a sensible attribution method to be spatially specific.}
{

\heading{Spatial specificity} We define \textit{spatial specificity}  $s^l \in [0,1]$ for a given attribution map $\mathbf{a^l}$ as the ratio
\[s^l = \frac{\mathbf{\tilde a}^l_l}{\sum_{i=1}^{12}{\mathbf{\tilde a}^l_i}},\] where $\mathbf{\tilde a}^l \in \mathbbm{R}^{12}$ is the euclidean norm of $\mathbf{a^l}$ along the temporal axis $\left[\|\mathbf{a^l_1}\| \ldots \|\mathbf{a^l_{12}}\|\right]$, i.e., as the ratio between temporally aggregated attribution in the target lead and all leads. This ratio is close to one if the attribution is specific to the lead in question. We expect a sensible attribution method to be spatially specific. Suppose the model is trained to predict the amplitude of the R-peak in a specific lead. A spatially specific attribution would mean that the model attributes the majority of the prediction to that particular lead. We visualize the median and quantiles over all samples and obtain in this way a spatial specificity statistic for each lead, which can be visualized in terms of boxplots as shown in \Cref{fig:sanitycheck}.
}

\REVIEWERTWO{
\heading{Temporal specificity} We also define a measure $(t^L)_\tau$ for the  \textit{temporal specificity} as described below:
We extract temporally aligned (beat-wise) attributions by cropping from 300ms before until 500ms after the R-peak $(\mathbf{b}^L)_{b\tau l}$, where $b=1,\ldots,B$ enumerates the number of beats in the sample and $\tau=1\ldots 80$.   
We compute spatially aggregated attributions $(\mathbf{\tilde b}^L)_{b\tau}$ by calculating the euclidean norm along the channel axis ($l$) of $\mathbf{b}^L_{b\tau l}$. Then we define the temporal specificity $(t^L)_\tau$ as the median of $(\mathbf{\tilde b}^L)_{b\tau}$ over all beats ($b$) from all samples entering the analysis. We additionally take the median across all leads $L$ as we are not interested in resolving the temporal specificity of single channels.}{

\heading{Temporal specificity} Complementarily, we also compute the \textit{temporal specificity} $\mathbf{t} \in \mathbbm{R}^{80}$ as the spatial mean over all lead-specific temporal means $\mathbf{t} = \frac{1}{12}\sum_{l=1}^{12}\mathbf{t}_l$,
where the lead-specific temporal mean $\mathbf{t}_l$ is computed as follows:
\begin{enumerate}
    \item For each sample- and lead-specific attribution $\mathbf{a^l}$, we crop all $B$ beats 300 ms before and 500 ms after each R-peak, yielding $\mathbf{\hat a^l}\in \mathbbm{R}^{B\times 80 \times 12}$.
    \item We remove the spatial information by first computing the euclidean norm along the spatial axis (i.e. the leads) for each beat, yielding  $\mathbf{\|\hat a^l\|_L}\in \mathbbm{R}^{B \times 80}$.
    \item Finally, we arrive at $\mathbf{t}_l$ by computing the mean attribution across beats $\mathbf{t}_l = \frac{1}{B}\sum_{i=1}^{B}{\mathbf{\|\hat a_i^l\|_L}}$
\end{enumerate}
This temporal specificity $\mathbf{t}$ is computed for all samples entering the analysis and is visualized as line-plots with the median as solid line and the 25\% and 75\% inter-quantile-range as transparent area below as in \Cref{fig:sanitycheck}. We expect a sensible attribution method to be temporally specific with respect to the regressed segment in question, even though context from other segments might be relevant. For instance, if the model is predicting T-wave amplitude, temporal specificity would mean that the attribution is concentrated around the relevant time segment of the T-wave.

In summary, spatial specificity ensures that the attribution is focused on the correct lead, while temporal specificity ensures that the attribution aligns with the relevant temporal segment associated with the regressed parameter in the ECG signal.}

\subsection{From local to glocal XAI}
\label{sec:glocal_xai}
\heading{ECG delineation}
We aim to align attribution maps based on beats, i.e. R-peaks, or alternatively based on ECG segments. Both require an ECG delineation as a first step. For this, we trained a model capable of segmenting 12-lead ECG samples into 24 different segments. For our purpose, we trained a 2d U-Net \cite{ronneberger2015u} with convolutional kernels that span the entire feature axis. The crucial advantage of this approach compared to conventional ECG delineation models is the ability to exploit the consistency of segmentations across several leads as opposed to segmenting all leads individually. As labels we used fiducial points extracted from ECGDeli \cite{Pilia2021} for PTB-XL (as segments of length \COMMENTEDIT{1}{one}) as well as the segments between them, see \Cref{fig:segmentation}, leading to a total of 24 different segments.
In general, given a 12 lead ECG sample of length \REVIEWERFOUR{L}{T} \REVIEWERFOUR{$\mathbf{x} \in\mathbb{R}^{L \times 12}$}{$\mathbf{x} \in\mathbb{R}^{T \times 12}$}, our segmentation model \REVIEWERFOUR{$S: \mathbb{R}^{L \times 12} \rightarrow \mathbb{R}^{L \times 12 \times M}$}{$S: \mathbb{R}^{T \times 12} \rightarrow \mathbb{R}^{T \times 12 \times M}$} computes for each timestamp a soft assignment score for all $M$ segment classes (which can be interpreted as a probability distribution over all \textit{M} segment classes). The model is trained with random patches minimizing the categorical cross-entropy with the Adam optimizer. \REVIEWERTWO{The model showed strong results and showed good overlap with the ground-truth annotations, but also showed more reliable segmentations based on a manual inspection in cases of disagreement with ECGDeli}{The model exhibited strong performance and displayed notable alignment with ground-truth annotations. Furthermore, during manual inspection, it demonstrated more reliable segmentations, especially in cases of disagreement with ECGDeli} \REVIEWERTWO{}{(as shown in \Cref{fig:third} via direct comparison)}. \REVIEWERTWO{We defer a}{A} detailed performance analysis \REVIEWERTWO{}{, both quantitatively and qualitatively,} of this model \REVIEWERTWO{to future work}{ is given in \Cref{sec:appendix_data}}. The ECG-features for the concept-based analysis were taken from the PTB-XL+ dataset \cite{Strodthoff:2023PTBXLplus}. We \REVIEWERTWO{provide}{make the} soft segmentation maps \cite{ptbxl_segmentations} for the entire PTB-XL dataset \REVIEWERTWO{}{underlying our analysis publicly available}. 

\heading{Beat-aligned attributions}
We perform a robust R-peak detection based on the soft predictions for the R-peak class in the segmentation \REVIEWERFOUR{. We perform a peak detection}{} (\REVIEWERFOUR{}{minimum} distance of 30 timestamps \REVIEWERFOUR{}{between two beats} and minimum output probability of 0.25) based on the spatial maximum of V1-V4. The R-peaks identified in this way are part of the data repository released with this paper. We then extract median beats from the signal by extracting the signal 300~ms before until 500~ms after each identified R-peaks \REVIEWERFOUR{}{(similar to previous approaches \cite{ Jones9288132,van2021discovering,bender2022analysis})}. We proceed similarly for the attribution maps to extract corresponding beat-centered attribution maps. We can then aggregate the signal to derive representative median beats across several beats within a given sample but also across entire subsets of patients, which share a common pathology. \REVIEWERFOUR{To illustrate the aggregation procedure, we provide an example of (a 5s segment of) one 12-lead ECG signal together with attributions. In this particular case, we want to demonstrate the following basic principles: 1. 12-lead ECG of arbitrary length (minimum of 2.5 seconds) is diagnosed as \texttt{NORM} with high confidence. 2. on top of the signal we visualize the attribution for this diagnosis. 3. Moreover we highlight the respective R-peaks which are used to crop and compute the median beat of this signal. The center of the plot shows the corresponding median beat superimposed with the corresponding attributions.}{}

For \Cref{fig:glocal_analysis_xresnet} and \Cref{fig:glocal_analysis_lenet}, we perform experiments on entire subgroups of pathological samples by filtering the top 100 model predictions per class as these most clearly express the patterns exploited by the model. 

\heading{Segment-aligned attributions}
For the segment-specific parts of \Cref{fig:glocal_analysis_xresnet} and \Cref{fig:glocal_analysis_lenet}, the idea is to aggregate attributions using the soft segmentation outputs as weighting factors. \REVIEWERTWO{}{We opt for soft segmentations (output probabilities), because they provide robustness in case of small segments (i.e. points at peaks in the signal), as opposed to hard segmentations, where the argmax sometimes fails to recover these small segments.} To this end,
we computed the aggregated maps $\mathbf{m} \in \mathbb{R}^{12 \times M}$ of the raw signal \REVIEWERFOUR{$\mathbf{x} \in\mathbb{R}^{L \times 12}$}{$\mathbf{x} \in\mathbb{R}^{T \times 12}$} and the segmentations \REVIEWERFOUR{$\mathbf{s} \in \mathbb{R}^{L \times 12 \times M}$}{$\mathbf{s} \in \mathbb{R}^{T \times 12 \times M}$} (similar for the attributions) as $\mathbf{m}_i=\mathbf{x}_{:,i}^{\top}\mathbf{s}_{:,i,:}$ for each i in {1,...,12} where each $\mathbf{m}_i\in\mathbb{R}^{M}$ is concatenated to $\mathbf{m}=\left[\mathbf{m}_1,\ldots,\mathbf{m}_{12}\right]$ and normalized by the temporal sum \REVIEWERFOUR{$\mathbf{\tilde s}=\sum_{i=0}^{L}{\mathbf{s}_{i,:,:}}$}{$\mathbf{\tilde s}=\sum_{i=0}^{T}{\mathbf{s}_{i,:,:}}$}. In other words, to compute the entry of the \REVIEWERFOUR{l}{i}-th channel and the m-th segment of the resulting aggregated map, we calculate the dot product of the attributions of the \REVIEWERFOUR{l}{i}-th channel with the segmentation outputs (normalized by their sum, so that they sum up to 1) corresponding to the m-th segment.\\
The two proposed aggregation approaches \REVIEWERFOUR{on beat and segment-level are summarized graphically in \mbox{\Cref{fig:test}}.}{at the beat and segment level, which facilitate ECG beat interpretation based on ECG analysis, are graphically summarized in \Cref{fig:test}.}

\subsection{Knowledge discovery}
\label{sec:discovery_methods}
\REVIEWERFOUR{}{Knowledge discovery using XAI enables a deeper understanding of the patterns underlying model decisions \cite{beer2020using,lu2024decoding,Tison2019,hicks2021explaining} but also insights into the data itself. In this study, we propose an explorative unsupervised approach by clustering aligned attributions to uncover patterns in the data. To evaluate the effectiveness of attributions, we compare them with baseline representations, providing evidence of their potential insights.}

\heading{Representations \REVIEWERTWO{and clustering metrics}{}}
We complement aligned median beats and attributions by hidden feature representations. Here, we use features from layers after the global pooling operation (present in all our considered models) to ensure translation invariance.
\REVIEWERTWO{As performance metrics we report (1) standard accuracy (ACC) after assigning the clusters to classes and (2) adjusted Rand score (ARS) suited for comparing similarities among clustering.}{}

\heading{Clustering procedure \REVIEWERTWO{}{and metrics}}
In order to avoid issues with dimensionality for clustering, all representations were dimensionally reduced via projecting onto the first $k$ principal components that manage to capture 75\% of the total variance. Since each representation differs in multiple unknown properties, there is no single clustering model, which is best-suited for all representations. Instead, we carefully selected models and hyper-parameters for each representation maximizing the clustering scores \REVIEWERTWO{.}{ via grid-search over models (k-\REVIEWERFOUR{mean}{Means}, Gaussian Mixture Models and Hierachial Clustering) and projections (with or without whitening). As performance metrics we report (1) standard accuracy (ACC) after assigning the clusters to classes and (2) adjusted Rand score (ARS) suited for comparing similarities among clustering.}

In this sense, our result is a confirmative study indicating the best possible score (within the range of considered clustering algorithms and hyperparameters) that can be achieved given the ground truth label assignments. The best clustering results were achieved for (1) attributions and (3) hidden features using Gaussian Mixture Models (GMMs), for (2) inputs using k-means.

\heading{Cluster insights}
We demonstrate a structured procedure to gain further insights into the identified clusters. To this end, we plot the (aligned) mean attributions of both clusters and identify temporally as well as spatially localized regions where both deviate most strongly. In a second step, we visualize these regions but this time superimposed on the cluster means in terms of the original signals, see \Cref{fig:cluster_means} for \COMMENTEDIT{MI}{clustering anterior and inferior myocardial infarction (AMI and IMI) as known sub-classes within general myocardial infarction (MI)} or \Cref{fig:discovery_means_resnet}/\Cref{fig:discovery_means_lenet} for \COMMENTEDIT{ASMI}{clustering novel and unknown sub-conditions within anteroseptal myocardial infarction (ASMI)}.


\subsection{Global XAI with concepts}
\label{sec:tcav} 

\heading{TCAV} We base our approach on TCAV \cite{kim2018interpretability}, which allows testing a trained model for alignment with human-comprehensible concepts. Concepts are defined through examples and can therefore include also abstract concepts. First, a \COMMENTEDIT{}{concept activation }vector (CAV) encoding the meaning of the concept is trained using the collected examples. Second, the gradient of pathology predictions in the direction of the concept vector is calculated to assess whether the trained model uses the concept for the prediction of those pathologies. The definition of a concept is in this case \REVIEWERTWO{}{is} always tied to a chosen hidden feature layer in the model architecture, where one expects that low-level concepts that directly relate to signal features are most clearly expressed in feature layers close to the input layers, whereas more abstract concepts at a higher semantic level are expected to be found in higher feature layers. Therefore we always include a range of different feature layers in our analysis.

\heading{Experimental setup}
First, we create a binary data set per concept \REVIEWERTWO{}{, as in \cite{finzel2022generating}}, each consisting of positive examples of the concept and randomly drawn data points that do not contain the concept.  Second, we use these datasets to train a linear classifier in the feature space of a hidden layer for each concept. We interpret the accuracy of the linear classifier as a measure of how well the concept can be defined in said feature space. The classifier then yields the concept activation vector (CAV), which we use to compute the TCAV scores via directional derivatives, see the technical description in App.~\ref{sec:appendix_xai_methods}). Although we computed the TCAV scores for each layer of each model, for visualization and comparison of both models, we picked six layers from each model (ranging from input layer to output layer).

\heading{Statistical significance} To circumvent accidental findings, the process is carried out for 10 CAVs for 10 trained models with different initializations. It is worth mentioning, that the negative samples are resampled at each CAV fitting. Based on the statistics of the resulting 100 TCAV scores per model, layer, concept and output class, we impose the following conditions to single out consistent model behavior: 
\begin{enumerate}
    \item The first condition relies on the linear separability of positive and negative samples in the feature space of a hidden layer, which we assess via the accuracy of the linear classifier on which the CAV is based. We consider a layer-concept pair, if the accuracy of the linear classifier is above 70\%. Otherwise, we omit any interpretation and highlight this condition as gray rows in \Cref{fig:tcav}.
    \item The second condition relates to the fluctuations of the TCAV scores. The corresponding confidence intervals, interquartile range (IQR) between the 75th and 25th percentiles, should not be too wide and bounded away from 0.5 to rule out coincidental findings.
    We therefore require the IQR to be less than 0.25 and that 0.5 is not contained in the confidence interval. We mark those combinations of model, layer, concept and class with a star symbol in \Cref{fig:tcav} if, and only if, all conditions are met.
\end{enumerate}
These conditions allow us to interpret consistent model behavior in terms of whether a concept is consistently used in a positive or negative way. While the first condition tells us whether a concept can be defined in a meaningful way in the respective feature space, the additional conditions tell us that the concept is exploited for a specific model prediction in a consistent way. We base our discussion on which concepts are consistently used in particular models on those concepts that fulfill all three prerequisites.

\subsection{Code and data availability}
The PTB-XL dataset underlying this work is publicly available \cite{Wagner:2020PTBXL,Wagner2020:ptbxlphysionet}. We provide soft segmentation maps \cite{ptbxl_segmentations} for the entire PTB-XL dataset as computed by our custom segmentation algorithm. 
We provide the source code to reproduce the main results of our investigations as part of the supplementary material \REVIEWERTWO{}{under \url{https://github.com/hhi-aml/xai4ecg}}. \REVIEWERTWO{It will eventually be provided in a publicly accessible code repository after acceptance.}{} The code builds on the XAI software packages captum \cite{captum2019github} and zennit \cite{anders2021software}.
\section{Results}

For most of our analysis, we consider a set of five conditions taken from the sub-diagnostic level (consisting of 23 labels) of the PTB-XL dataset, which covers the NORM class (representing healthy subjects) and a broad range of pathologies: These are prior/chronic, as opposed to acute\cite{thygesen2018fourth}, \textit{anterior and inferior myocardial infarction (AMI/IMI)}, which represent two localizations of the myocardial infarction, \textit{left ventricular hypertrophy (LVH)}, which describes the pathological increase in left ventricular mass, and \textit{complete left bundle branch block (CLBBB)} as a form of conduction disturbance, which causes a delayed activation of the left ventricle. The corresponding model performances can be found in the supplementary material.

\subsection{Sanity checks: Necessary conditions for attribution methods}
\label{sec:results_sanity}
\begin{figure*}[ht!]
    \centering
    \begin{subfigure}[b]{0.45\textwidth}
        \centering
        \includegraphics[width=\textwidth]{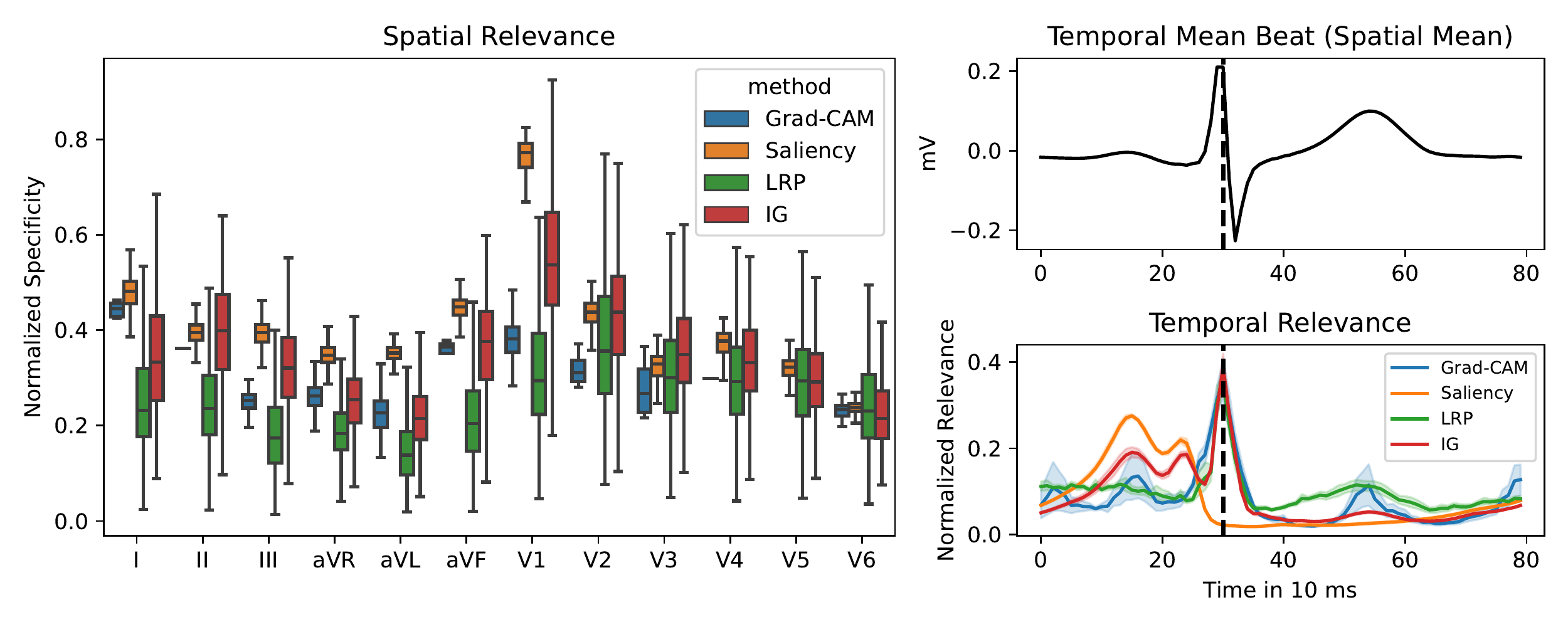}
        \caption{P-wave amplitude (\xresnet{})}
        \label{fig:p_wave_XResNet}
    \end{subfigure}\hfill
    \begin{subfigure}[b]{0.45\textwidth}
        \centering
        \includegraphics[width=\textwidth]{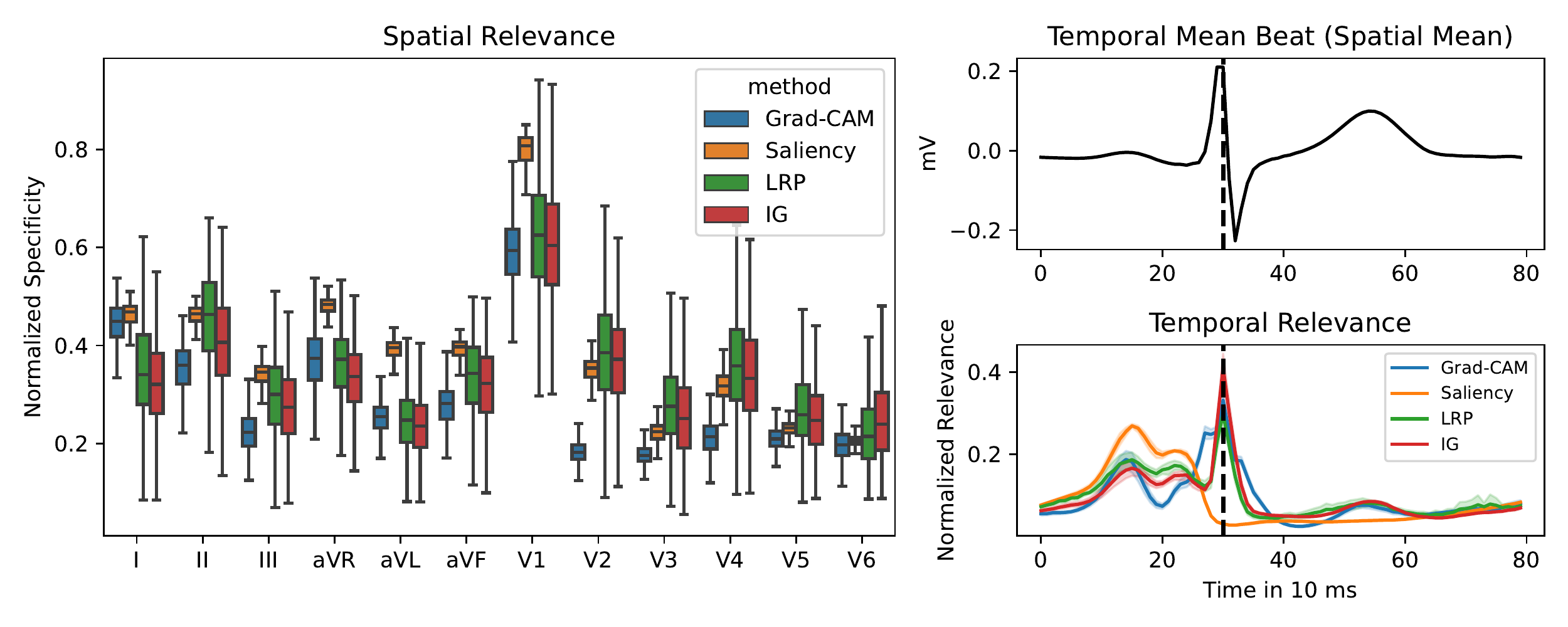}
        \caption{P-wave amplitude (\lenet{})}
        \label{fig:p_wave_lenet}
    \end{subfigure}
    \\
    \begin{subfigure}[b]{0.45\textwidth}
        \centering
        \includegraphics[width=\textwidth]{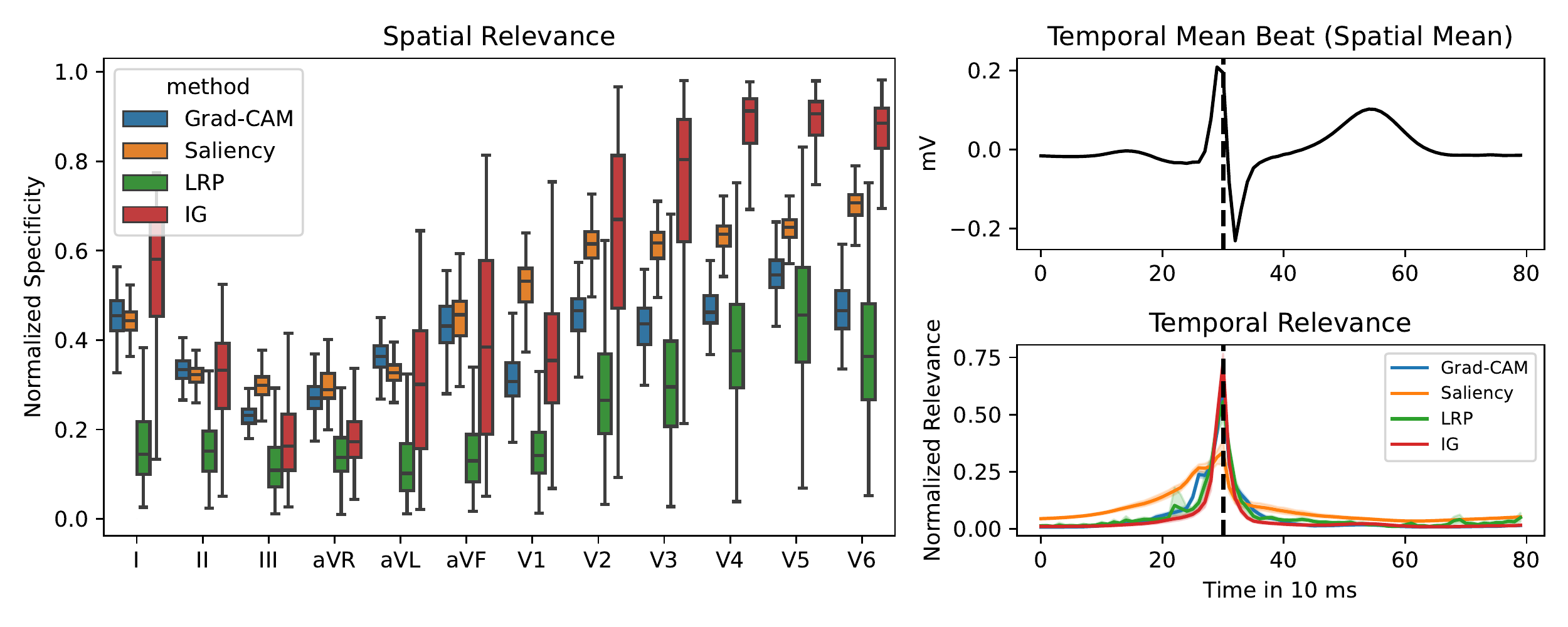}
        \caption{R-peak amplitude (\xresnet{})}
        \label{fig:r_peak_XResNet}
    \end{subfigure}\hfill
    \begin{subfigure}[b]{0.45\textwidth}
        \centering
        \includegraphics[width=\textwidth]{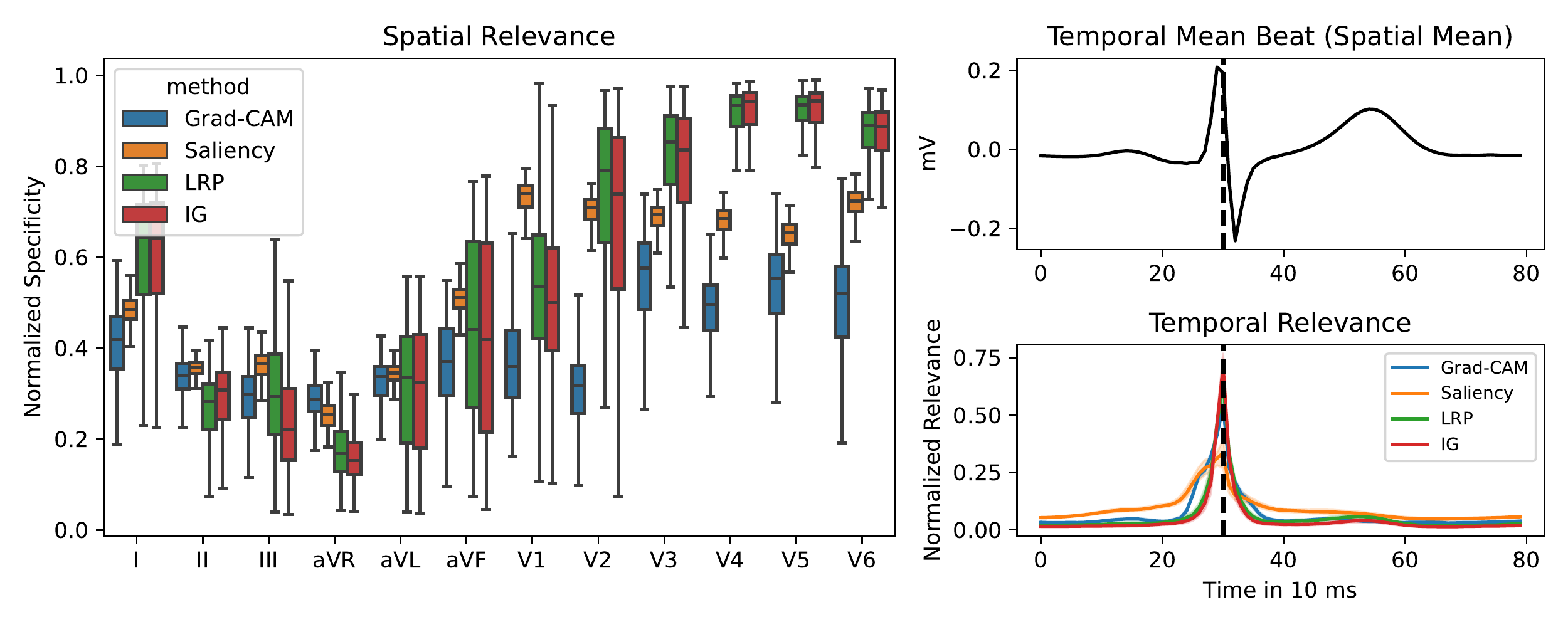}
        \caption{R-peak amplitude (\lenet{})}
        \label{fig:r_peak_lenet}
    \end{subfigure}
    \\
    \begin{subfigure}[b]{0.45\textwidth}
        \centering
        \includegraphics[width=\textwidth]{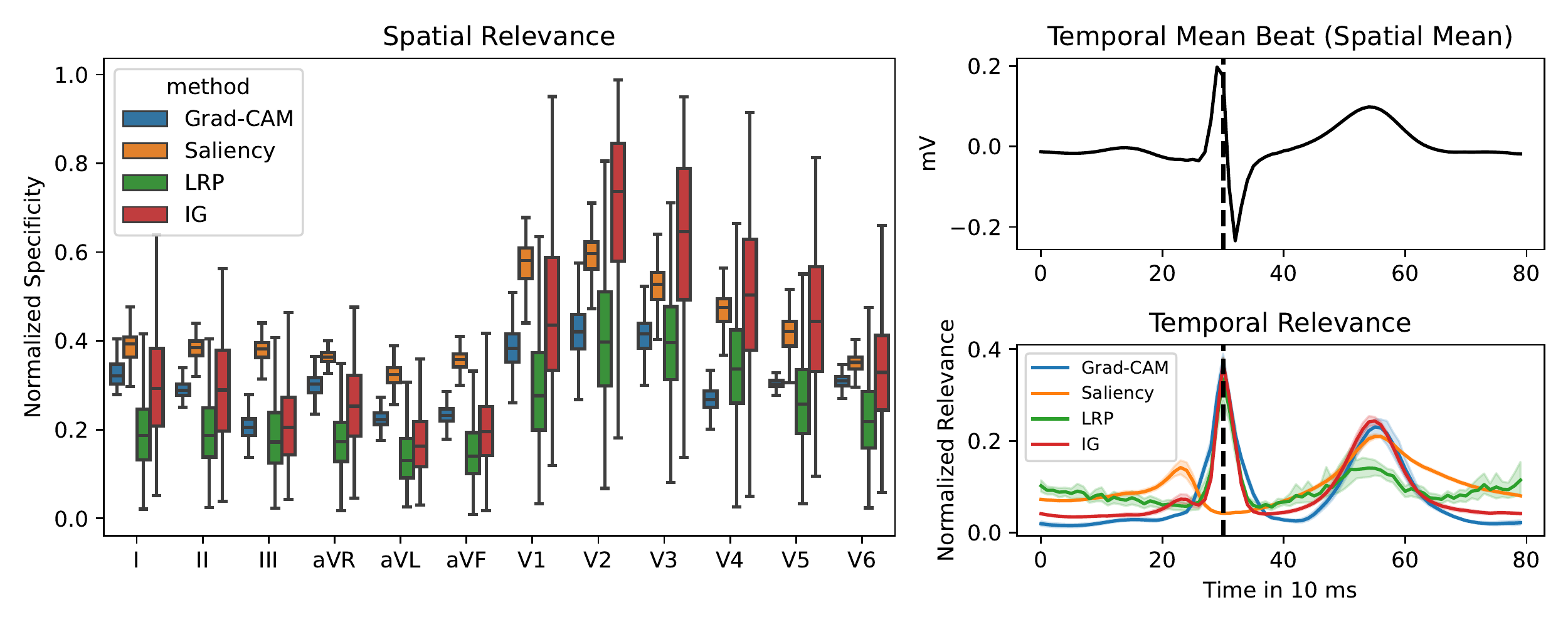}
        \caption{T-wave amplitude (\xresnet{})}
        \label{fig:t_wave_XResNet}
    \end{subfigure}\hfill
    \begin{subfigure}[b]{0.45\textwidth}
        \centering
        \includegraphics[width=\textwidth]{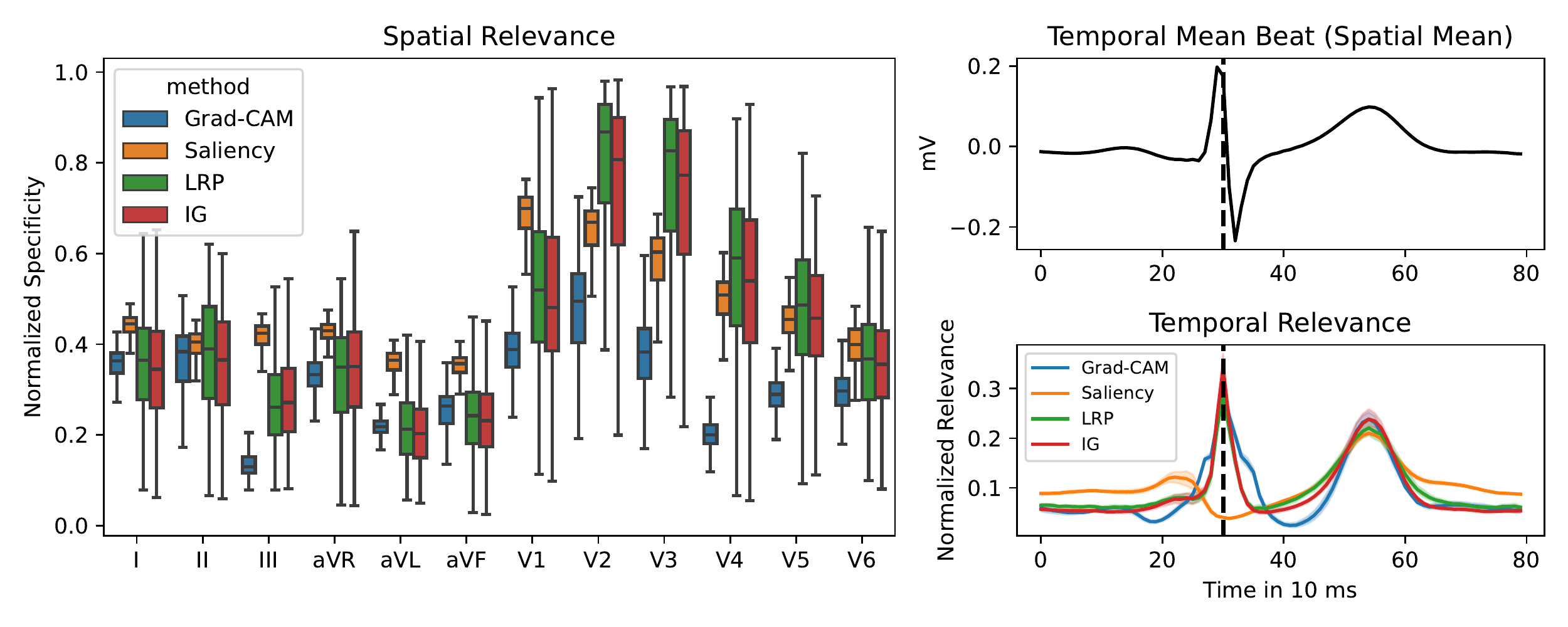}
        \caption{T-wave amplitude (\lenet{})}
        \label{fig:t_wave_lenet}
    \end{subfigure}
    \caption{Results of the three experiments described in \Cref{sec:sanity}: P-wave amplitude (\Cref{fig:p_wave_XResNet}, \Cref{fig:p_wave_lenet}), R-peak amplitude (\Cref{fig:r_peak_XResNet}, \Cref{fig:r_peak_lenet} ) and T-wave amplitude (\Cref{fig:t_wave_lenet}) for \lenet{} and \xresnet{}, respectively. Each subplot is organized in the same way: On the left, we show spatial specificities with different attribution methods color-coded. In the lower right plot, we show temporal specificities with different attribution methods color-coded. These properties are computed for all samples and their attributions are concatenated to allow for analysis across the whole dataset. For spatial specificity, we consider boxplots, where the leads are on the x-axis and the specificity is on the y-axis. For temporal specificity, we consider continuous line-plots, where time is on the x-axis and the temporal specificity is on the y-axis. In the upper right plot, we provide a median beat as a reference for better localization of time steps in the temporal specificity plot. We compute the median beat attributions across the whole dataset and scale them to have a norm of one. If only the segment in the lead under consideration was relevant, the spatial specificity would be one, and the temporal specificity strongly peaked around the corresponding segment. In terms of spatial specificity, saliency shows the highest specificity among the limb leads with a comparably small variance. For temporal specificity, all methods attribute more relevance to the QRS complex \REVIEWERTWO{}{(comprising the Q-, R-, and S-peaks) rather} than to the interval in question, which questions their validity.
    } 
    \label{fig:sanitycheck}
\end{figure*}

\heading{XAI methods} The XAI community distinguishes between inherently interpretable models \cite{rudin2019stop} and post-hoc attributions. 
There is an ongoing debate in the community, about whether post-hoc attributions represent explanations from the perspective of human end-users. We refrain from addressing this question and focus on the latter category, first, due to its widespread use and broad applicability to commonly used models in the field and, second, with an auditing scenario in mind, where one cannot enforce the exclusive use of inherently interpretable model architectures. In particular, we consider four different post-hoc attribution methods that are widely used either in the ECG domain or in the broader XAI community, in particular in the context of image data:
\textit{Layer-wise Relevance Propagation (LRP)} \cite{Bach2015}, \textit{Integrated Gradients (IG)} \cite{sundararajan2017axiomatic}, \textit{Gradient-weighted Class Activation Mapping (GradCAM)} \cite{Selvaraju2019}, and \textit{Saliency maps} \cite{simonyan2013deep}.  Interestingly, applying them to an identical trained model leads to qualitatively different attribution maps, which is a well-known issue in the XAI community \cite{krishna2022disagreement}. A central requirement for attribution methods to be used for auditing or knowledge discovery is that \textit{the attribution method is supposed to be faithful}, i.e., it should accurately reflect the model behavior. \REVIEWERTWO{This is non-trivial to assess since the features the model attributes to are in general unknown. Therefore, we rely on sanity checks that directly relate to specific input features.}{However, it is non-trivial to assess the accuracy of the model's attributions since the features it considers important are not visible to us. This opacity is commonly referred to as the 'black-box nature' of deep neural networks. Consequently, we are compelled to rely on indirect methods (such as sanity checks) linked to specific input features to evaluate these attribution methods.}

\heading{Sanity checks} We use ECG parameter regression tasks as sanity checks, i.e., train regression models for both model architectures to predict P-, R- or T-wave amplitudes from the raw signal. This does not preclude, that the model could exploit information from correlated leads, but the expectation behind this task is that the attributions corresponding to this model should show high temporal and spatial specificity. For example, the prediction of the P-wave amplitude in lead V1 should be strongly influenced by the P-wave in V1, which should be reflected in the corresponding attribution map. We define two measures, the \textit{spatial specificity} and the \textit{temporal specificity}, to quantify how effectively the method prioritizes the correct lead or temporal segment in the ECG. The results of the sanity check are compiled in \Cref{fig:sanitycheck}.

\heading{Spatial specificity} Comparing spatial specificities across all experiments, we observe a largely consistent ranking among methods and leads: while LRP and IG are most specific in the precordial leads (V1-V6) followed by saliency, in the limb leads (I, II, III, aVR, aVL, aVF) saliency is the most specific method. Furthermore, we observed significantly less variance in the attributions for saliency, particularly in comparison to LRP and IG. These observations provide a first argument in favor of saliency, as we prefer methods that provide reliable results across all samples rather than only on average.
Generally, attributions \REVIEWERTWO{are significantly more specific}{exhibit a notably higher degree of specificity} in the precordial leads than in the limb leads (especially in V1, V2 and V3). \REVIEWERTWO{We relate this to the fact that only two of the six limb leads are independent, whereas the remaining four can be computed as linear combinations of any two given limb leads.}{This is largely due to the distinct and unique data presented by the precordial leads. On the other hand, for the limb leads, only two of the six leads are independent, with the remaining four being calculated as linear combinations of these two.} This leads to strong \REVIEWERTWO{correlations}{inter-lead correlations}\REVIEWERTWO{, in particular}{} among the limb leads\REVIEWERTWO{, which cannot be disentangled without any further assumptions}{}.\REVIEWERTWO{}{In such cases, changes in one variable have effects on others, making it intricate to isolate and distribute attributions independently due to the inherent dependent nature of their relationships}.\\
The analysis of the cross-correlations among leads in \Cref{fig:data_correlations} in the supplementary material, \REVIEWERTWO{shows significantly less cross-correlations in V1 and V2 as compared to the other leads}{indicates a notably lower level of total cross-correlation (sum of off-diagonals) in leads V1 and V2 compared to others}, which is also reflected in the overall largest absolute spatial specificities in \Cref{fig:sanitycheck}. \REVIEWERTWO{}{Without claiming universality, we observe higher spatial specificity for leads with lower total cross-correlations between all other leads.}

\heading{Temporal specificity} In terms of temporal specificity, IG and LRP\REVIEWERTWO{, which are both constrained by completeness properties (i.e. the sum of attributions equals the output prediction up to approximation effects)}{} attribute the largest fraction of relevance to the R-peak even if the R-peak is not directly relevant for the given task (e.g., for predicting the P-wave or T-wave amplitude). Interestingly, a similar effect is observed for GradCAM. It is not completely implausible that the model could use the R-peak for localization within each beat and therefore could also attribute relevance to the R-peak. If we impose as a necessary condition for sensible attributions that when predicting a particular amplitude, the region around the corresponding segment should show the overall largest attribution across the entire temporal attribution plot, saliency is the single XAI method, which satisfies this condition in all six experiments. 

\heading{Summary of sanity checks} To summarize \REVIEWERTWO{}{the results of our sanity checks}, spatial specificity, but in particular temporal specificity, clearly favors saliency over the \REVIEWERTWO{}{other} three competing attribution methods\REVIEWERTWO{, which we, therefore,}{. Therefore, we} consider \REVIEWERTWO{}{saliency} as \REVIEWERTWO{only}{the} reference attribution method in the following experiments. 

\subsection{Glocal XAI: Insights from aggregated aligned attributions}
\label{sec:glocal_xai_exp}
\begin{figure*}[ht!]
    \centering
    \includegraphics[width=\textwidth]{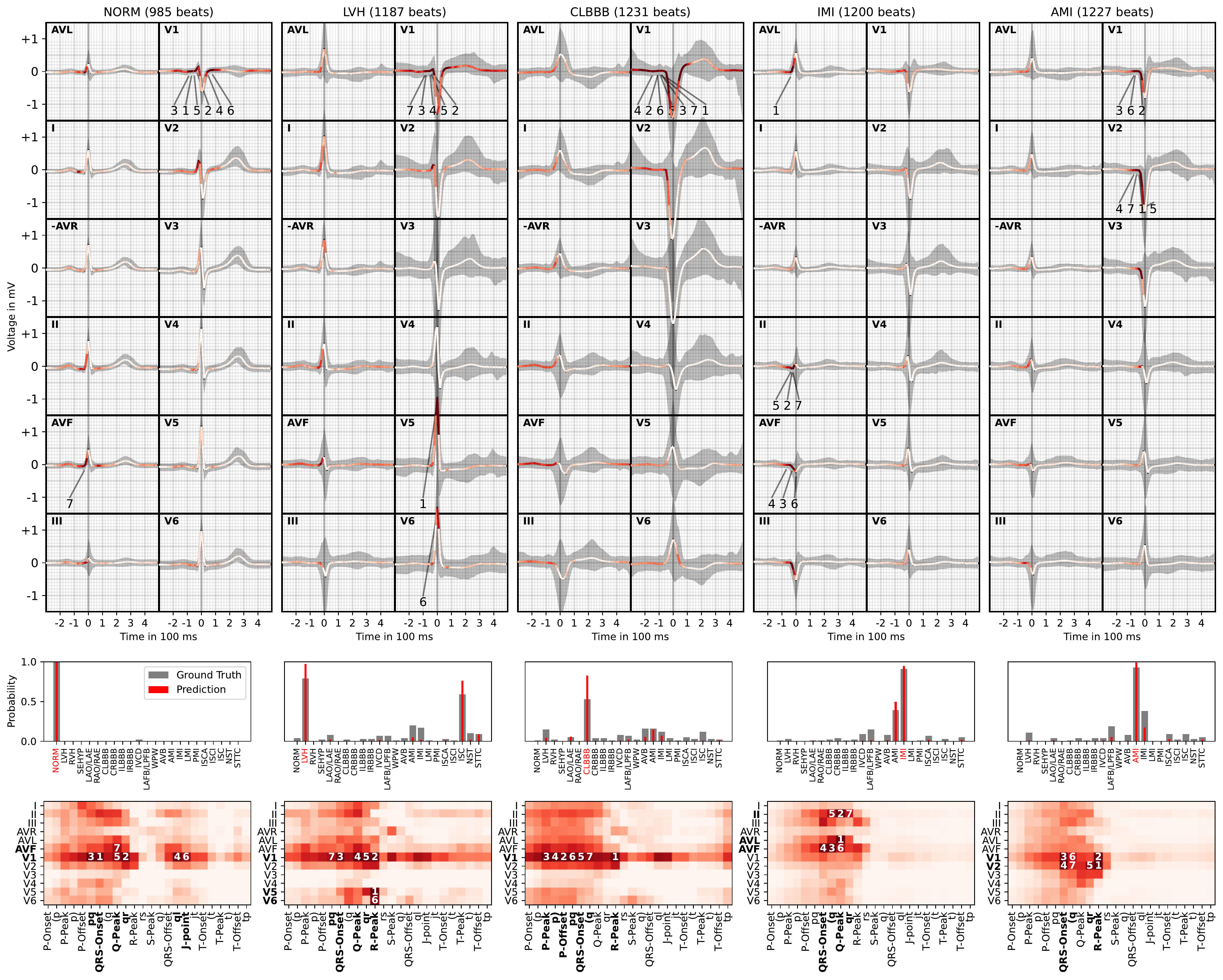}
    \caption{Results of the glocal (dataset-wide) analysis for saliency as an attribution method for a \xresnet{} model. Here, we consider five classes: 1. \texttt{NORM} normal ECG as reference 2. \texttt{LVH} left ventricular hypertrophy 3. \texttt{CLBBB} complete left branch bundle block 4. \texttt{IMI} (prior) inferior myocardial infarction and 5. \texttt{AMI} (prior) anterior myocardial infarction. For each class, we aggregate a median beat for the top 100 predictions per class and also provide the mean of ground truth labels (gray bars) as compared to the prediction (red bars) below each plot (see \texttt{LVH} for inter dependencies with \texttt{\_ISC}). On top of each plot, we visualize the absolute attributions color-coded, where deep red indicates high attribution for the respective diagnosis (e.g. the QS-complex in V1 and V2 for \texttt{AMI} is highly relevant). At the bottom, we show the relevance distribution broken down according to ECG segments with the top 7 segments with thehighest relevance per segment length marked, which allows for quantitative statements about the relevance distribution. These show good agreement with the relevant segments used in decision rules from clinical literature.
    }
    \label{fig:glocal_analysis_xresnet}
\end{figure*}
\begin{figure}[ht!]
    \centering
    \includegraphics[width=0.45\textwidth]{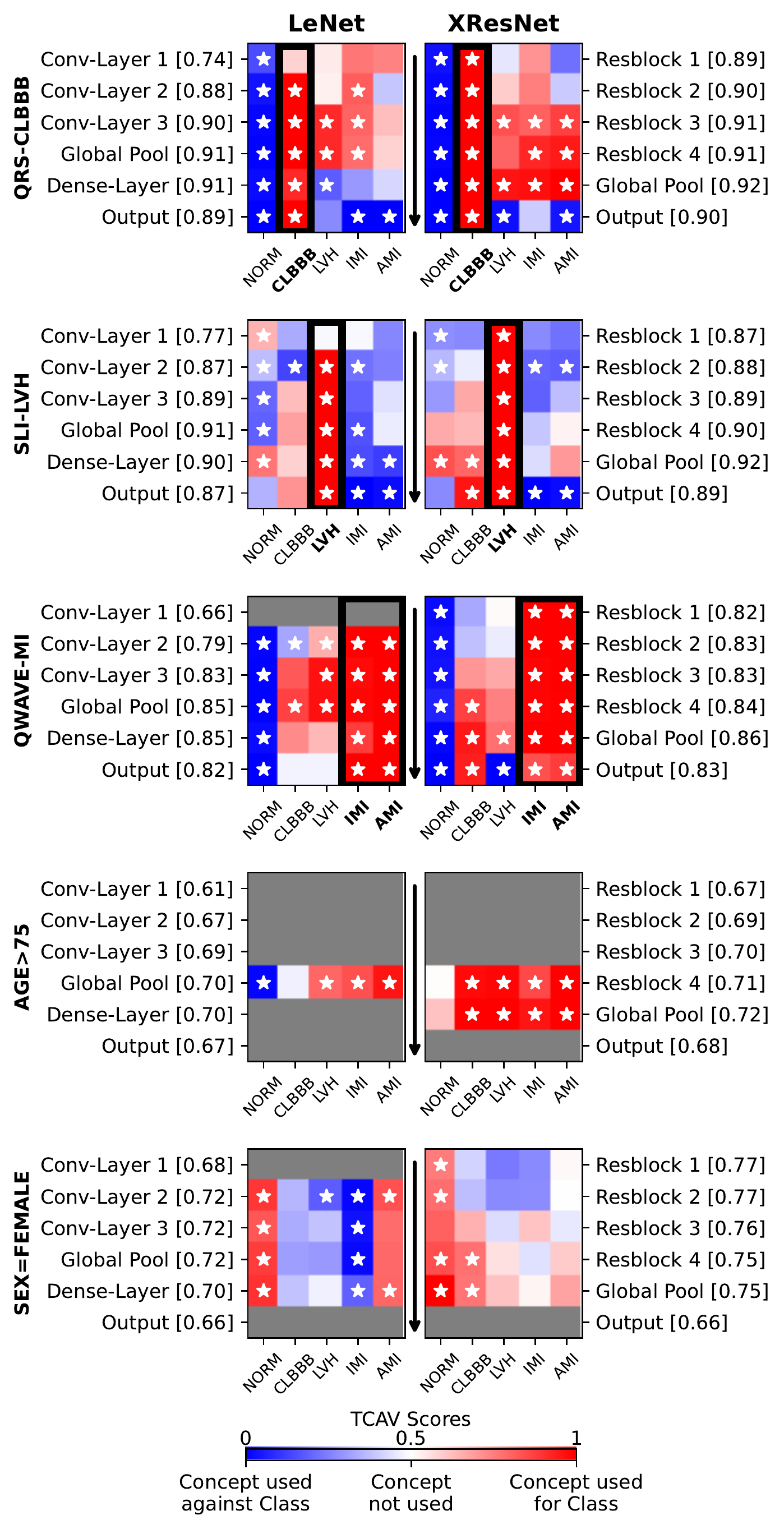}
    \caption{Concept-based analysis: Investigating which of five concepts (from top to bottom: \texttt{QRS-CLBBB}, \texttt{SLI-LVH}, \texttt{QWAVES-MI}, \texttt{AGE>75}, \texttt{SEX=FEMALE}) are used \REVIEWERTWO{consistently}{} for the prediction of a certain class in \lenet{}(left) vs. \xresnet{}(right). Within a block (i.e., a particular concept-model combination to be tested) rows denote different layers in the model and columns represent the different output classes. Each block is color-coded according to the corresponding mean TCAV score indicating whether the concept is used for/against the class under consideration. Stars indicate confidence intervals for the TCAV score that are sufficiently narrow and do not overlap with 0.5, see the text description for details, i.e., correspond to cases where concepts are consistently exploited.
    The numbers in brackets are the respective CAV accuracies, which describe how well a concept can be linearly separated, where blocks with insufficient accuracy are grayed out.}
    \label{fig:tcav}
\end{figure}
\begin{figure}[ht!]
    \centering
    \begin{subfigure}[b]{.45\textwidth}
        \centering
        \includegraphics[width=\textwidth]{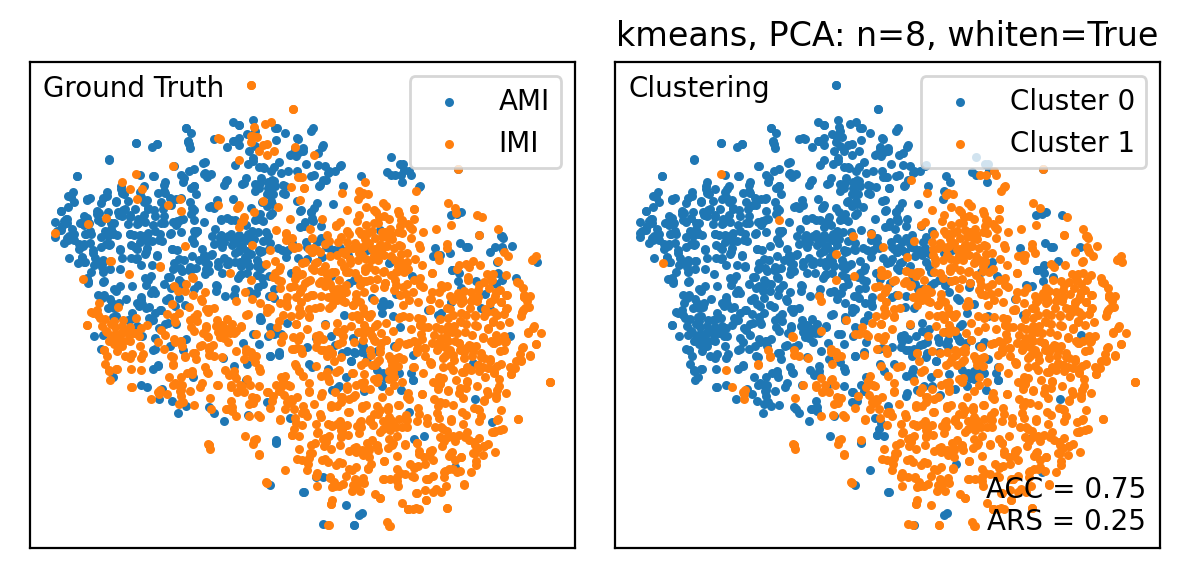}
        \caption{Input}
        \label{fig:discovery_input}
    \end{subfigure}\\
    \begin{subfigure}[b]{.45\textwidth}
        \centering
        \includegraphics[width=\textwidth]{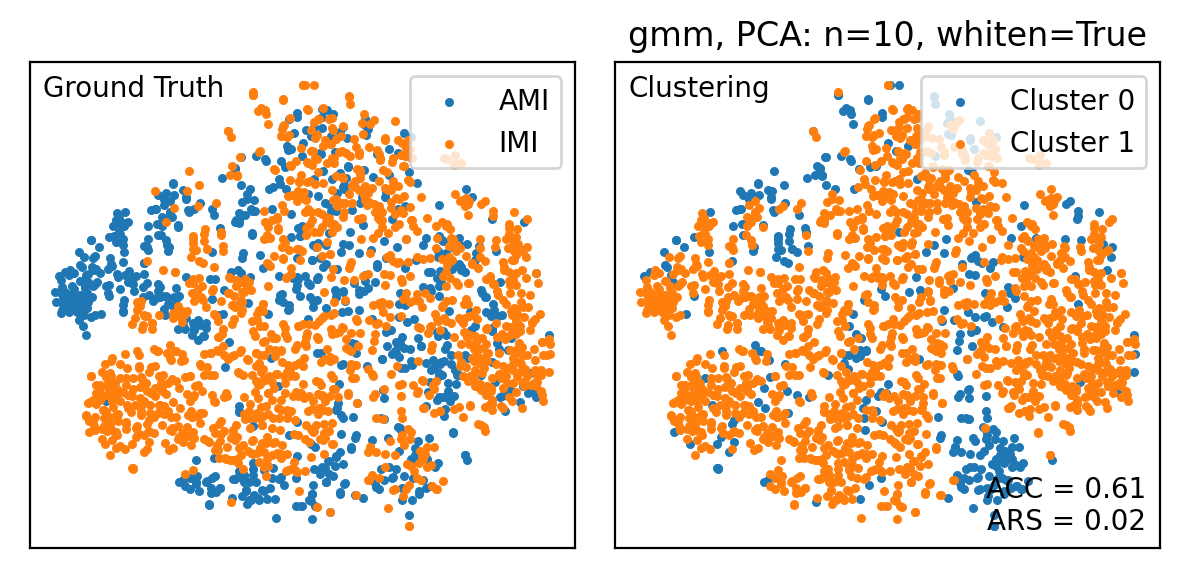}
        \caption{Hidden Features}
        \label{fig:discovery_features}
    \end{subfigure}\\
    \begin{subfigure}[b]{.45\textwidth}
        \centering
        \includegraphics[width=\textwidth]{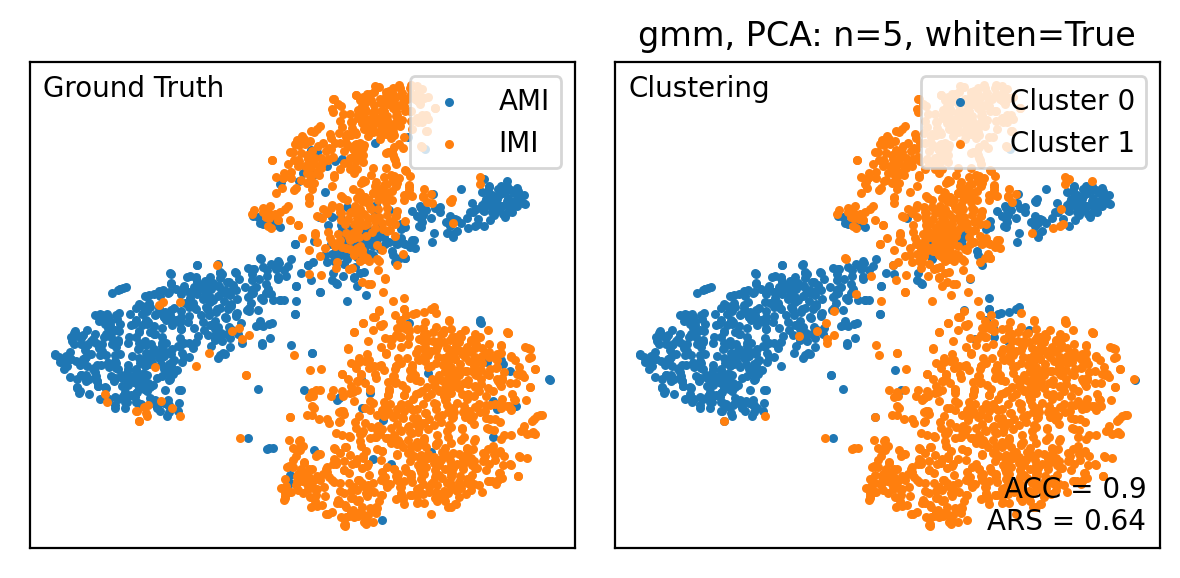}
        \caption{Attribution}
        \label{fig:discovery_relevance}
    \end{subfigure}\\
    \caption{Results of the experiments as described in \Cref{sec:discovery_confirmative}. \COMMENTEDIT{TSNE embeddings}{T-distributed stochastic neighbor embeddings (TSNE)} (with default parameters each) of three different representations extracted from a \xresnet{}  model: \Cref{fig:discovery_input} for the input median beats, \Cref{fig:discovery_features} for the hidden features after global pooling and \Cref{fig:discovery_relevance} for the saliency attributions. In each subplot, we color-coded the ground truth labels left and the resulting clustering on the right. The plots reveal that (aligned) attributions are the most effective input representation for subclass discovery.}
    \label{fig:discovery}
\end{figure}

\heading{Aggregated attribution maps} In the XAI literature, one distinguishes local and global attributions, where the former are sample-specific and the latter characterize the entire model. It is worth stressing, that while local attributions can potentially aid human decision-making, they are not directly applicable for auditing, which focuses on model behavior across entire patient groups. We propose to compute \textit{aggregated beat-aligned attributions} across entire subgroups with common pathologies to infer global model insights and refer to this approach as \textit{glocal XAI}.

\Cref{fig:glocal_analysis_xresnet} summarizes the outcome of this analysis for the five subclasses under consideration in terms of median beats and attributions (top panel), average predictions (middle panel), and in terms of attributions aggregated on segment level for a \xresnet{} model. A  \REVIEWERTWO{largely consistent}{similar} plot for the \lenet{} can be found in \Cref{fig:glocal_analysis_lenet} in the supplementary material. In the following sections, we analyze the three main diagnostic conditions under consideration in detail.

\subsubsection{Left ventricular hypertrophy (LVH)}

\heading{Condition}
The early sign of LVH (left ventricular hypertrophy) is an increase in R-amplitude, which is caused by the increased left ventricular mass. That is why the Sokolow-Lyon index \REVIEWERTWO{}{\cite{sokolow1949ventricular}} is a frequently used, but relatively unspecific diagnostic tool to assist in the diagnosis of LVH. The Sokolow-Lyon index is calculated as the sum of the R-amplitude in V5 and the S-amplitude in V1. The sum must exceed 3.5 mV to be considered an indicator of LVH.

\heading{Attributions} 
Both models agree on assigning the highest attributions to the R-peaks in V5, V1, and V6 (in that order). They also attribute relevance to the ST-segment in V1. However, the two models differ in the rest of their attributions: the \lenet{} places slightly more relevance on the R-peak in V2, while the \xresnet{} additionally focuses on the QR-interval and the beginning of the Q-wave (all in V1).

\heading{Discussion} The strong emphasis on the R-peak in V5 and V6 by both models aligns well with its presence in various LVH concepts and rules, such as the Sokolow-Lyon-Index \texttt{SLI-LVH}. However, except for some attribution in V1, there are no indications of S-peak amplitudes, which are part of some LVH decision rules (e.g., \texttt{S12-LVH},\texttt{LI-LVH-}).

\subsubsection{Complete left bundle branch block (CLBBB)}
\heading{Condition} In the case of CLBBB, the consequence of the caused conduction disturbance and at the same time the dominant ECG finding is a widening of the QRS complex to at least 120 ms \cite{surawicz2009aha}. Further criteria include, for example, a QS-complex or rS-complex in V1 \cite{nikoo2013lbbb}.

\heading{Attributions}  
First of all, both models focus strongly on V1, with the R-peak in V2 as the only minor exception (for \lenet{}). Both models put most attribution on the R-peak in V1 \REVIEWERFOUR{}{, in line with previous findings \cite{bender2022analysis}}. Further relevance is attributed to the second half of the P-wave and the PQ-segment, as well as the beginning of the ST-segment. The models differ in their emphasis: \xresnet{} focuses more on the Q-wave, while \lenet{} places more emphasis on the ST-segment (all in V1).

\heading{Discussion}
One might speculate that the attribution maps reveal a focus on the smoother morphology of the onset of the QRS complex compared to normal samples. On the other hand, it is difficult to determine the significance of the QRS width in an attribution map (which is found to be \REVIEWERTWO{consistently}{} exploited by both models, as shown in the global XAI analysis in Section~\ref{sec:results_tcav}). The strong focus on V1 is \REVIEWERTWO{ consistent}{aligned} with decision rules that are based on large QS or rS-complexes in this lead, V1.

\subsubsection{Anterior/inferior myocardial infarction (AMI/IMI)}
\label{sec:imi_ami}
\heading{Condition}
For the characterization of ECG changes related to prior myocardial infarction, we use the consensus definition \cite{thygesen2018fourth}, which focuses particularly on pathological Q-waves. For the manifestation of the localization of anterior vs. inferior myocardial infarction, we refer to \cite{rautaharju2007investigative} which suggests that AMIs are predominantly detected through QS waves in V1–V3 and IMIs through longer and deeper Q-waves in II, III and aVF.

\heading{Attributions}  
For IMI and AMI, there is a strong similarity between the aggregated attribution maps of both models. In the case of IMI, both models focus on the Q-peak and the first half of the Q-wave in aVL, II, aVF and III with minor differences in terms of ranking. For AMI, both models focus almost exclusively on leads V1 and V2 and the area between the QRS-onset and R-peak. Both put most relevance on the R-peak and slightly less attribution on the QRS-onset in V1 and V2. 

\heading{Discussion} The attributions for both localizations of the prior MI align very well with the corresponding diagnostic criteria applied in the clinical setting, both in terms of spatial and temporal localization. The \REVIEWERTWO{clarity and consistency}{alignment} of these patterns \REVIEWERTWO{}{with domain knowledge} across different models is \REVIEWERTWO{surprising}{convincing}.

\subsection{Global XAI: Does the model exploit cardiologists' expert concepts?}
\label{sec:results_tcav}
\heading{Shortcomings of attribution maps}
The aggregated attribution maps are an effective tool to highlight which parts of the signal are most relevant for certain decisions/predictions. However, it remains unclear whether, for example, low or high voltage amplitudes or the particular morphology at the location are decisive for the attribution of a signal, let alone how cardiologists’ decision rules can be related to the model behavior. Nevertheless, prominent features in aggregated attribution maps, such as Q-waves for MI in \Cref{fig:glocal_analysis_xresnet}, can provide guidance for the development of formalized rules, for which we can check if they are exploited by the model, as described below.

\heading{Concept-based XAI}
We draw on \textit{Testing with concept activation vectors (TCAV)}\cite{kim2018interpretability} as a complementary approach, which, in contrast to the attribution methods discussed so far, allows to test a trained model for its alignment with human-comprehensible concepts.
Our aim is to use it to provide insights that generalize across different training runs for a given model architecture. This goes beyond the usual TCAV setting, where only a single model instance is considered. For the three diagnostic classes CLBBB, LVH and MI, we select a single discriminative concept \REVIEWERTWO{}{for each class} from a selection of well-established concepts\REVIEWERTWO{, see \hbox{\Cref{tab:concepts}} for a brief description and \hbox{\Cref{fig:mcc}} in the supplementary material for details on selecting most discriminative concepts}{Specifically, we choose the concepts \texttt{CLBBB-QRS} for CLBBB, \texttt{LVH-SLI} for LVH, and \texttt{MI-QWAVES} for MI. These concepts are not just theoretical constructs; they are grounded in well-established clinical decision rules that practicing cardiologists use in diagnosing these conditions. For a concise overview of these concepts, we refer to \hbox{\Cref{tab:concepts}} and for a discussion on the criteria for selecting the most discriminative concepts, see \hbox{\Cref{fig:mcc}} in the supplementary material.} We stress again that the approach is easily applicable to any concept that can be framed in terms of conventional ECG features. To demonstrate the versatility of our approach, we also investigate age and sex as two concepts that are not pathology-specific. \REVIEWERTWO{\hbox{\Cref{tab:concepts}} provides an overview of all concepts used.}{}
\begin{table}
	\centering
	\definecolor{mygrey}{RGB}{192, 192, 192}
	\begin{tabular}{|c|c|}
		\hline
		\rowcolor{mygrey}
		 \textbf{Concept} & \textbf{Description}  \\
   \texttt{QRS-CLBBB} & QRS width $>$ 120~ms \cite{surawicz2009aha}\\
 \texttt{SLI-LVH} & Sokolow-Lyon index \cite{sokolow1949ventricular} $>$ 35~mm   \\
 \texttt{QWAVES-MI} & 3 criteria on pathological Q-waves \cite{thygesen2018fourth}  \\
 \texttt{AGE>75} & age $>$ 75  \\
 \texttt{SEX=FEMALE} & female sex  \\
		\hline
	\end{tabular}
	\caption{Summary of the concepts used for concept-based analysis. The first three concepts represent typical ECG criteria used to diagnose complete left bundle branch block (CLBBB), left ventricular hyperthrophy (LVH), and myocardial infarction (MI), respectively. In addition, we consider two concepts that are not pathology-specific.}
    \label{tab:concepts}
\end{table}

\heading{Disease-specific concepts}
\Cref{fig:tcav} illustrates the results of the concept-based analysis and shows that the pathology-specific concepts (\texttt{CLBBB-QRS}, \texttt{LVH-SLI} and \texttt{MI-QWAVES}) are \textit{consistently} \REVIEWERTWO{}{(i.e. marked with stars) and \textit{consequently} (i.e. in all layers)} used \REVIEWERTWO{of}{by} both models and \REVIEWERTWO{}{thus} have a strong positive impact on the model prediction of their associated pathologies. This is indicated by the red squares marked with asterisks, which almost entirely fill the bold black rectangles, which highlights the TCAV scores of the pathologies matching the corresponding concepts. Interestingly, both \texttt{QRS-CLBBB} and \texttt{QWAVES-MI} show a consistently \REVIEWERTWO{}{and consequently} negative effect \REVIEWERTWO{}{(i.e. TCAV values close to zero)} on the prediction of the normal class, which we consider as a \REVIEWERTWO{consistency check confirming}{confirmation of} the validity of our analysis. In the case of \texttt{SLI-LVH}, there is no \REVIEWERTWO{consistent}{consequent} \REVIEWERTWO{}{negative} effect on the prediction of the normal class, which aligns with a sizable number of normal sample\REVIEWERTWO{§}{s} satisfying the criterion \cite{Samesima2017}. 

Summarizing the investigation of the first three pathology-specific concepts, this experiment shows in a so far unseen, consistent fashion that across both model architectures, the concepts are exploited and have a positive effect on the classes they were designed for. The fact that concepts, in a few cases, are also exploited for other classes simply reflects the fact that the rules are not perfectly specific, see \hbox{\Cref{fig:mcc}} in the supplementary material, and are additionally influenced by co-occurring diseases. For example, the negative impact of the concept \texttt{SLI-LVH} on MI predictions aligns with decreasing R-amplitudes for MIs, as opposed to increasing R-amplitudes for LVH.

\heading{Disease-unspecific concepts} For \lenet{}, the \texttt{AGE>75} concept exhibits on one layer negative influence on the prediction of the normal class and positive influences for LVH, IMI and AMI. In contrast, for \xresnet{}, the concept is exploited exclusively positively, on two layers, with effects on all the pathologies considered. This aligns with the expectation that age is an important covariate for pathologies and the models exploit this fact to a certain degree. Finally, the concept \texttt{SEX=FEMALE} has for both models a consistent\REVIEWERTWO{ly}{} positive effect for NORM. \REVIEWERTWO{}{\Cref{fig:mcc} shows that this can at least partially explained by the different prevalence of the NORM class comparing men and women. The TCAV evaluation further substantiates that the model uses this relationship in its predictions.} Further, \REVIEWERTWO{}{for} the \lenet{}\REVIEWERTWO{}{, the concept \texttt{SEX=FEMALE}} seems to have a strong negative influence on IMI and some negative influence on LVH \REVIEWERTWO{, while}{The positive effect of the concept for AMI aligns very well with potentially misdiagnosed female ASMI samples, see the discussion in \Cref{sec:explorative}. For }the \xresnet{}\REVIEWERTWO{}{, it }only has some positive influence on CLBBB, speaking of significant influences, which is most likely related to the different prevalence of pathologies in the PTB-XL dataset comparing men and women. These results highlight the ability to check if demographic attributes are implicitly exploited by the model, which is important in the context of fairness.

\subsection{Knowledge discovery: Identifying subclasses using glocal XAI and unsupervised learning}
\label{sec:discovery}
In this final section, we revisit aggregated attribution maps and demonstrate the effectiveness of attributions in revealing sub-structures within a model, beyond the granularity of label information.  To do this, we first present evidence in a controlled environment using the hierarchical labels provided by PTB-XL. Then, we use this framework to uncover intriguing insights into a shift in the interpretation of female ECGs and the possibility of ASMIs.
\subsubsection{Confirmative analysis: Discovery of sub-diagnostic classes of myocardial infarction}
\label{sec:discovery_confirmative}
\heading{Experiment} In this section, we propose an experiment to discover sub-diagnostic classes within a super-diagnostic class using aggregated attribution maps as a source for knowledge discovery, instead of input or hidden feature representations. Building on the insights from the \REVIEWERTWO{previous section (Glocal XAI)}{analysis on glocal XAI in \Cref{sec:glocal_xai_exp}}, we focus on the differentiation of myocardial infarctions (MI) into anterior (AMI) and inferior (IMI) myocardial infarctions (see \Cref{sec:imi_ami}). For this, we train a model capable of classifying into five super-diagnostic classes (including MI). 
Once trained, we fit clustering models on all samples labelled as \texttt{MI} (specifically we filter for samples that were labeled with \texttt{MI} as the exclusive super-diagnostic label and either uniquely \texttt{AMI} ($n=702$) or \texttt{IMI} ($n=1250$)).
In order to provide evidence for the efficiency of (1) attribution maps in discovering sub-diagnostics, we consider the following representations as baselines: (2) median input beats and (3) features from deeper layers of the trained model.

\heading{Clustering results} In \Cref{fig:discovery}, we show the results of this experiment, where we observed the best clustering performance on attributions, i.e.\ an accuracy of 90\% for attributions (\Cref{fig:discovery_relevance}) compared to 75\% for input (\Cref{fig:discovery_input}) and only 61\% for hidden features (\Cref{fig:discovery_features}). 

\begin{figure}
    \centering
    \includegraphics[width=.45\textwidth]{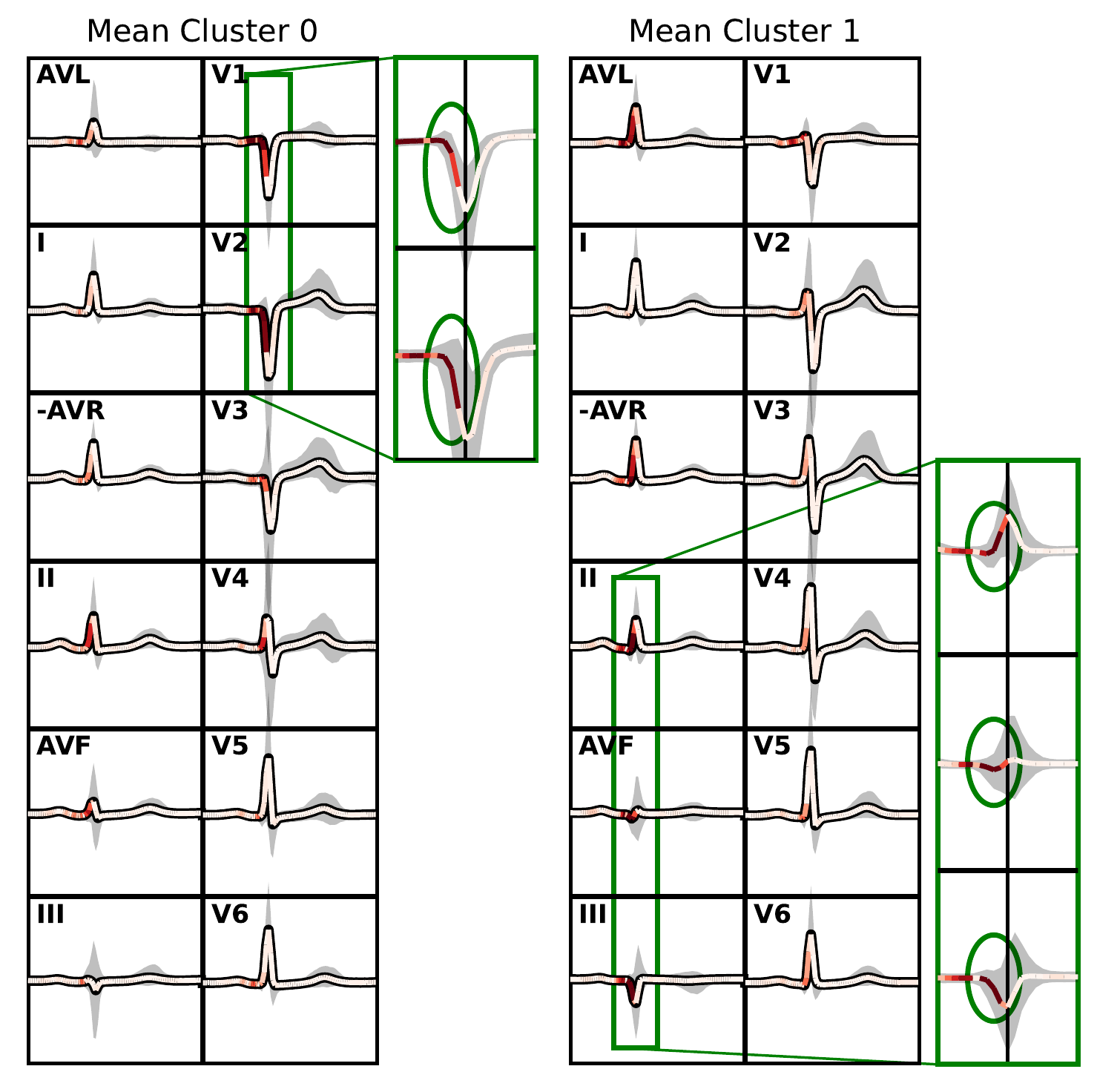}
    \caption{Analysis \COMMENTEDIT{}{and visualization} of the cluster means of the Gaussian Mixture Model fitted on attributions \COMMENTEDIT{}{for super-class MI} as described in \Cref{sec:discovery} \COMMENTEDIT{}{revealing two clusters corresponding to sub-classes AMI (cluster 0) and IMI (cluster 1)}.}
    \label{fig:cluster_means}
\end{figure}

\heading{Cluster visualization} In \Cref{fig:cluster_means}, both clusters are visualized, with cluster 0 predominantly corresponding to AMI and cluster 1 predominantly to IMI. The median beats for both clusters were computed by aggregating the corresponding beats, based on the attribution clusters, along with the corresponding mean attributions on top. The resulting clusters agree with the glocal analysis of supervised models in \Cref{fig:glocal_analysis_xresnet}. 

\heading{Conclusion} \REVIEWERTWO{Given the fact, that best ability for discovery is given via attributions,}{The fact that the most effective discovery capability is achieved through attributions} provides supporting evidence of the usefulness of XAI for multiple applications\REVIEWERTWO{}{,} including debugging, interpretation and discovery.  It is important to note, that those observations \REVIEWERTWO{are not possible given only}{cannot be made solely based on} input data or feature activations of intermediate layers. This finding aligns very well with prior work \cite{lapuschkin2019unmasking} where clustered attribution maps were used to identify shortcuts exploited by the model.  

\subsubsection{Explorative analysis: Identifying subgroups within anteroseptal myocardial infarction}
\label{sec:explorative}
\begin{figure*}[ht!]
    \centering
    \begin{subfigure}[b]{.45\textwidth}
        \centering
        \includegraphics[width=\textwidth]{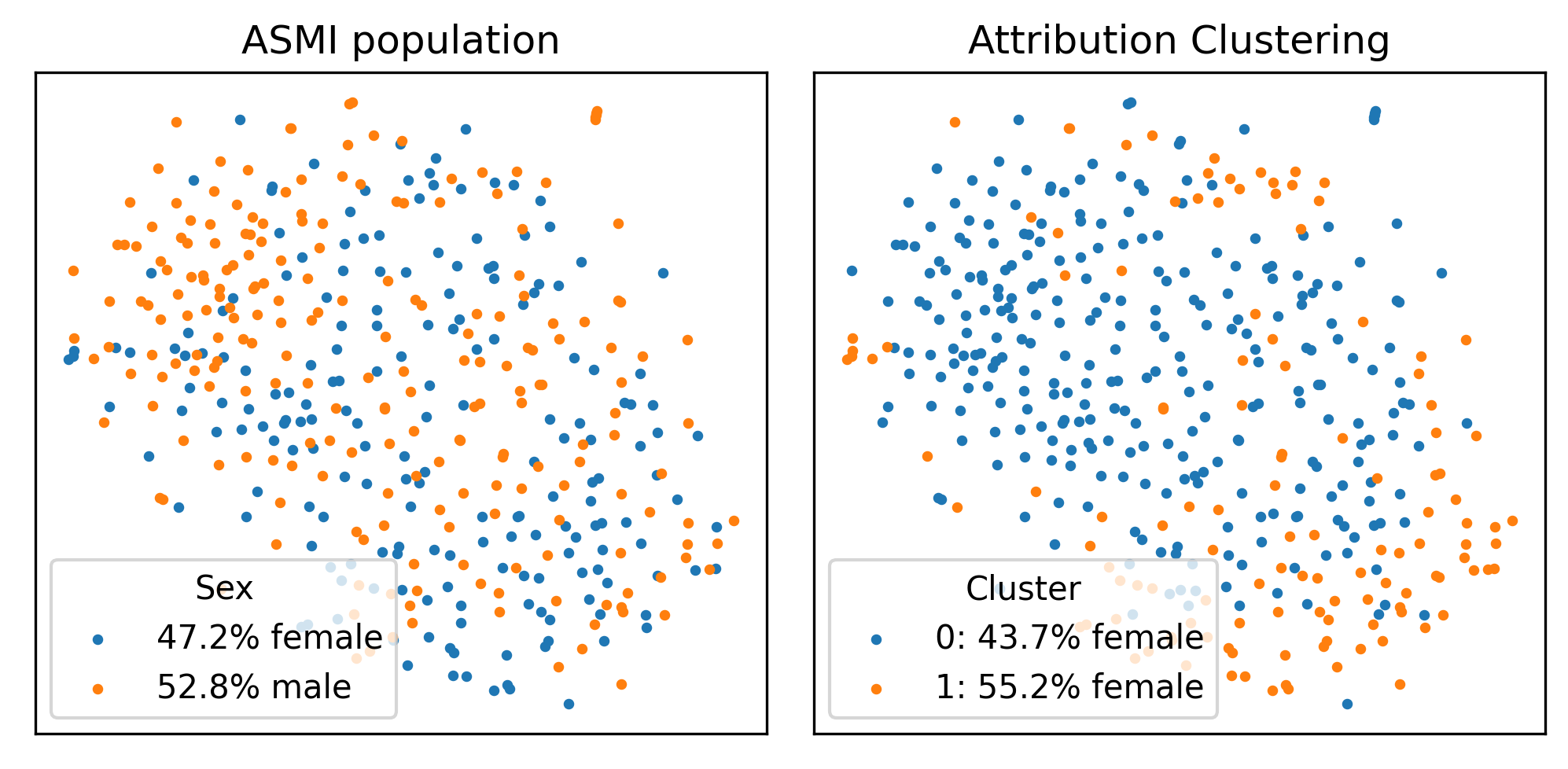}
        \caption{TSNE embeddings for the \xresnet{} model.}
        \label{fig:discovery_embeddings_resnet}
    \end{subfigure}
    \hfill
    \begin{subfigure}[b]{.45\textwidth}
        \centering
        \includegraphics[width=\textwidth]{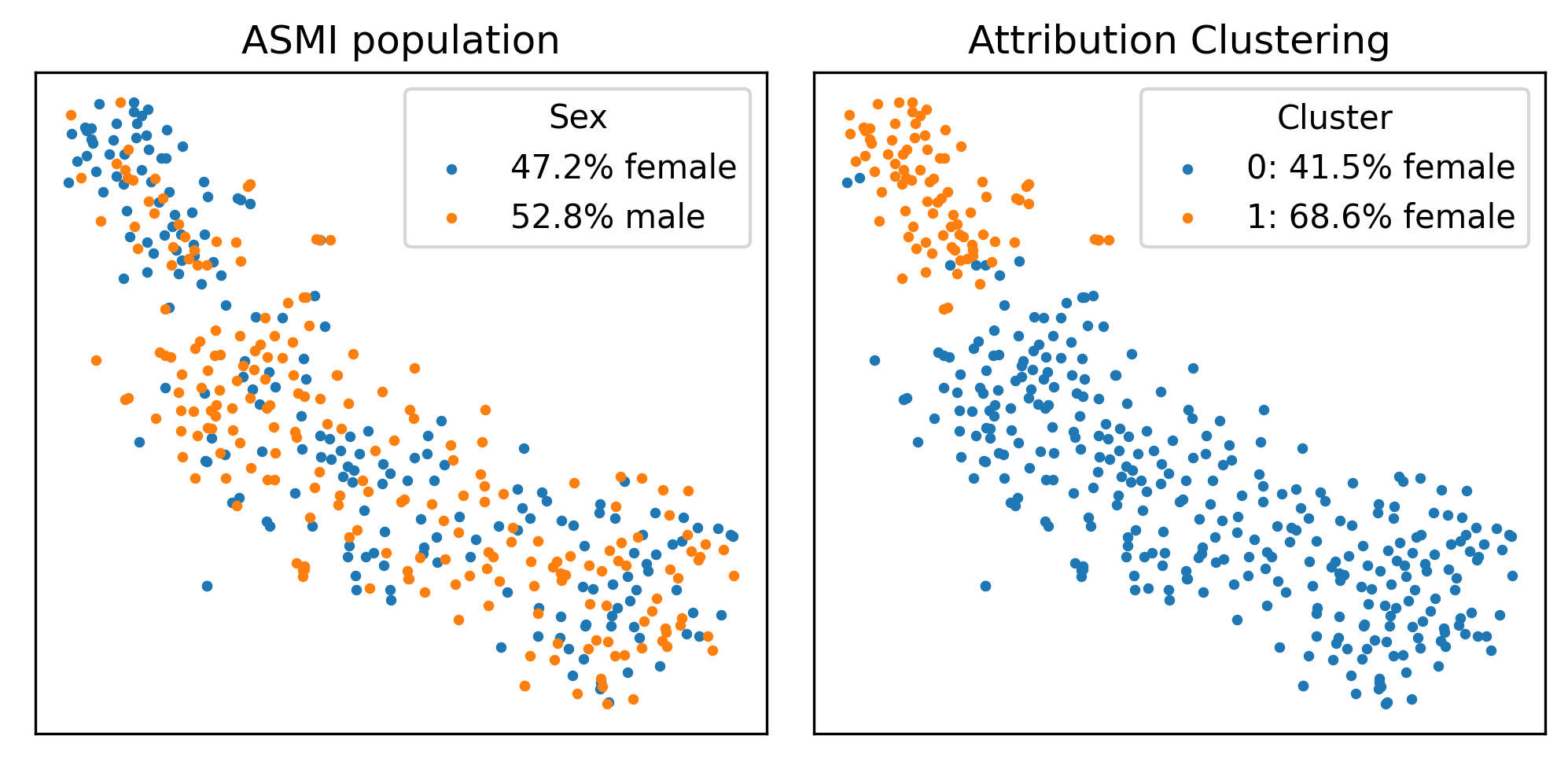}
        \caption{TSNE embeddings for the \lenet{} model.}
        \label{fig:discovery_embeddings_lenet}
    \end{subfigure}\\
    \begin{subfigure}[b]{.45\textwidth}
        \centering
        \includegraphics[width=\textwidth]{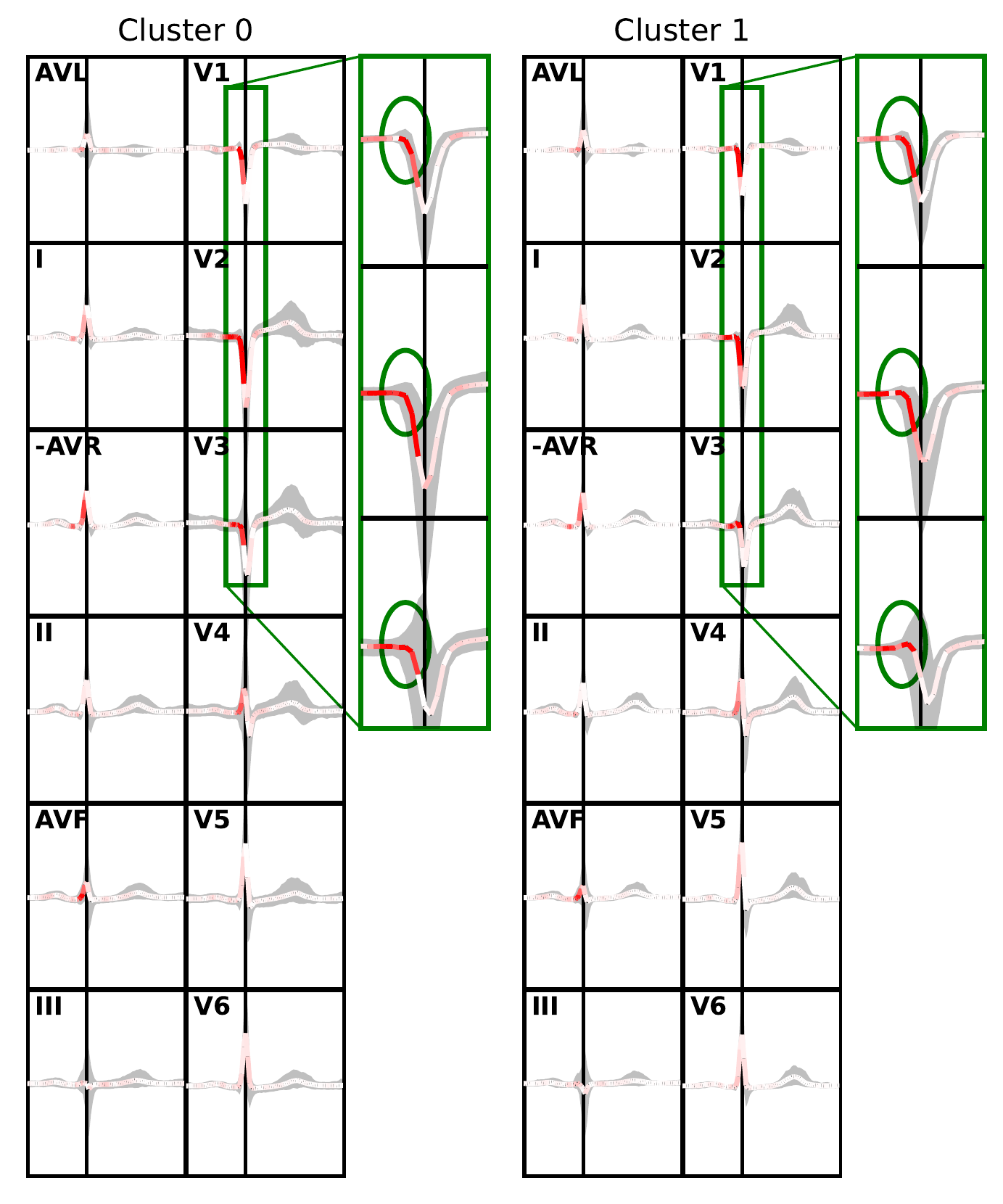}
        \caption{Cluster means based on \xresnet{}.}
        \label{fig:discovery_means_resnet}
    \end{subfigure}
    \hfill
    \begin{subfigure}[b]{.45\textwidth}
        \centering
        \includegraphics[width=\textwidth]{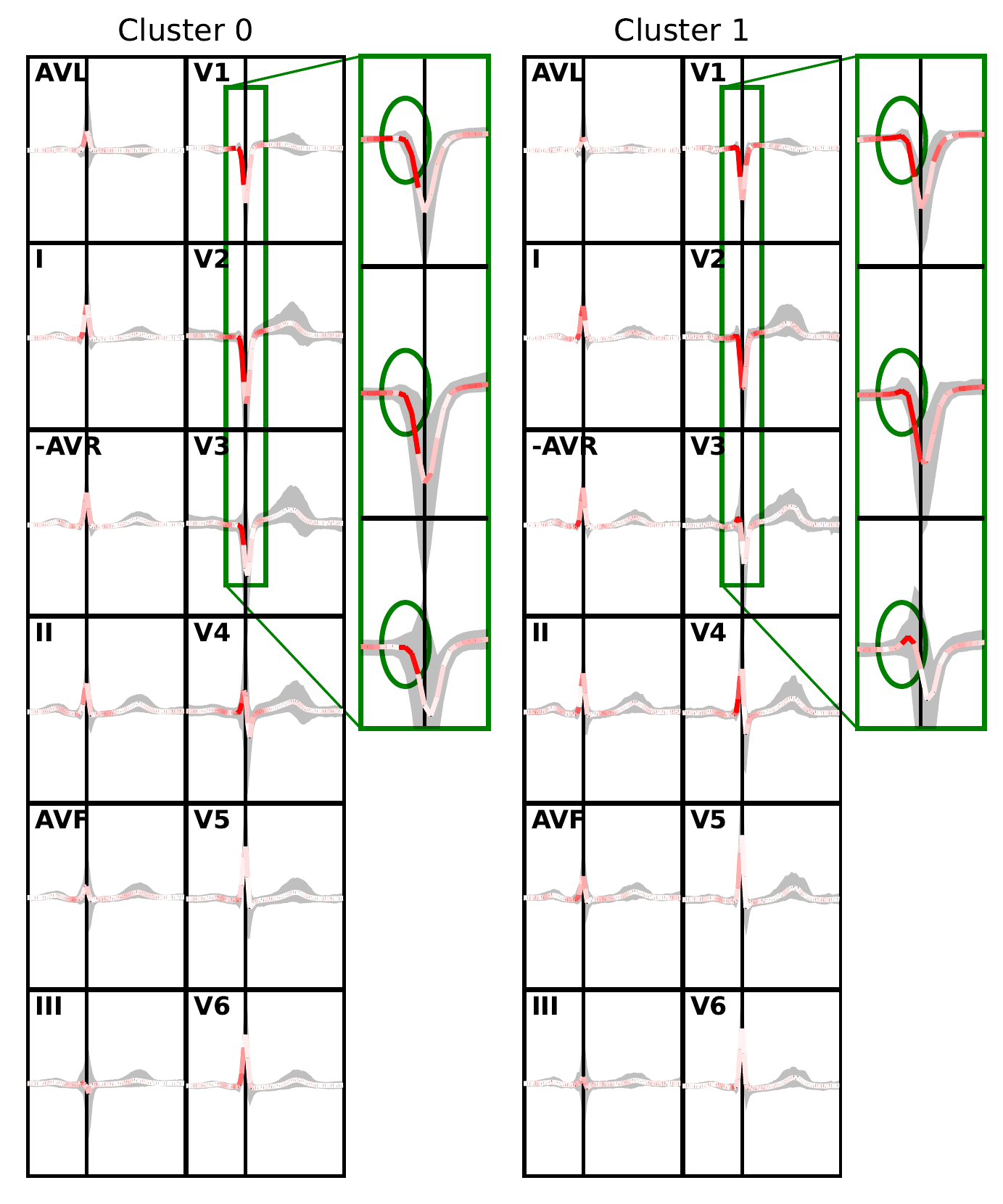}
        \caption{Cluster means based on \lenet{}.}
        \label{fig:discovery_means_lenet}
    \end{subfigure}
    \caption{Explorative analysis of the substructure of the ASMI (anteroseptal myocardial infarction) for the \xresnet{} model on the left (\Cref{fig:discovery_embeddings_resnet} and \Cref{fig:discovery_means_resnet}) and \lenet{} model on the right (\Cref{fig:discovery_embeddings_lenet} and \Cref{fig:discovery_means_lenet}). Top plots \Cref{fig:discovery_embeddings_resnet} and \Cref{fig:discovery_embeddings_lenet}: TSNE embeddings (based on attributions) for ASMI colored by sex (left) and cluster assignment (right, as in \Cref{fig:discovery_relevance}) reveals two distinct clusters. Bottom plots \Cref{fig:discovery_means_resnet} and \Cref{fig:discovery_means_lenet}: Corresponding cluster means and median beats for ASMI attribution clusters reveal distinct morphological differences, which can be associated with different degrees of severity of the myocardial infarction. Encircled regions in the plots indicate regions where the difference between the corresponding attributions, which are used for clustering, are significantly different from zero, i.e.\ the main regions where one expects to find differences between both clusters. These reveal a cluster of transmural ASMIs (cluster 1) and a cluster of less severe ASMIs and according to modern diagnostic standards even normal variants (cluster 0).}
    \label{fig:discovery_asmi}
\end{figure*}

\heading{Experiment} We go one step beyond confirmation and demonstrate how the proposed approach can be used to identify clinically relevant structures for pathologies, where no further substructure is known. Here, we focus on  \COMMENTEDIT{a particular sub-condition, the anteroseptal myocardial infarction (ASMI),}{ASMI,} which is the most frequently annotated subclass of anterior myocardial infarction in PTB-XL. We deliberately chose a statement at the most finegrained level of the diagnostic label hierarchy to avoid rediscovering already known subclasses.

\heading{Clustering results} Using the method from the previous section, we cluster the aligned attribution maps for all ASMI samples and visualize the corresponding embeddings in \Cref{fig:discovery_embeddings_resnet} and \Cref{fig:discovery_embeddings_lenet} for \xresnet{} and \lenet{}, respectively. In \Cref{fig:discovery_means_resnet} and \Cref{fig:discovery_means_lenet} we visualize the mean beats corresponding to the respective clusters, where we highlight the regions of interest via ellipses. Interestingly, both model types identify similar clusters highlighting the area before the vertical line in leads V1, V2, V3. 

The main difference between the clusters is the presence or absence of an R-peak.\REVIEWERTWO{}{The R-peak refers to the peak of the R-wave within the QRS complex, which represents ventricular depolarization. Its presence, size, and shape provide important information about heart function and health.} In the case of the \lenet{}, see \Cref{fig:discovery_means_lenet}, a remnant R-peak is \REVIEWERTWO{clearly}{} visible in cluster 1 in all three leads V1, V2, V3, in the case of the \xresnet{} only in V3 and V2. In both cases, cluster 0 shows \REVIEWERTWO{clear}{} signs of a transmural anteroseptal myocardial infarction, in particular, no visible R-peaks in leads V1, V2, V3.\REVIEWERTWO{}{A transmural infarction represents a severe type of heart attack, characterized by involving the entire thickness of the heart muscle in the affected area. }As the size of the R-peak is indicative for extension of the scar, cluster 0 hence encompasses ASMIs of a less severe degree. We stress again that the two clusters were both comprised of samples that were labeled as ASMI and were only differentiated based on their respective attribution maps.

\heading{Clinical interpretation} Going beyond a descriptive analysis, we aim to explore the potential clinical meaning of these results and identify sex as \REVIEWERTWO{covariate discriminating both clusters.}{a covariate that discriminates between both clusters.} \REVIEWERTWO{And indeed}{Indeed}, for women\REVIEWERTWO{such}{, certain} ECG changes are in some cases even considered as normal variants \cite{rautaharju2007investigative}\REVIEWERTWO{, which aligns very well with the majority}{. This observation aligns with the high proportion} of female samples in \REVIEWERTWO{this}{} cluster 1 (\xresnet{}: 55\% \lenet{}:69\% all ASMIs: 53\%). This is \REVIEWERTWO{also supported}{further evidenced} by the mean predicted probabilities for the non-transmural cluster 1\REVIEWERTWO{, b}{. B}oth models show a \REVIEWERTWO{reduced AMI probability}{lower probability of AMI} (compared to the set of all ASMI predictions) and an increased \REVIEWERTWO{NORM probability}{probability of NORM}. \REVIEWERTWO{Again, t}{T}his pattern is most clearly exposed in the case of the \lenet{} (AMI 0.27 and NORM 0.41 on cluster 1 as compared to AMI 0.72 and NORM 0.09 on all ASMI samples). \REVIEWERTWO{Indeed, the explicit inspection of a subset of 10 female samples of the \lenet{} cluster 1 revealed that}{A detailed analysis of 10 female samples from \lenet{} cluster 1 supports this finding.} \REVIEWERTWO{these}{These cases} would be diagnosed as normal variants according to modern standards, whereas they would have been diagnosed as ASMIs due to R-progression according to the common diagnostic standards at the time the PTB-XL dataset was created. 
\section{Discussion}

\noindent\heading{Sanity checks}
We propose sanity checks based on an ECG parameter regression task\REVIEWERFOUR{}{, informed by methodologies previously considered and documented in other domains, as described in \cite{schwalbe2023comprehensive, doumard2023quantitative}, however, to the best of our knowledge investigated for the first time in the context of ECG analysis}. Surprisingly, three out of the four considered attribution methods do not show a sensible degree of temporal or spatial specificity and show a strong bias toward attributing relevance to the QRS-complex, which typically contains the highest signal amplitudes. In particular, GradCAM, the most widely used attribution method in the field of ECG analysis, fails to satisfy the sanity check. Saliency maps, the only method that passed the sanity check, show a high degree of temporal specificity. Gradient noise, a major drawback of saliency methods, can be very effectively reduced by aggregating attributions across beats and samples. The outcome of the sanity check not only call into question results achieved using attribution methods that did not pass the sanity check in the field of ECG analysis, and serves as a warning against blindly applying of off-the-shelf attribution methods in any domain. 

\heading{Glocal XAI}
Most of the existing applications of attribution methods in the field demonstrate insights based on single hand-picked examples. 
Our analysis provides strong arguments in favor of aligned attribution maps \REVIEWERFOUR{}{, in line with arguments and experimental evidence put forward in the literature \cite{Jones9288132,van2021discovering,bender2022analysis,goodfellow2018towards}}, which can be defined on beat level or even on the level of individual ECG segments composing the beat, that can be aggregated across entire patient subgroups. To summarize, our analysis shows an unexpectedly strong similarity between the attribution maps of both model architectures, in particular also in terms of ranking of the most relevant segments. The quantified attribution distribution onto segments and leads can be used to compare to cardiologists' decision rules \REVIEWERFOUR{}{\cite{sokolow1949ventricular, surawicz2009aha, nikoo2013lbbb, thygesen2018fourth, rautaharju2007investigative}}. We find good agreement for the most relevant parts both in terms of spatial and temporal localization. This technique finds its boundaries when decision rules can no longer be directly related to single ECG features but to more abstract concepts, which could be analysed in the global XAI section. \REVIEWERFOUR{To the best of our knowledge, we are the first to make such statements in a quantitative manner broken down according to leads and segments in a dataset-wide analysis as opposed to anecdotal evidence based on single hand-picked examples.}{Prior studies \cite{van2021discovering,bender2022analysis} have leveraged visualizations and relevance score analyses to explore ECG features and disease classifications, emphasizing specific leads and features. However, to the best of our knowledge, our study is the first to quantitatively analyze these aspects across the entire dataset. We provide a more comprehensive breakdown by leads and segments, allowing for a deeper understanding and broader generalization beyond the anecdotal evidence typically based on isolated examples. This methodological advancement facilitates a more nuanced insight into ECG data segmentation and its clinical implications.} This technology could serve as a complementary assessment of the internal validity during a certification process but can also be used for knowledge discovery. 

\heading{Global XAI}
We see the ability to test if deep learning models exploit a certain (abstract) concept as a decisive advance in the global analysis of ML models. It provides crucial new insights into the model that are difficult to attain with conventional attribution methods \REVIEWERFOUR{}{(as suggested by \cite{kim2018interpretability,finzel2022generating,crabbe2022concept})}. The ECG domain is particularly well-suited for this purpose due to the availability of large rulebooks comprising decades of cardiologists' expert knowledge, which are mostly formulated in terms of ECG features that can be automatically extracted from the signal. \REVIEWERFOUR{We demonstrated}{To the best of our knowledge, we are the first to demonstrate} the potential of this technology for selected concepts and pathologies and found a compelling alignment with expert knowledge. We envision that this approach could serve as a component to assess the internal validity of a machine learning algorithm during the certification process for automatic ECG analysis algorithms. On the one hand, it allows for the verification of certain concepts that are considered to be mandatory for every model deployed in clinical environments. On the other hand, it can also be used to check whether concepts related to sensitive attributes are systematically exploited. This provides a complementary perspective to the fairness literature, which mostly relies only on model performance, see e.g., \cite{SeyyedKalantari2021}, and touches on the fundamental question of whether the model should be allowed to exploit sensitive attributes for its decision.

Finally, we would like to emphasize the strong consistency of the glocal and global analysis. Our selection of concepts and the resulting TCAV values are consistent with the observations made in the glocal analysis, where we observed the same concepts per diagnostic class in the input using aggregated aligned attribution maps. In particular, compare \Cref{fig:glocal_analysis_xresnet} and \Cref{fig:glocal_analysis_lenet}: For LVH, the global analysis reveals that the \texttt{SLI-LVH} concept is clearly exploited and, correspondingly, the glocal analysis shows highest relevance on R-peaks in V1 and V5. For MI, the Q-wave concepts {QWAVES-MI} are exploited and the glocal analysis highlights Q-waves for AMI and IMI. This finding supports our claim that glocal methods can be used to arrive at global insights, which underlines their dual nature.

\heading{Knowledge discovery}
\REVIEWERFOUR{}{In the realm of deep learning applied to 12-lead ECG, previous studies \cite{beer2020using,lu2024decoding,Tison2019,hicks2021explaining} have demonstrated the capability of machine learning models to capture and interpret known clinical features and differences in ECG data. While these studies have enhanced our understanding of ECG interpretation, our research builds upon this foundation by introducing explorative clustering. This novel method not only analyzes decision processes but also identifies previously undetectable subclasses within the data, pushing the boundaries of clinical knowledge discovery in ECG data.} We demonstrate that aligned attribution maps represent the most reliable way of identifying subclasses within a given superclass (compared to aligned raw signals or hidden features), which is a surprising observation. It crucially relies on the alignment, which is non-trivial to be achieved in other data domains. However, similar techniques have been used in the literature to identify spatially localized artifacts compromising image classifiers \cite{lapuschkin2019unmasking}. Concept-based methods that directly relate to structures in the models' hidden representations \cite{vielhaben2023multi} might be a way to overcome the limitations of the requirement of aligned samples and might provide a characterization on the level of individual segments rather than entire beats. 

Applying the same approach to anteroseptal myocardial infarctions, we demonstrate that the method can distinguish transmural from non-transmural myocardial infarctions as clinically meaningful subgroups. This goes as far as providing hints on the internal consistency of diagnostic classes. We see the ability to question diagnostic knowledge through data-driven insights as a promising path to advance the field of ECG analysis providing insights that even extend to diagnostic criteria underlying the different conditions. 
Here, the most accurate model does not necessarily allow the deepest insights. The more complex \xresnet{} has sufficient model capacity to learn essentially arbitrary input-output patterns. However, the shallower \lenet{}, due to its smaller model capacity, reveals the tension between the NORM condition and the diagnostic criterion used to diagnose ASMI.
Extending this approach to further pathologies to deepen the understanding of ECG signs through such a combination of data- and model-driven techniques represents a promising perspective for future research.

\heading{Limitations and future work} \REVIEWERFOUR{In the context of our research, it is essential to acknowledge the existing literature on post-hoc local XAI for ECG \mbox{\cite{Kwon2020,van2021discovering,goodfellow2018towards,hicks2021explaining,lu2024decoding,kwon2020deep,Jones9288132,lima2021deep,cho2020artificial}} and glocal XAI (in general and specific to ECG) \mbox{\cite{lapuschkin2019unmasking, Jones9288132,van2021discovering,bender2022analysis}}. It has laid the foundation for our exploration of explaining deep learning ECG analysis suitable for future directions in auditing such models prior to clinical deployment.}{} While our study makes significant contributions to the field by applying both concept-based and glocal XAI to the domain of ECG, it is not without limitations. One notable constraint is the reliance on PTB-XL as the primary dataset, which may introduce biases and impact the generalizability of our findings due to its potential limitations in encompassing the full spectrum of demographic diversity, clinical conditions, and ECG signal variability.

While we identified saliency as the most reliable attribution method through a defined sanity check, comparing it against integrated gradients, LRP, and Grad-CAM, we acknowledge the limitations inherent in relying on a single attribution method for the proposed experiments. This reliance potentially narrows our perspective, as various attribution methods might provide different insights.

\REVIEWERFOUR{}{Furthermore, it is important to stress that the proposed glocal XAI approach is only able to identify consistent patterns that equally impact all beats, which is also the idea of basing ECG feature extraction on median beats. In particular, this approach will not be able to reveal temporal variations of features across several beats such as RR variability or spectral characteristics, even if the model can potentially leverage them internally. However, the global XAI approach with respective concepts can be applied here, emphasizing the importance of a comprehensive choice of views on this topic and highlighting the pros and cons of each approach.} Additionally, the methodological approach \REVIEWERFOUR{}{of global XAI}, has inherent constraints such as translation of domain expert concepts into formal conditions, which may encounter challenges in other domains, such as medical imaging, where often no commonly accepted features are available, which can be extracted in an automated fashion, that can be used as part of expert concept definitions. 

Further investigations incorporating alternative methodologies for automatic concept discovery, as suggested by \cite{achtibat2023attribution, vielhaben2023multi}, could provide a complementary perspective on this topic. Recognizing these limitations, our work serves as a valuable step forward, offering insights that should be considered in conjunction with the broader discourse in the field.
\section{Conclusion}
In this work we provide evidence for (1) the need for sanity checks to ensure the faithfulness of post-hoc attribution methods \COMMENTEDIT{}{(see \Cref{sec:results_sanity})}, (2) the prospects of using segmentation maps to \COMMENTEDIT{}{align and} aggregate attribution maps both temporally (in terms of ECG segments) and spatially (in terms of leads) to \COMMENTEDIT{characterize}{explain} model behavior across entire patient populations \COMMENTEDIT{}{(see \Cref{sec:glocal_xai_exp})}, (3) the power of using concept-based XAI methods to verify if expert concepts based on ECG features are consistently exploited by a model \COMMENTEDIT{}{(see \Cref{sec:results_tcav})}, and (4) the ability to discover \COMMENTEDIT{subcategories}{previously unknown sub-conditions} from aggregated attributions that remain hidden from raw signals or hidden model representations \COMMENTEDIT{}{see \Cref{sec:discovery}}. \COMMENTEDIT{Both the glocal analysis (2) and the global analysis (3) reveal hints for a quantitative alignment with cardiologists' decision rules and represent two complementary but consistent views on the model behavior.}{Both, the glocal and global analyses offer complementary yet consistent insights into the model's behavior, suggesting a quantitative alignment with cardiologists' decision rules. In particular, both methods agree on that the model is exploiting following pathology-concept-pairs: (1) for LVH an increase in R-Amplitude (Sokolow-Lyon-Index), (2) for CLBBB a widening of QRS complex to at least 120 ms and (3) for AMI/IMI pathological Q-waves in specific leads (specific to the localization).} We believe that our four main findings are of utmost importance for the future use of XAI methods in the ECG domain with auditing or knowledge discovery applications in mind.

\bibliography{bibfile}

\begin{thebibliography}{10}

\bibitem{NACMS2016}
CDC, ``{National Ambulatory Medical Care Survey: 2016 National Summary
  Tables},'' tech. rep., Centers for Disease Control and Prevention, 2019.

\bibitem{schlapfer2017computer}
J.~Schl{\"a}pfer and H.~J. Wellens, ``Computer-interpreted electrocardiograms:
  benefits and limitations,'' {\em Journal of the American College of
  Cardiology}, vol.~70, no.~9, pp.~1183--1192, 2017.

\bibitem{TOPOL2021785}
E.~J. Topol, ``What's lurking in your electrocardiogram?,'' {\em The Lancet},
  vol.~397, no.~10276, p.~785, 2021.

\bibitem{strodthoff2018detecting}
N.~Strodthoff and C.~Strodthoff, ``Detecting and interpreting myocardial
  infarction using fully convolutional neural networks,'' {\em Physiological
  Measurement}, vol.~40, p.~015001, jan 2019.

\bibitem{Strodthoff:2020Deep}
N.~Strodthoff, P.~Wagner, T.~Schaeffter, and W.~Samek, ``Deep learning for
  {ECG} analysis: Benchmarks and insights from {PTB}-{XL},'' {\em {IEEE}
  Journal of Biomedical and Health Informatics}, pp.~1--1, 2020.

\bibitem{Kashou2020}
A.~H. Kashou, W.-Y. Ko, Z.~I. Attia, M.~S. Cohen, P.~A. Friedman, and P.~A.
  Noseworthy, ``A comprehensive artificial intelligence{\textendash}enabled
  electrocardiogram interpretation program,'' {\em Cardiovascular Digital
  Health Journal}, vol.~1, pp.~62--70, Sept. 2020.

\bibitem{Hannun2019}
A.~Y. Hannun, P.~Rajpurkar, M.~Haghpanahi, G.~H. Tison, C.~Bourn, M.~P.
  Turakhia, and A.~Y. Ng, ``Cardiologist-level arrhythmia detection and
  classification in ambulatory electrocardiograms using a deep neural
  network,'' {\em Nature Medicine}, vol.~25, pp.~65--69, Jan. 2019.

\bibitem{Tison2019}
G.~H. Tison, J.~Zhang, F.~N. Delling, and R.~C. Deo, ``Automated and
  interpretable patient {ECG} profiles for disease detection, tracking, and
  discovery,'' {\em Circulation: Cardiovascular Quality and Outcomes}, vol.~12,
  Sept. 2019.

\bibitem{Attia2019}
Z.~I. Attia, P.~A. Friedman, P.~A. Noseworthy, F.~Lopez-Jimenez, D.~J. Ladewig,
  G.~Satam, P.~A. Pellikka, T.~M. Munger, S.~J. Asirvatham, C.~G. Scott, R.~E.
  Carter, and S.~Kapa, ``Age and sex estimation using artificial intelligence
  from standard 12-lead {ECGs},'' {\em Circulation: Arrhythmia and
  Electrophysiology}, vol.~12, Sept. 2019.

\bibitem{Attia2019b}
Z.~I. Attia, S.~Kapa, F.~Lopez-Jimenez, P.~M. McKie, D.~J. Ladewig, G.~Satam,
  P.~A. Pellikka, M.~Enriquez-Sarano, P.~A. Noseworthy, T.~M. Munger, S.~J.
  Asirvatham, C.~G. Scott, R.~E. Carter, and P.~A. Friedman, ``Screening for
  cardiac contractile dysfunction using an artificial
  intelligence{\textendash}enabled electrocardiogram,'' {\em Nature Medicine},
  vol.~25, pp.~70--74, Jan. 2019.

\bibitem{attia2019artificial}
Z.~I. Attia, P.~A. Noseworthy, F.~Lopez-Jimenez, S.~J. Asirvatham, A.~J.
  Deshmukh, B.~J. Gersh, R.~E. Carter, X.~Yao, A.~A. Rabinstein, B.~J.
  Erickson, {\em et~al.}, ``An artificial intelligence-enabled ecg algorithm
  for the identification of patients with atrial fibrillation during sinus
  rhythm: a retrospective analysis of outcome prediction,'' {\em The Lancet},
  vol.~394, no.~10201, pp.~861--867, 2019.

\bibitem{Kwon2020}
J.~myoung Kwon, Y.~Cho, K.-H. Jeon, S.~Cho, K.-H. Kim, S.~D. Baek, S.~Jeung,
  J.~Park, and B.-H. Oh, ``A deep learning algorithm to detect anaemia with
  {ECGs}: a retrospective, multicentre study,'' {\em The Lancet Digital
  Health}, vol.~2, pp.~e358--e367, July 2020.

\bibitem{Kulkarnibmjinnov-2021-000759}
A.~R. Kulkarni, A.~A. Patel, K.~V. Pipal, S.~G. Jaiswal, M.~T. Jaisinghani,
  V.~Thulkar, L.~Gajbhiye, P.~Gondane, A.~B. Patel, M.~Mamtani, and
  H.~Kulkarni, ``Machine-learning algorithm to non-invasively detect diabetes
  and pre-diabetes from electrocardiogram,'' {\em BMJ Innovations}, 2022.

\bibitem{Ahn2021}
J.~C. Ahn, Z.~I. Attia, P.~Rattan, A.~F. Mullan, S.~Buryska, A.~M. Allen, P.~S.
  Kamath, P.~A. Friedman, V.~H. Shah, P.~A. Noseworthy, and D.~A. Simonetto,
  ``Development of the {AI}-cirrhosis-{ECG} score: An electrocardiogram-based
  deep learning model in cirrhosis,'' {\em American Journal of
  Gastroenterology}, vol.~117, pp.~424--432, Dec. 2021.

\bibitem{covert2021explaining}
I.~Covert, S.~Lundberg, and S.-I. Lee, ``Explaining by removing: A unified
  framework for model explanation,'' {\em Journal of Machine Learning
  Research}, vol.~22, no.~209, pp.~1--90, 2021.

\bibitem{lundberg2017unified}
S.~M. Lundberg and S.-I. Lee, ``A unified approach to interpreting model
  predictions,'' {\em Advances in Neural Information Processing Systems},
  vol.~30, 2017.

\bibitem{Samek2021}
W.~Samek, G.~Montavon, S.~Lapuschkin, C.~J. Anders, and K.-R. Müller,
  ``Explaining deep neural networks and beyond: A review of methods and
  applications,'' {\em Proceedings of the IEEE}, vol.~109, no.~3, pp.~247--278,
  2021.

\bibitem{saporta2022benchmarking}
A.~Saporta, X.~Gui, A.~Agrawal, A.~Pareek, S.~Q. Truong, C.~D. Nguyen, V.-D.
  Ngo, J.~Seekins, F.~G. Blankenberg, A.~Y. Ng, {\em et~al.}, ``Benchmarking
  saliency methods for chest x-ray interpretation,'' {\em Nature Machine
  Intelligence}, vol.~4, no.~10, pp.~867--878, 2022.

\bibitem{DeGrave2021}
A.~J. DeGrave, J.~D. Janizek, and S.-I. Lee, ``{AI} for radiographic {COVID}-19
  detection selects shortcuts over signal,'' {\em Nature Machine Intelligence},
  vol.~3, pp.~610--619, May 2021.

\bibitem{lapuschkin2019unmasking}
S.~Lapuschkin, S.~W{\"a}ldchen, A.~Binder, G.~Montavon, W.~Samek, and K.-R.
  M{\"u}ller, ``Unmasking clever hans predictors and assessing what machines
  really learn,'' {\em Nature communications}, vol.~10, no.~1, pp.~1--8, 2019.

\bibitem{Ayano2022}
Y.~M. Ayano, F.~Schwenker, B.~D. Dufera, and T.~G. Debelee, ``Interpretable
  machine learning techniques in {ECG}-based heart disease classification: A
  systematic review,'' {\em Diagnostics}, vol.~13, p.~111, Dec. 2022.

\bibitem{YAO2020174}
Q.~Yao, R.~Wang, X.~Fan, J.~Liu, and Y.~Li, ``Multi-class arrhythmia detection
  from 12-lead varied-length ecg using attention-based time-incremental
  convolutional neural network,'' {\em Information Fusion}, vol.~53,
  pp.~174--182, 2020.

\bibitem{Elul2021}
Y.~Elul, A.~A. Rosenberg, A.~Schuster, A.~M. Bronstein, and Y.~Yaniv, ``Meeting
  the unmet needs of clinicians from {AI} systems showcased for cardiology with
  deep-learning{\textendash}based {ECG} analysis,'' {\em Proceedings of the
  National Academy of Sciences}, vol.~118, June 2021.

\bibitem{Selvaraju2019}
R.~R. Selvaraju, M.~Cogswell, A.~Das, R.~Vedantam, D.~Parikh, and D.~Batra,
  ``Grad-{CAM}: Visual explanations from deep networks via gradient-based
  localization,'' {\em International Journal of Computer Vision}, vol.~128,
  pp.~336--359, Oct. 2019.

\bibitem{Raghunath2020}
S.~Raghunath, A.~E.~U. Cerna, L.~Jing, D.~P. vanMaanen, J.~Stough, D.~N.
  Hartzel, J.~B. Leader, H.~L. Kirchner, M.~C. Stumpe, A.~Hafez, A.~Nemani,
  T.~Carbonati, K.~W. Johnson, K.~Young, C.~W. Good, J.~M. Pfeifer, A.~A.
  Patel, B.~P. Delisle, A.~Alsaid, D.~Beer, C.~M. Haggerty, and B.~K. Fornwalt,
  ``Prediction of mortality from 12-lead electrocardiogram voltage data using a
  deep neural network,'' {\em Nature Medicine}, vol.~26, pp.~886--891, May
  2020.

\bibitem{van2021discovering}
R.~R. van~de Leur, K.~Taha, M.~N. Bos, J.~F. van~der Heijden, D.~Gupta, M.~J.
  Cramer, R.~J. Hassink, P.~van~der Harst, P.~A. Doevendans, F.~W. Asselbergs,
  {\em et~al.}, ``Discovering and visualizing disease-specific
  electrocardiogram features using deep learning: proof-of-concept in
  phospholamban gene mutation carriers,'' {\em Circulation: Arrhythmia and
  Electrophysiology}, vol.~14, no.~2, p.~e009056, 2021.

\bibitem{goodfellow2018towards}
S.~D. Goodfellow, A.~Goodwin, R.~Greer, P.~C. Laussen, M.~Mazwi, and D.~Eytan,
  ``Towards understanding ecg rhythm classification using convolutional neural
  networks and attention mappings,'' in {\em Machine learning for healthcare
  conference}, pp.~83--101, PMLR, 2018.

\bibitem{hicks2021explaining}
S.~A. Hicks, J.~L. Isaksen, V.~Thambawita, J.~Ghouse, G.~Ahlberg, A.~Linneberg,
  N.~Grarup, I.~Str{\"u}mke, C.~Ellervik, M.~S. Olesen, {\em et~al.},
  ``Explaining deep neural networks for knowledge discovery in
  electrocardiogram analysis,'' {\em Scientific reports}, vol.~11, no.~1,
  pp.~1--11, 2021.

\bibitem{lu2024decoding}
L.~Lu, T.~Zhu, A.~H. Ribeiro, L.~Clifton, E.~Zhao, J.~Zhou, A.~L.~P. Ribeiro,
  Y.-T. Zhang, and D.~A. Clifton, ``Decoding 2.3 million ecgs: Interpretable
  deep learning for advancing cardiovascular diagnosis and mortality risk
  stratification,'' {\em European Heart Journal-Digital Health}, p.~ztae014,
  2024.

\bibitem{simonyan2013deep}
K.~Simonyan, A.~Vedaldi, and A.~Zisserman, ``Deep inside convolutional
  networks: Visualising image classification models and saliency maps,'' {\em
  arXiv preprint arXiv:1312.6034}, 2013.

\bibitem{kwon2020deep}
J.-M. Kwon, S.~Y. Lee, K.-H. Jeon, Y.~Lee, K.-H. Kim, J.~Park, B.-H. Oh, and
  M.-M. Lee, ``Deep learning--based algorithm for detecting aortic stenosis
  using electrocardiography,'' {\em Journal of the American Heart Association},
  vol.~9, no.~7, p.~e014717, 2020.

\bibitem{Jones9288132}
Y.~Jones, F.~Deligianni, and J.~Dalton, ``Improving ecg classification
  interpretability using saliency maps,'' in {\em 2020 IEEE 20th International
  Conference on Bioinformatics and Bioengineering (BIBE)}, pp.~675--682, 2020.

\bibitem{lima2021deep}
E.~M. Lima, A.~H. Ribeiro, G.~M. Paix{\~a}o, M.~H. Ribeiro, M.~M. Pinto-Filho,
  P.~R. Gomes, D.~M. Oliveira, E.~C. Sabino, B.~B. Duncan, L.~Giatti, {\em
  et~al.}, ``Deep neural network-estimated electrocardiographic age as a
  mortality predictor,'' {\em Nature communications}, vol.~12, no.~1,
  pp.~1--10, 2021.

\bibitem{cho2020artificial}
Y.~Cho, J.-m. Kwon, K.-H. Kim, J.~R. Medina-Inojosa, K.-H. Jeon, S.~Cho, S.~Y.
  Lee, J.~Park, and B.-H. Oh, ``Artificial intelligence algorithm for detecting
  myocardial infarction using six-lead electrocardiography,'' {\em Scientific
  reports}, vol.~10, no.~1, pp.~1--10, 2020.

\bibitem{Hughes2021}
J.~W. Hughes, J.~E. Olgin, R.~Avram, S.~A. Abreau, T.~Sittler, K.~Radia,
  H.~Hsia, T.~Walters, B.~Lee, J.~E. Gonzalez, and G.~H. Tison, ``Performance
  of a convolutional neural network and explainability technique for 12-lead
  electrocardiogram interpretation,'' {\em {JAMA} Cardiology}, vol.~6, p.~1285,
  Nov. 2021.

\bibitem{bender2022analysis}
T.~Bender, J.~M. Beinecke, D.~Krefting, C.~M\"{u}ller, H.~Dathe, T.~Seidler,
  N.~Spicher, and A.-C. Hauschild, ``Analysis of a deep learning model for
  12-lead {ECG} classification reveals learned features similar to diagnostic
  criteria,'' {\em {IEEE} Journal of Biomedical and Health Informatics},
  pp.~1--12, 2023.

\bibitem{kim2018interpretability}
B.~Kim, M.~Wattenberg, J.~Gilmer, C.~Cai, J.~Wexler, F.~Viegas, {\em et~al.},
  ``Interpretability beyond feature attribution: Quantitative testing with
  concept activation vectors (tcav),'' in {\em International conference on
  machine learning}, pp.~2668--2677, PMLR, 2018.

\bibitem{Wagner:2020PTBXL}
P.~Wagner, N.~Strodthoff, R.-D. Bousseljot, D.~Kreiseler, F.~I. Lunze,
  W.~Samek, and T.~Schaeffter, ``{PTB}-{XL}, a large publicly available
  electrocardiography dataset,'' {\em Scientific Data}, vol.~7, no.~1, p.~154,
  2020.

\bibitem{Wagner2020:ptbxlphysionet}
P.~Wagner, N.~Strodthoff, R.-D. Bousseljot, W.~Samek, and T.~Schaeffter,
  ``{PTB-XL, a large publicly available electrocardiography dataset},'' {\em
  PhysioNet}, 2020.

\bibitem{Goldberger2020:physionet}
A.~L. Goldberger, L.~A.~N. Amaral, L.~Glass, J.~M. Hausdorff, P.~C. Ivanov,
  R.~G. Mark, J.~E. Mietus, G.~B. Moody, C.-K. Peng, and H.~E. Stanley,
  ``{PhysioBank, PhysioToolkit, and PhysioNet},'' {\em Circulation}, vol.~101,
  no.~23, pp.~e215--e220, 2000.

\bibitem{Glasgow}
P.~Macfarlane, B.~Devine, and E.~Clark, ``The university of {G}lasgow {(Uni-G)}
  {ECG} analysis program,'' in {\em Computers in Cardiology, 2005},
  pp.~451--454, 2005.

\bibitem{Strodthoff:2023PTBXLplus}
N.~Strodthoff, T.~Mehari, C.~Nagel, P.~J. Aston, A.~Sundar, C.~Graff, J.~K.
  Kanters, W.~Haverkamp, O.~D{\"o}ssel, A.~Loewe, M.~B{\"a}r, and
  T.~Schaeffter, ``{PTB-XL+, a comprehensive electrocardiographic feature
  dataset},'' {\em Scientific Data}, vol.~10, no.~1, 2023.

\bibitem{Ribeiro2020}
A.~H. Ribeiro, M.~H. Ribeiro, G.~M.~M. Paix{\~{a}}o, D.~M. Oliveira, P.~R.
  Gomes, J.~A. Canazart, M.~P.~S. Ferreira, C.~R. Andersson, P.~W. Macfarlane,
  W.~Meira, T.~B. Sch\"{o}n, and A.~L.~P. Ribeiro, ``Automatic diagnosis of the
  12-lead {ECG} using a deep neural network,'' {\em Nature Communications},
  vol.~11, Apr. 2020.

\bibitem{Mehari:2021Self}
T.~Mehari and N.~Strodthoff, ``Self-supervised representation learning from
  12-lead {ECG} data,'' {\em Computers in Biology and Medicine}, vol.~141,
  p.~105114, 2022.

\bibitem{mehari2022advancing}
T.~Mehari and N.~Strodthoff, ``Towards quantitative precision for ecg analysis:
  Leveraging state space models, self-supervision and patient metadata,'' {\em
  IEEE Journal of Biomedical and Health Informatics}, 2023.

\bibitem{lecun1989backpropagation}
Y.~LeCun, B.~Boser, J.~S. Denker, D.~Henderson, R.~E. Howard, W.~Hubbard, and
  L.~D. Jackel, ``Backpropagation applied to handwritten zip code
  recognition,'' {\em Neural computation}, vol.~1, no.~4, pp.~541--551, 1989.

\bibitem{he2016deep}
K.~He, X.~Zhang, S.~Ren, and J.~Sun, ``Deep residual learning for image
  recognition,'' in {\em Proceedings of the IEEE conference on computer vision
  and pattern recognition}, pp.~770--778, 2016.

\bibitem{he2019bag}
T.~He, Z.~Zhang, H.~Zhang, Z.~Zhang, J.~Xie, and M.~Li, ``Bag of tricks for
  image classification with convolutional neural networks,'' in {\em
  Proceedings of the IEEE Conference on Computer Vision and Pattern
  Recognition}, pp.~558--567, 2019.

\bibitem{sundararajan2017axiomatic}
M.~Sundararajan, A.~Taly, and Q.~Yan, ``Axiomatic attribution for deep
  networks,'' in {\em International Conference on Machine Learning},
  pp.~3319--3328, PMLR, 2017.

\bibitem{Bach2015}
S.~Bach, A.~Binder, G.~Montavon, F.~Klauschen, K.-R. M\"{u}ller, and W.~Samek,
  ``On pixel-wise explanations for non-linear classifier decisions by
  layer-wise relevance propagation,'' {\em {PLOS} {ONE}}, vol.~10, p.~e0130140,
  July 2015.

\bibitem{schwalbe2023comprehensive}
G.~Schwalbe and B.~Finzel, ``A comprehensive taxonomy for explainable
  artificial intelligence: a systematic survey of surveys on methods and
  concepts,'' {\em Data Mining and Knowledge Discovery}, pp.~1--59, 2023.

\bibitem{doumard2023quantitative}
E.~Doumard, J.~Aligon, E.~Escriva, J.-B. Excoffier, P.~Monsarrat, and
  C.~Soul{\'e}-Dupuy, ``A quantitative approach for the comparison of additive
  local explanation methods,'' {\em Information Systems}, vol.~114, p.~102162,
  2023.

\bibitem{ronneberger2015u}
O.~Ronneberger, P.~Fischer, and T.~Brox, ``U-net: Convolutional networks for
  biomedical image segmentation,'' in {\em International Conference on Medical
  image computing and computer-assisted intervention}, pp.~234--241, Springer,
  2015.

\bibitem{Pilia2021}
N.~Pilia, C.~Nagel, G.~Lenis, S.~Becker, O.~D\"{o}ssel, and A.~Loewe,
  ``{ECGdeli}~- an open source {ECG} delineation toolbox for {MATLAB},'' {\em
  {SoftwareX}}, vol.~13, p.~100639, Jan. 2021.

\bibitem{ptbxl_segmentations}
P.~Wagner, T.~Mehari, W.~Haverkamp, and N.~Strodthoff, ``{PTB-XL (v1.0.1) Soft
  Segmentations (Delineation)}.'' \url{https://doi.org/10.5281/zenodo.7610236},
  Feb. 2023.

\bibitem{beer2020using}
T.~Beer, B.~Eini-Porat, S.~Goodfellow, D.~Eytan, and U.~Shalit, ``Using deep
  networks for scientific discovery in physiological signals,'' {\em arXiv
  preprint arXiv:2008.10936}, 2020.

\bibitem{finzel2022generating}
B.~Finzel, A.~Saranti, A.~Angerschmid, D.~Tafler, B.~Pfeifer, and A.~Holzinger,
  ``Generating explanations for conceptual validation of graph neural networks:
  An investigation of symbolic predicates learned on relevance-ranked
  sub-graphs,'' {\em KI-K{\"u}nstliche Intelligenz}, vol.~36, no.~3-4,
  pp.~271--285, 2022.

\bibitem{captum2019github}
N.~Kokhlikyan, V.~Miglani, M.~Martin, E.~Wang, J.~Reynolds, A.~Melnikov,
  N.~Lunova, and O.~Reblitz-Richardson, ``Pytorch captum.''
  \url{https://github.com/pytorch/captum}, 2019.

\bibitem{anders2021software}
C.~J. Anders, D.~Neumann, W.~Samek, K.-R. Müller, and S.~Lapuschkin,
  ``Software for dataset-wide xai: From local explanations to global insights
  with {Zennit}, {CoRelAy}, and {ViRelAy},'' {\em arxiv preprint
  arxiv:2106.13200}, 2021.

\bibitem{thygesen2018fourth}
K.~Thygesen, J.~S. Alpert, A.~S. Jaffe, B.~R. Chaitman, J.~J. Bax, D.~A.
  Morrow, H.~D. White, and E.~G. on~behalf of the Joint European Society of
  Cardiology (ESC)/American College of Cardiology (ACC)/American Heart
  Association (AHA)/World Heart Federation (WHF) Task Force for the Universal
  Definition~of Myocardial~Infarction, ``Fourth universal definition of
  myocardial infarction (2018),'' {\em Journal of the American College of
  Cardiology}, vol.~72, no.~18, pp.~2231--2264, 2018.

\bibitem{rudin2019stop}
C.~Rudin, ``Stop explaining black box machine learning models for high stakes
  decisions and use interpretable models instead. nat mach intell 1:
  206--215,'' {\em DOI: https://doi. org/10.1038/s42256-019-0048-x}, 2019.

\bibitem{krishna2022disagreement}
S.~Krishna, T.~Han, A.~Gu, J.~Pombra, S.~Jabbari, S.~Wu, and H.~Lakkaraju,
  ``The disagreement problem in explainable machine learning: A practitioner's
  perspective,'' {\em arXiv preprint arXiv:2202.01602}, 2022.

\bibitem{sokolow1949ventricular}
M.~Sokolow and T.~P. Lyon, ``The ventricular complex in left ventricular
  hypertrophy as obtained by unipolar precordial and limb leads,'' {\em
  American heart journal}, vol.~37, no.~2, pp.~161--186, 1949.

\bibitem{surawicz2009aha}
B.~Surawicz, R.~Childers, B.~J. Deal, and L.~S. Gettes, ``Aha/accf/hrs
  recommendations for the standardization and interpretation of the
  electrocardiogram: part iii: intraventricular conduction disturbances a
  scientific statement from the american heart association electrocardiography
  and arrhythmias committee, council on clinical cardiology; the american
  college of cardiology foundation; and the heart rhythm society endorsed by
  the international society for computerized electrocardiology,'' {\em Journal
  of the American College of Cardiology}, vol.~53, no.~11, pp.~976--981, 2009.

\bibitem{nikoo2013lbbb}
M.~H. Nikoo, A.~Aslani, and M.~V. Jorat, ``Lbbb: state-of-the-art criteria,''
  {\em International Cardiovascular Research Journal}, 2013.

\bibitem{rautaharju2007investigative}
P.~Rautaharju and F.~Rautaharju, {\em Investigative electrocardiography in
  epidemiological studies and clinical trials}.
\newblock Springer Science \& Business Media, 2007.

\bibitem{Samesima2017}
N.~Samesima, L.~F. Azevedo, L.~D. N. J.~D. Matos, L.~S. Echenique, C.~E.
  Negrao, and C.~A. Pastore, ``Comparison of electrocardiographic criteria for
  identifying left ventricular hypertrophy in athletes from different sports
  modalities,'' {\em Clinics}, vol.~72, no.~6, pp.~343--350, 2017.

\bibitem{crabbe2022concept}
J.~Crabb{\'e} and M.~van~der Schaar, ``Concept activation regions: A
  generalized framework for concept-based explanations,'' {\em Advances in
  Neural Information Processing Systems}, vol.~35, pp.~2590--2607, 2022.

\bibitem{SeyyedKalantari2021}
L.~Seyyed-Kalantari, H.~Zhang, M.~B.~A. McDermott, I.~Y. Chen, and M.~Ghassemi,
  ``Underdiagnosis bias of artificial intelligence algorithms applied to chest
  radiographs in under-served patient populations,'' {\em Nature Medicine},
  vol.~27, pp.~2176--2182, Dec. 2021.

\bibitem{vielhaben2023multi}
J.~Vielhaben, S.~Bl{\"u}cher, and N.~Strodthoff, ``{Multi-dimensional concept
  discovery (MCD): A unifying framework with completeness guarantees},'' {\em
  Transactions on Machine Learning Research}, 2023.

\bibitem{achtibat2023attribution}
R.~Achtibat, M.~Dreyer, I.~Eisenbraun, S.~Bosse, T.~Wiegand, W.~Samek, and
  S.~Lapuschkin, ``From attribution maps to human-understandable explanations
  through concept relevance propagation,'' {\em Nature Machine Intelligence},
  vol.~5, no.~9, pp.~1006--1019, 2023.

\bibitem{lewis1913heart}
T.~Lewis, {\em Heart}, ch.~5, pp.~367--402.
\newblock Shaw, 1914.

\bibitem{Romhilt1969}
D.~W. Romhilt, K.~E. Bove, R.~J. Norris, E.~Conyers, S.~Conrada, D.~T.
  Rowlands, and R.~C. Scott, ``A critical appraisal of the electrocardiographic
  criteria for the diagnosis of left ventricular hypertrophy,'' {\em
  Circulation}, vol.~40, pp.~185--196, Aug. 1969.

\bibitem{ancona2018towards}
M.~Ancona, E.~Ceolini, C.~Öztireli, and M.~Gross, ``Towards better
  understanding of gradient-based attribution methods for deep neural
  networks,'' in {\em International Conference on Learning Representations},
  2018.

\end{thebibliography}
\bibliographystyle{ieeetr}
\vfill

\clearpage
\appendix
\setcounter{page}{1}
\pagenumbering{roman}
\section{Supplementary Material}
\subsection{Data and Models}
\label{sec:appendix_data}
\heading{Lead correlations}
\begin{figure}
    \centering
    \includegraphics[width=.5\textwidth]{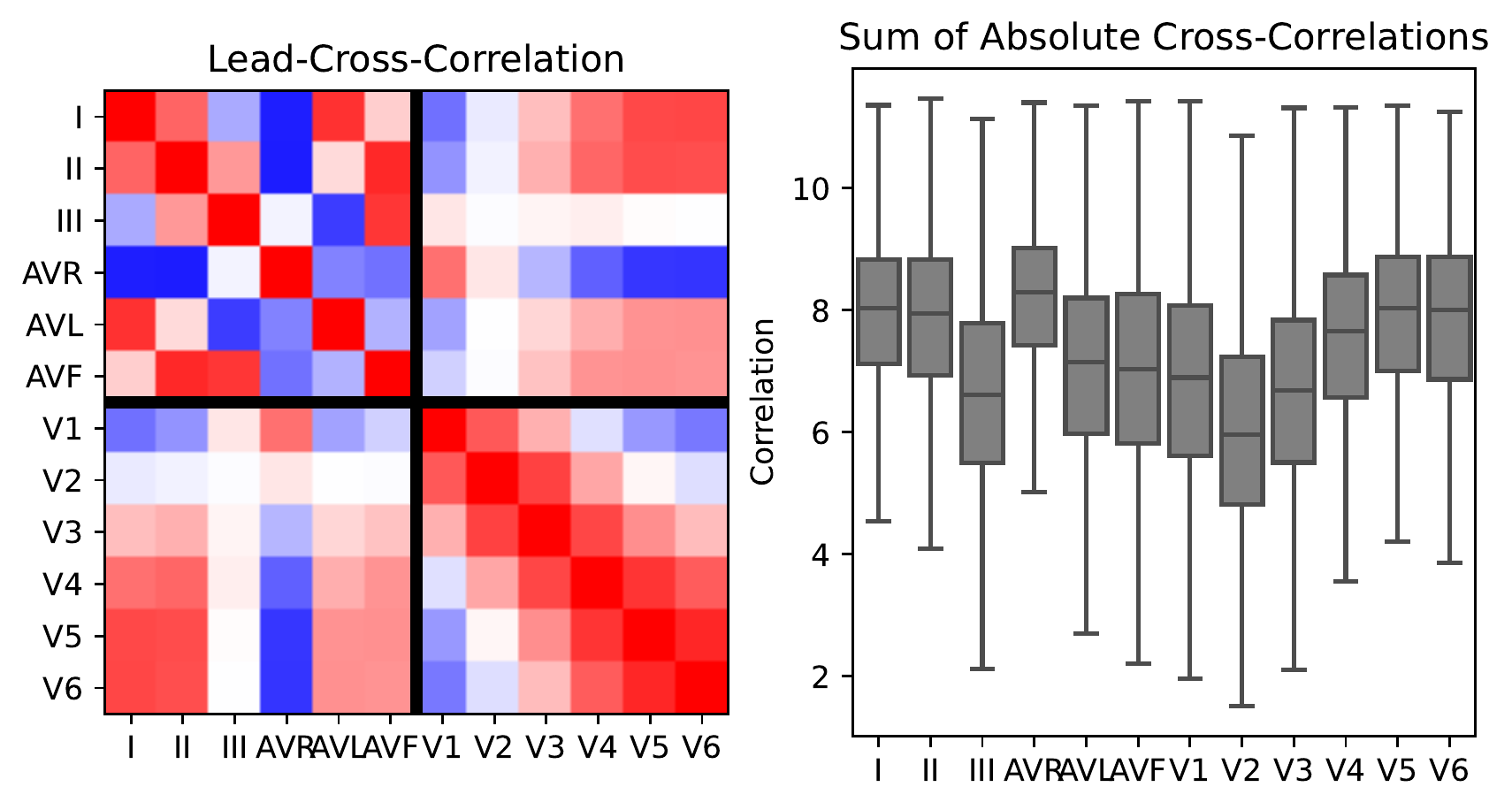}
    \caption{Cross-Correlations among leads showing leads V2, V3 and III have less cross-correlation compared to other leads.}
    \label{fig:data_correlations}
\end{figure}
In \Cref{fig:data_correlations} we provide cross-correlations among leads highlighting how much the leads in 12 lead ECG signals correlate with each other. This cross-correlation is most prominent in the limb  leads, where four out of six leads are synthetic, i.e. a linear combination of the remaining two leads, which are linearly independent. This fact provides evidence for the drop in the performance of attributions methods in terms of spatial specificity in some leads.

\heading{Model performance on PTB-XL}
For the task of predicting the sub-diagnostic labels of PTB-XL, in terms of macro AUC, i.e. the mean across all individual label AUCs, we observe that shallow convolutional models like \lenet{} (macro AUC  $0.9245\pm 0.0030$) are almost competitive with deep models like \xresnet{} (macro AUC $0.9292\pm 0.0043$). To put this into perspective, we also report the performance on the more comprehensive set of all 71 labels at the most finegrained level in PTB-XL, where we notice a bigger gap between the \xresnet{} model (macro AUC $0.9286\pm 0.0028$) and the  \lenet{} (macro AUC $0.9022\pm 0.0043$). Both values should be set into perspective with the performance of the best-performing models in \cite{Strodthoff:2020Deep} (sub-diagnostic: macro AUC $0.93$ and all: macro AUC $0.925$). 

\begin{figure}[ht!]
\centering
\begin{subfigure}{0.24\textwidth}
    \includegraphics[width=\textwidth]{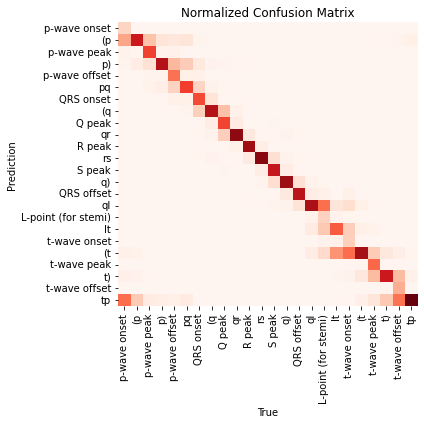}
    \caption{\REVIEWERTWO{}{Hard}}
    \label{fig:first}
\end{subfigure}
\begin{subfigure}{0.23\textwidth}
    \includegraphics[width=\textwidth]{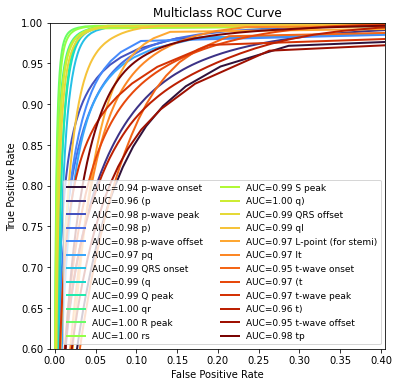}
    \caption{\REVIEWERTWO{}{Soft}}
    \label{fig:second}
\end{subfigure}
\begin{subfigure}{0.4\textwidth}
    \includegraphics[width=\textwidth]{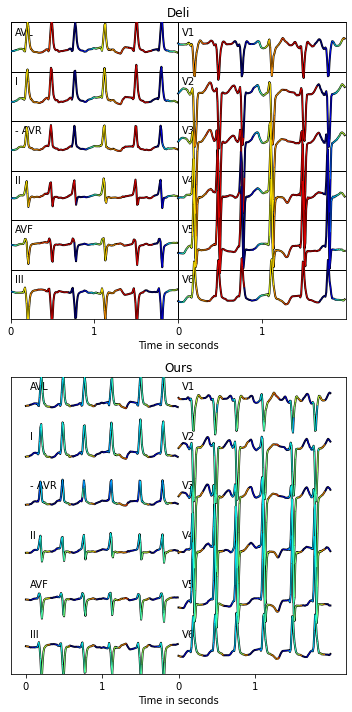}
    \caption{\REVIEWERTWO{}{Comparison between ECGDeli segmentations (upper) and the proposed approach (lower).}}
    \label{fig:third}
\end{subfigure}

\caption{\REVIEWERTWO{}{Evaluation results our proposed segmentation model which was trained on ECGDeli segmentations. \Cref{fig:first} evaluates the hard predictions after argmax based on confusion matrix. \Cref{fig:second} evaluates soft predictions based on class-wise ROC. \Cref{fig:third} shows a direct comparison of delination/segmentation between Deli and our model as described in \Cref{sec:glocal_xai} and evaluated in \Cref{sec:appendix_data}.}}
\label{fig:segmentation_performance}
\end{figure}

\REVIEWERTWO{}{\heading{Segmentation Model Performance}}
\REVIEWERTWO{}{In evaluating our proposed segmentation model, we conducted a comprehensive analysis considering both soft predictions, as measured by ROC curves, and hard predictions obtained after applying argmax, utilizing confusion matrices and classification reports (precision, recall, and F1 scores). The results are highly promising, showcasing a macro AUC of $0.98$ and a accuracy of $0.75$. Notably, our evaluation focused on the 6 seconds around the center of each sample (disregarding 2s at the beginning and the end of each of sample), as ECGDeli, representing the ground truth segmentation, exhibited limitations at signal borders. \Cref{fig:segmentation_performance} visually summarizes the quantitative outcomes for hard predictions (\Cref{fig:first}), soft predictions (\Cref{fig:second}), and provides a qualitative comparison that highlights potential weaknesses in ECGDeli's performance (\Cref{fig:third}), where in case of sinus tachycardia (STACH), i.e. more than 100 beats per minute, ECGDeli provides poor segmentations and fails to identify individual beats.  Additionally, \Cref{tab:segmentation_classification_report} offers detailed classification reports for each segment, providing a comprehensive overview of our model's efficacy according to different metrics. These findings highlights the robustness and reliability of our proposed segmentation approach.}

\begin{table}[ht!]
    \centering
    \REVIEWERTWO{}{
    \begin{tabular}{lrrrr}
        \toprule
        {} &  precision &  recall &  f1-score &      support \\
        \midrule
        p-wave onset        &       0.15 &    0.66 &      0.25 &    148240 \\
        (p                  &       0.70 &    0.73 &      0.71 &   4941068 \\
        p-wave peak         &       0.57 &    0.70 &      0.63 &    508346 \\
        p)                  &       0.77 &    0.76 &      0.77 &   3519907 \\
        p-wave offset       &       0.44 &    0.61 &      0.51 &    338247 \\
        pq                  &       0.58 &    0.69 &      0.63 &   1299086 \\
        QRS onset           &       0.54 &    0.65 &      0.59 &    537188 \\
        (q                  &       0.77 &    0.70 &      0.73 &   1134861 \\
        Q peak              &       0.56 &    0.71 &      0.63 &    495888 \\
        qr                  &       0.86 &    0.82 &      0.84 &   1326771 \\
        R peak              &       0.82 &    0.82 &      0.82 &    636917 \\
        rs                  &       0.86 &    0.84 &      0.85 &   1214691 \\
        S peak              &       0.72 &    0.78 &      0.75 &    590824 \\
        q)                  &       0.83 &    0.78 &      0.80 &   1089137 \\
        QRS offset          &       0.75 &    0.66 &      0.70 &    694432 \\
        ql                  &       0.79 &    0.67 &      0.73 &   2613817 \\
        L-point             &       0.16 &    0.50 &      0.25 &    197520 \\
        lt                  &       0.50 &    0.63 &      0.56 &   1688194 \\
        t-wave onset        &       0.17 &    0.54 &      0.26 &    202333 \\
        (t                  &       0.81 &    0.74 &      0.77 &   7615251 \\
        t-wave peak         &       0.47 &    0.65 &      0.55 &    443966 \\
        t)                  &       0.70 &    0.73 &      0.71 &   4583645 \\
        t-wave offset       &       0.27 &    0.64 &      0.38 &    244892 \\
        tp                  &       0.93 &    0.81 &      0.87 &  15825179 \\
        \midrule
        accuracy            &       0.75 &    0.75 &      0.75 &      0.75 \\
        macro avg           &       0.61 &    0.70 &      0.64 &  51890400 \\
        weighted avg        &       0.78 &    0.75 &      0.76 &  51890400 \\
        \bottomrule
    \end{tabular}}
    \caption{\REVIEWERTWO{}{Classification report in terms of class-wise precision, recall, f1-score and support for our proposed segmentation U-Net as described in \Cref{sec:glocal_xai} and evaluated in \Cref{sec:appendix_data}.}}
    \label{tab:segmentation_classification_report}
\end{table}

\pagebreak

\subsection{Sanity checks}
\label{sec:appendix_regression}
\heading{Performance evaluation}
In \Cref{tab:regression}, we describe the result of the regression experiment introduced in \Cref{sec:sanity}. The results in terms of mean absolute error and coefficient of determination are provided in \Cref{tab:regression}, showing slight advantages of the \lenet{} models in two of the three experiments. Overall, each considered model performs reasonably well such that the analysis is not affected significantly by the choice of the model architecture.
\begin{table}[ht!]
\centering
\definecolor{mygrey}{RGB}{192, 192, 192}
 \begin{tabular}{|l | c c |} 
 \hline
 \rowcolor{mygrey}
 Model & \text{MAE}$\downarrow$ & $\text{r}^2\uparrow$\\ \hline
 Task & \multicolumn{2}{c|}{P-Wave Amp.} \\
 \hline
\lenet{} & $0.0199\pm 0.0004$ & $0.7693\pm 0.0247$ \\

\xresnet{} & $\mathbf{0.0187\pm 0.0004}$ & $\mathbf{0.7822\pm 0.0219}$\\ \hline
Task & \multicolumn{2}{c|}{R-Peak Amp.}\\ \hline
 
\lenet{} & $\mathbf{0.0602\pm 0.0010}$ & $\mathbf{0.9742\pm 0.0021}$ \\

\xresnet{} & $0.0801\pm 0.0013$ & $0.9491\pm 0.0026$\\\hline
Task & \multicolumn{2}{c|}{T-Wave Amp.} \\  \hline
 
\lenet{} & $\mathbf{0.0352\pm 0.0008}$ & $\mathbf{0.9402\pm 0.0079}$\\

\xresnet{} & $0.0385\pm 0.0005$ & $0.9308\pm 0.0070$\\ \hline
 \end{tabular}
 \caption{Results of regression experiments as described in \Cref{sec:sanity}, where we report mean and standard deviation (obtained by bootstrapping) mean absolute error (MAE) and coefficient of determination ($r^2$) for each pair of models (\lenet{} and \xresnet{}) and task (P- and T-wave and R-peak amplitudes in all 12 leads).}
 \label{tab:regression}
\end{table}

\subsection{Global XAI}
\label{sec:appendix_concepts}
\heading{Concept definitions}
\begin{figure}
    \centering
    \includegraphics[width=.5\textwidth]{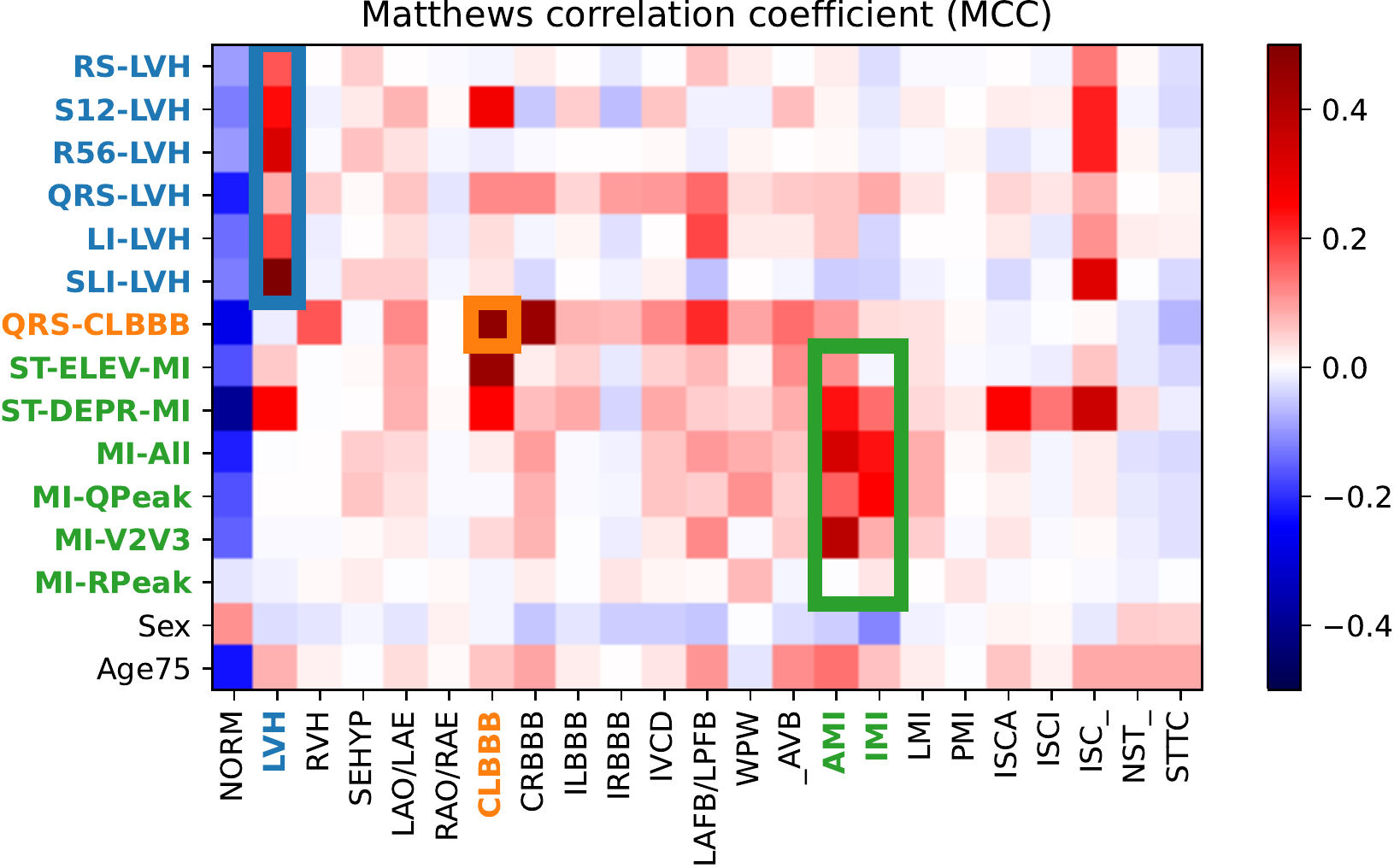}
    \caption{Matthews correlation coefficient (MCC) for all pairs of concepts and pathology.}
    \label{fig:mcc}
\end{figure}
We summarize commonly used concepts/decision rules for the pathologies considered in this work \Cref{Tab:rules} along with formalized versions of them in terms of automatically extracted ECG parameters. In \hbox{\Cref{fig:mcc}}, we analyze the overlap between annotations for particular pathologies and specific concepts in terms of Matthews correlation coefficients (MCCs). For the selection of the most discriminative concept for each pathology, we use the concept with the highest MCC value for the given pathology. Therefore, we picked \texttt{SLI-LVH} for \texttt{LVH}, \texttt{QRS-CLBBB} for \texttt{CLBBB} and \texttt{QWAVES-MI} for \texttt{IMI/AMI}.

\begin{table*}[ht!]
    \scriptsize
	\centering
 \definecolor{mygrey}{RGB}{192, 192, 192}
	\begin{tabular}{|ll|l|l|}
		\hline 
  \rowcolor{mygrey}
		&name & description & formalized condition \\
	\hline	
		
		\multirow{16}{*}{\rotatebox[origin=c]{90}{AMI/IMI}}
		
		& \texttt{V2V3-MI} & Any Q Wave in leads $V_2$-$V_3$ $\geq 0.02s$  &\hspace*{0.5mm} ((Q\_Dur\_V2 $>$0.02) \& (Q\_Dur\_V3 $>$0.02))  \\
		&&or QS complex in leads $V_2$-$V_3$ \cite{thygesen2018fourth}&   |  ((R\_Amp\_V2 $=$ 0) \& (R\_Amp\_V3 $=$0))            \\
		&&&\\
		
		& \texttt{RPEAK-MI} & R wave  $ \geq 0.04s$ in $V_1$-$V_2$ and $R/S \geq 1$  with  a& \hspace*{2mm}	(R\_Dur\_V1 $>$ 0.04)  \& (R\_Dur\_V2 > 0.04)     \\
		&  				&  concordant positive T wave in absence of conduction defect\cite{thygesen2018fourth}  & \& (R\_Amp\_V1 $>$ 0)  \& (R\_Amp\_V2 > 0)            \\
		&              &                                                                                                    &         \& (T\_Amp\_V1 $>$ 0)  \& (T\_Amp\_V2 > 0)       \\
			&              &                                                                                                    &      \& (abs(R\_Amp\_V1) $>$ abs(S\_Amp\_V1))      \\
				&              &                                                                                                    &  \& (abs(R\_Amp\_V2) $>$ abs(S\_Amp\_V2))     \\
		&&&\\
		
		& \texttt{QPEAK-MI} & Q wave  $ \geq 0.03s$ and $ \geq 1$mm or QS complex    & \hspace*{3mm}(Q\_Dur\_X$\geq 0.03$) \& (abs(Q\_Amp\_X)$\geq 0.1$)  \\
		&& in leads I, II, aVL, aVF or $V_4$-$V_6$ in any 2 leads  & \& (Q\_Dur\_Y$\geq 0.03$) \& (abs(Q\_Amp\_Y)$\geq 0.1$)  \\
		&&of a contiguous lead grouping (I, aVL; $V_1$-$V_6$; II, III, aVF) \cite{thygesen2018fourth} & \& $(X\neq Y)$ \& ($X,Y \in$ \{I, aVL\} \\
		&& & \hspace*{15mm} | $X,Y \in$ \{V4, V5, V6\}  | $X,Y \in$ \{II, aVF\}) \\
		&&&\\
		
		& \texttt{QWAVES-MI} & Combination of all three \cite{thygesen2018fourth} & \texttt{MI-V2V3} $\vee$ \texttt{MI-RPEAK} $\vee$ \texttt{QPEAK-MI} \\
		&&&\\

		\hline 
		\multirow{3}{*}{\rotatebox[origin=c]{90}{CLBBB}}
		&&&\\
		& \texttt{QRS-CLBBB} & QRS duration $\geq 0.12s$ \cite{surawicz2009aha} & QRS\_Dur\_Global  $\geq 0.12$\\
		&&&\\
		
		\hline
  		\multirow{18}{*}{\rotatebox[origin=c]{90}{ISC}}
		&&&\\
				& \texttt{ST-ELEV-ISC} & New ST-elevation at the J-Point in two contiguous leads & ($X\neq Y$)   \\
		& &with the cut-point: $\geq$ 1mm  in all leads other than leads $V_2$-$V_3$  &  \& (($X,Y \in$ \{I, aVL\}   | $X,Y \in$ \{V1, V4, V5, V6\}  | $X,Y \in$ \{II, III, aVF\}) \\
		& & where the following cut-points apply:  &\hspace*{5mm} \& ((ST\_Amp\_X $\geq$ 0.1)   \& (ST\_Amp\_Y $\geq$ 0.1))\\
		&&$\geq$ 2mm in men  $\geq$ 40 years; $\geq$ 2.5mm in men  $\leq$ 40 years& \hspace*{1.5mm} | ($X,Y \in$ \{V1, V2\} \\ 
		&&, or $\geq$ 1.5mm in women  regardless of age \cite{thygesen2018fourth}&   \hspace*{5mm} \& ((SEX=0 \& AGE$\geq$40 \& ST\_Amp\_X$\geq$0.15 \& ST\_Amp\_Y$\geq$0.15)  \\
		&&& \hspace*{7.5mm} | (SEX=0 \& AGE<40 \& ST\_Amp\_X$\geq$0.15 \& ST\_Amp\_Y$\geq$0.15)  \\
		&&& \hspace*{7.5mm} | (SEX=0 \& ST\_Amp\_X$\geq$0.15 \& ST\_Amp\_Y$\geq$0.15)))) \\ 
		&&&\\

		& \texttt{ST-DEPR-ISC} & New horizontal or downsloping ST-depression  $\geq$ 0.5mm  & ($X\neq Y$)    \\
		& & in two contiguous leads and/or T inversion > 1mm  &\& ($X,Y \in$ \{I, aVL\}  | $X,Y \in$ \{V1, V4, V5, V6\}  | $X,Y \in$ \{II, III, aVF\}) \\
		&&in two contiguous leads with prominent R wave  & \& ((ST\_Amp\_X $\leq$ -0.5   \& ST\_Amp\_Y $\leq$ -0.5)  \\
		&&or R/S ratio > 1 \cite{thygesen2018fourth}& \hspace*{3.5mm} |  ((T\_Morph\_X $= -1$ \& R\_Amp\_X$>2$  \\
		&&&  \hspace*{7.5mm} | abs(R\_Amp\_X) > abs(S\_Amp\_X))   \\
		&&& \hspace*{5.5mm} \&  T\_Morph\_Y $= -1$ \& R\_Amp\_Y$>2$)   \\
		&&&  \hspace*{8.7mm} | abs(R\_Amp\_Y) > abs(S\_Amp\_Y)))) \\
		&&&\\
		&&&\\
		
		\hline
		\multirow{9}{*}{\rotatebox[origin=c]{90}{LVH}}
		
		&\texttt{LI-LVH}&\emph{Lewis-Index} > 16mm \cite{lewis1913heart}&  R\_Amp\_I + S\_Amp\_III \\
		&           &                                         &   - R\_Amp\_III - S\_Amp\_I > 1.6                    \\
		&&&\\
		
		&\texttt{SLI-LVH}&\emph{Sokolow-Lyon-Index} > 35mm \cite{sokolow1949ventricular}& R\_Amp\_V5 + S\_Amp\_V1 > 3.5 \\
		&&&\\
		
		& & \emph{Romhilt-Estes Score} \cite{Romhilt1969} &\\
		
		&\texttt{RS-LVH} & includes \texttt{RS-LVH}, \texttt{S12-LVH}, \texttt{R56-LVH}: & ((R\_Amp\_I> 2)  | (R\_Amp\_II> 2) | \\

		&& Largest R or S Wave in the limb leads $\geq 20$mm  & (R\_Amp\_III> 2) |  (S\_Amp\_I> 2) | \\

		&& & (S\_Amp\_II> 2)   | (S\_Amp\_III> 2))   \\
		&&&\\
		
		&\texttt{S12-LVH}& S wave in $V_1$ or $V_2 \geq$ 30mm &   ((S\_Amp\_V1> 3) | (S\_Amp\_V2> 3))     \\
		&&&\\
		
		&\texttt{R56-LVH}& R wave in $V_5$ or $V_6  \geq$ 30mm & ((R\_Amp\_V5 > 3) | (R\_Amp\_V6 > 3)) \\
		&&&\\
		
		\hline 
		\multirow{3}{*}{\rotatebox[origin=c]{90}{Other}}
		& \texttt{SEX=FEMALE} & sex of the patient (female=True, male=False) &-\\
		&&&\\
		
		& \texttt{AGE>75} & age $\geq 75$ years & - \\
		
		\hline 
		
	\end{tabular}
	\caption{Formalized concept descriptions }
     \label{Tab:rules}
\end{table*}

\subsection{Glocal XAI}
\begin{figure}
    \centering
    \includegraphics[width=.5\textwidth]{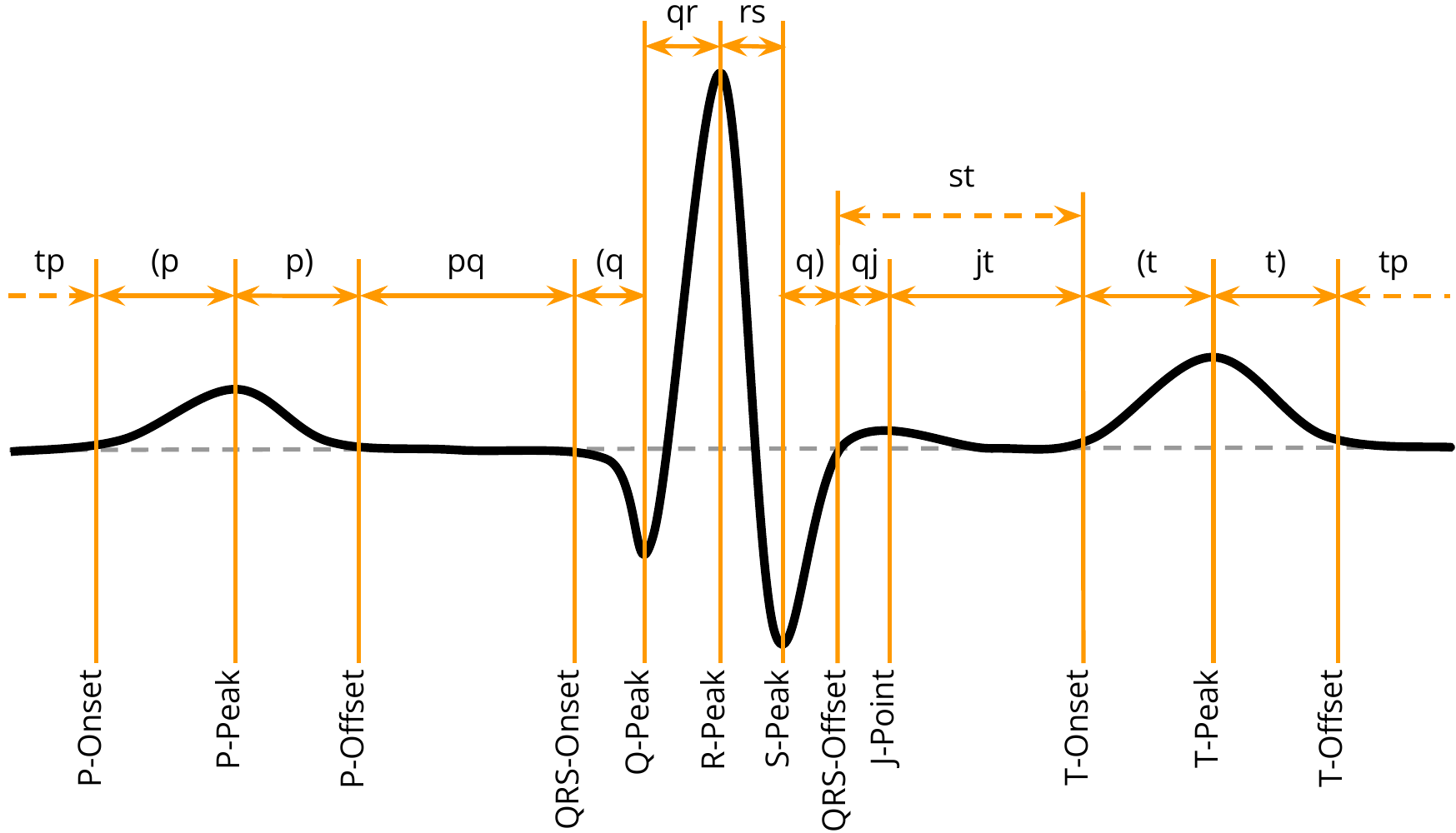}
    \caption{Our used segmentation/annotation schema based on ECGDeli \cite{Pilia2021} fiducial points (lower annotations), enriched by segments (upper annotations).}
    \label{fig:segmentation}
\end{figure}

\heading{ECG segments} In \Cref{fig:segmentation}, we visualize the ECG segments that were used to train the segmentation model.

\begin{figure*}[ht!]
    \centering
    \includegraphics[width=\textwidth]{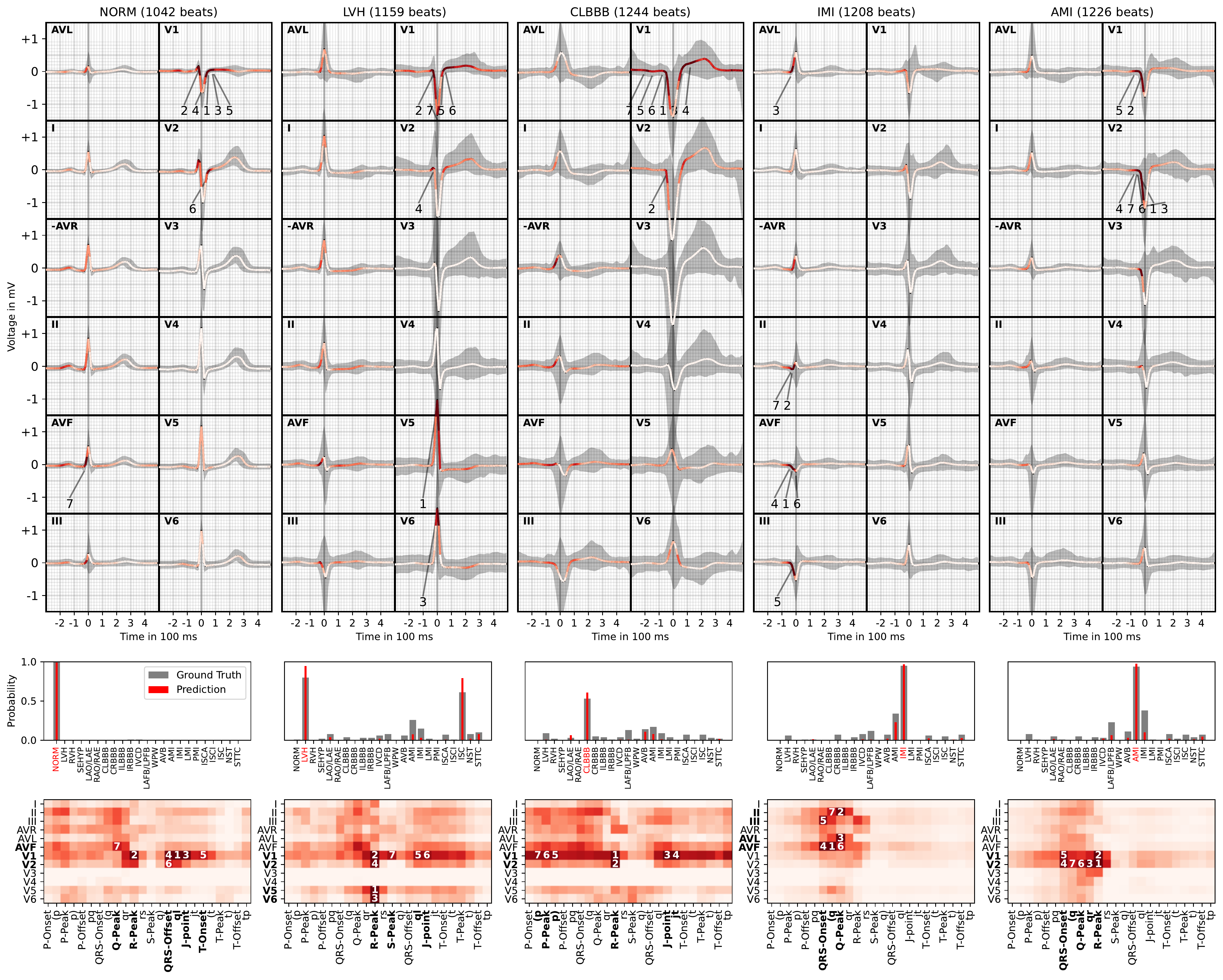}
    \caption{Results of the glocal (dataset-wide) analysis for the saliency as attribution method and \lenet{} as model architecture (analogous to \Cref{fig:glocal_analysis_xresnet}).}
    \label{fig:glocal_analysis_lenet}
\end{figure*}

\heading{Glocal XAI results for the LeNet model}
 \Cref{fig:glocal_analysis_lenet} shows the results of the glocal analysis (corresponding to \Cref{fig:glocal_analysis_xresnet} in the main text) for a \lenet{} instead of a \xresnet{} model architecture. 

\subsection{XAI methods}
\label{sec:appendix_xai_methods}
In order to make the paper self-contained, we provide short technical descriptions of the XAI methods used in this study. For more details, we refer the reader to the original publications. 

\begin{figure}
	\includegraphics[width=\columnwidth]{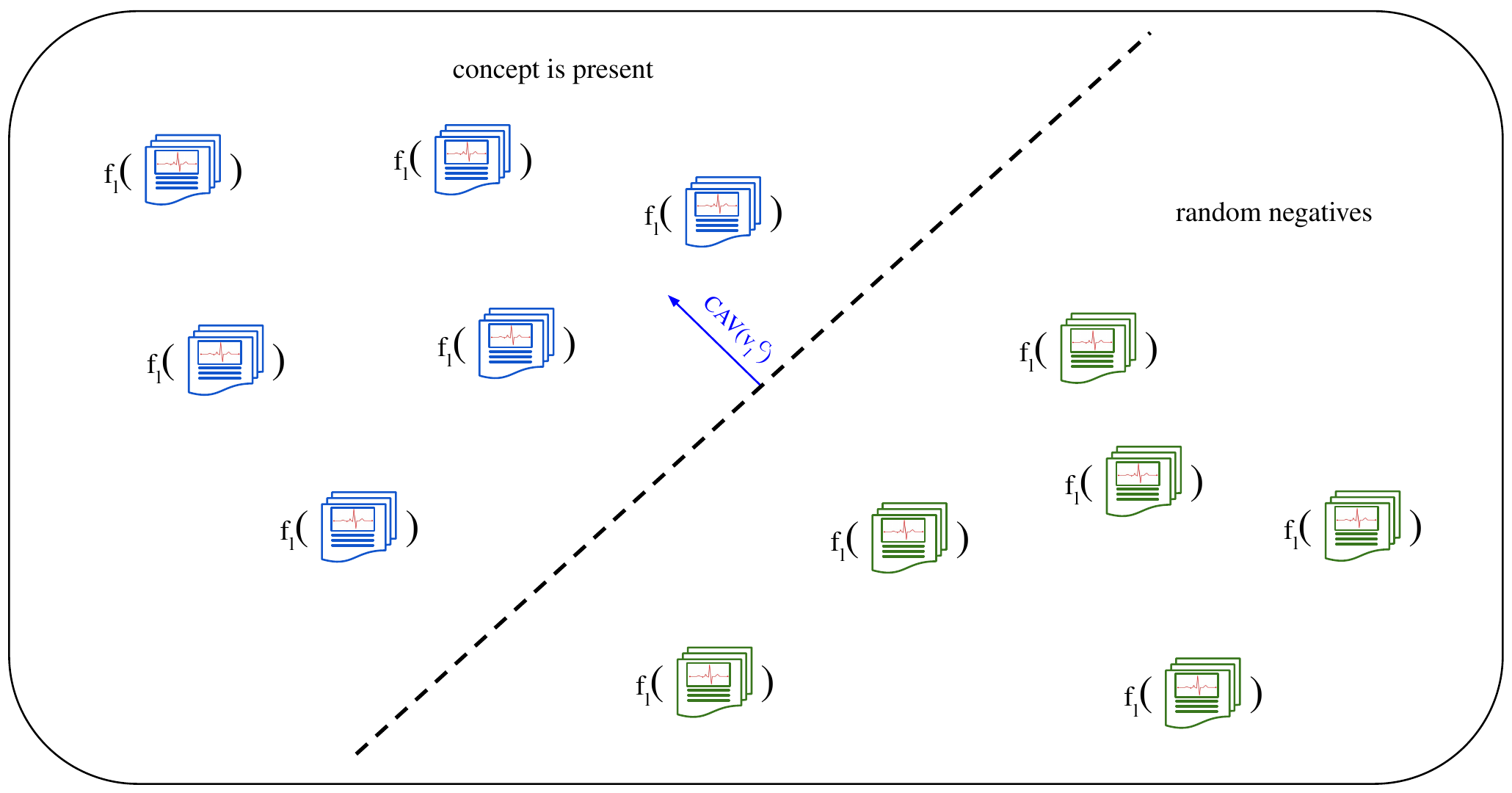}
	\caption{\small The concept activation vector ($v_l^C$) of concept $C$ at layer $l$ is a vector orthogonal to the classifier's decision boundary, which differentiates data points in which the concept occurs from random negative samples. All data points are projected into the feature space of layer $l$ of the given model. In other words, the linear classifier learns to detect for a data sample, whether the given concept C is present in the feature space at layer $l$ of the model.}
	\label{fig:cav}
\end{figure}

\heading{TCAV}
With TCAV \cite{kim2018interpretability} it is possible to test a trained model against \textit{human comprehensible} concepts. These concepts are implicitly defined by data. In contrast to the attribution maps, with TCAV we can explicitly determine, whether e.g. a high magnitude at the R-Peak of a signal is of high importance for the prediction of a certain pathology. In general, the TCAV approach works as follows. First, a binary data set is compiled per concept, where the concept occurs in the positive samples and the negative samples represent a random composition of data points in which the concept does not occur. Second, for every dataset, we train a linear classifier that tries to differentiate the positive samples from the negative samples in feature space. The orthogonal vector of the linear classifier is called \textit{concept activation vector (CAV)}. It provides the direction in which the concept lies in the feature space, as depicted in Figure \ref{fig:cav}. Using the CAV, we can determine the TCAV score. First, we calculate for every data point $x_i$ belonging to class \textit{k} the sensitivity $S_{C, k, l}(x_i)$ which constitutes the directional derivative of a class \textit{k} in the direction of the CAV $v_C^l$ : 

\begin{equation}
\begin{split}
	S_{C, k, l}(x_i) & =   \frac{\partial h_{l, k}(x_i)}{\partial v_C^l} \\
						&  =   \nabla h_{l, k}(f_l(x_i)) * v_C^l,
\end{split}
\end{equation}
In a second step, the TCAV score is built as the fraction of data points $x_i$ that belong to class $k$, that have a positive sensitivity  $S_{C, k, l}(x_i)$: 

\begin{equation}
TCAV_{C, k, l} = \frac{ \lvert  \{x \in X_k : S_{C, k, l}(x) > 0 \} \rvert }{ \lvert X_k \rvert } 
\end{equation}
It provides a global score that allows assessing the importance of concept $C$ at layer $l$ of the model for the prediction of class $k$.

\heading{Saliency maps}
Saliency maps \cite{simonyan2013deep} infer attribution simply from input gradients of the corresponding class outputs. Denoting the model's output logits for class $k$ by $F_k(x)$, the saliency for input sample $x$, we define saliency maps as \cite{simonyan2013deep}
\begin{equation}
S_k(x)=\left|\frac{\partial F_k}{\partial x}\right|
\end{equation}

\heading{GradCAM}
GradCAM \cite{Selvaraju2019} takes a different approach in that it does not use the gradient information directly as attribution  maps, but rather to calculate weights for the activation feature maps. To this end, on a given convolutional layer, for every $K$ feature maps, it computes the activation gradient $\frac{\delta y^c}{\delta A^k}$ of the class $c$ we try to explain. Then, it calculates for every feature map $A^k$ a weight $\alpha^k$ by simply applying a global mean pooling to the gradient:
\begin{equation}
	\alpha^k_c = \frac{1}{Z} \sum_{i} \sum_{j}  \frac{\delta y^c}{\delta A^k_{i, j}}\,,
\end{equation}
where $Z$ corresponds to the number of entries in the feature map. The attribution map is then given by the ReLU activation of a weighted mean of the feature maps:
\begin{equation}
	G =\text{ReLU}( \frac{1}{K} \sum_{k}\alpha^k_c A^k)
\end{equation}

\heading{Layer-wise relevance  propagation}
Layer-wise relevance propagation (LRP) \cite{Bach2015} propagates the model prediction $f(x)$ from output to input, assigning attributions to each neuron in the network, which are computed based on activations and weights of the same layer and attributions of connected neurons. A key idea of the method is that attributions are \textit{conserved} across layers, i.e., the sum of the attributions at each layer $k$ yields the model prediction $f(x)$:
\begin{equation}
\sum_i R^k_i = f(x)    
\end{equation}
The exact update procedure is dependent on the rule that is used. We use the \textit{$\epsilon$-rule} for all layers, which is formally equivalent to taking the product of input and input gradient (Gradient * Input) for models with only ReLU activations \cite{ancona2018towards} as those considered in this work:

\begin{equation}
    R_j = \sum_{k} \frac{a_j w_{jk}}{\epsilon \sum_{j}{a_jw_{jk}}}R_k
\end{equation}

We also tested the \textit{Z-Plus rule} for convolutional layers, which represents a default choice in the imaging domain:
\begin{equation}
    R_j = \sum_k \frac{z_{jk}^+}{ \sum_{k}{z_{jk}^+}}R_k
\end{equation}
with $z_{jk}^+=x_jw_{jk}^+$ and $w_{jk}^+=1_{w_{jk\geq 0}}w_{jk}$. Due to worse results in the sanity check, we omit this rule and use the \textit{$\epsilon$-rule} throughout the model.

\heading{Integrated Gradients}
Integrated Gradients (IG) \cite{sundararajan2017axiomatic} is a model-agnostic explainability method that computes attribution maps by integrating over multiple gradients. It starts with a baseline $\boldsymbol{b}$, which is chosen as an input with neutral prediction, e.g. a null tensor, which would in our case represent an ECG with no electrical activity. To explain the prediction of class $k$ for a given input ECG $x \in \mathbb{R}^{m \times n}$, IG integrates the gradients of the model prediction for class $k$ w.r.t to all interpolated data points between the baseline $\boldsymbol{b}$  and $\boldsymbol{x}$: 
\begin{equation}
	IG_{k, i}(\boldsymbol{x}) = (x_{i} - b_{i}) \times \int_{0 }^{1} \frac{\partial F_k(\boldsymbol{b} + \alpha \times  (\boldsymbol{x}-\boldsymbol{b}))}{\partial x_{i}} d \alpha 
\end{equation}
where $F_k(\boldsymbol{x})$ is the logit for a data point $\boldsymbol{x}$ for class $k$ and $IG_{k, i, j}(\boldsymbol{x})$ the explanation of class $k$ for the $i$-th timestep in input $\boldsymbol{x}$. The full explanation is then given by:
$IG_k(\boldsymbol{x})$, which is also called \textit{attribution map}. It assigns an attribution score to each timestep $i$ in the input $\boldsymbol{x}$. By superimposing the attribution map on the ECG, we can mark the relevant areas in the ECG and thus simplify the interpretation of the diagnosis.

\end{document}